%% file: main.tex
\pdfoutput=1

\documentclass[11pt]{article}
\usepackage{jheppub}

\usepackage{bm}

\usepackage[usenames,dvipsnames,svgnames,table]{xcolor}

\usepackage{amsmath}
\usepackage{amssymb}
\usepackage{graphicx}
\usepackage{grffile}    
\usepackage{slashed}
\usepackage{soul}
\usepackage{multirow}
\usepackage{pdflscape}
\usepackage[section]{placeins} 
\usepackage{lineno}

\graphicspath{{plots/}}

\title{
\begin{center}
New Technologies For Discovery \\ 
\qquad \\ 
A report of the 2018 DPF Coordinating Panel for
Advanced Detectors (CPAD) Community Workshop
\end{center}
}

\author[a]{Z.~Ahmed,}
\author[b]{A.~Apresyan,}
\author[c]{M.~Artuso,}
\author[d]{P.~Barry,}
\author[e]{E.~Bielejec,}
\author[ah]{F.~Blaszczyk,}
\author[f]{T. Bose,}
\author[b]{D.~Braga,}
\author[ai]{S.A.~Charlebois,}
\author[j]{A.~Chatterjee,}
\author[g]{A.~Chavarria,}
\author[a]{H.-M.~Cho,}
\author[h]{S.~Dalla Torre,}
\author[ar]{M.~Demarteau$^*$,}
\author[v]{D.~Denisov,}
\author[ag]{M.~Diefenthaler,}
\author[a]{A.~Dragone,}
\author[b]{F.~Fahim,}
\author[af]{C.~Gee,}
\author[d]{S.~Habib,}
\author[a]{G.~Haller,}
\author[i]{J.~Hogan,}
\author[j]{B.J.P.~Jones,}
\author[k]{M.~Garcia-Sciveres,}
\author[v]{G.~Giacomini,}
\author[an,p]{K.~Gilmore,}
\author[al]{G.K.~Giovanetti,}
\author[b]{D.~Glenzinski,}
\author[l]{S.~Gleyzer,}
\author[ad] {A.H.~Goldan,}
\author[m]{S.~Gollapinni,}
\author[k]{C.~Grace,}
\author[n]{R.~Guenette,}
\author[b]{O.~Gutsche,}
\author[o]{U.~Heintz,}
\author[aj]{S.A.~Hertel,} 
\author[ac]{N.~R.~Hutzler,}
\author[f]{S.~Kolkowitz,}
\author[ak]{T.~Kovachy,}
\author[am]{F.~Leonard,}
\author[b]{R.~Lipton,}
\author[b]{M.~Liu,}
\author[l]{J.F.~Low,}
\author[s,k]{P.~Madigan,}
\author[aq]{S.~Malik,}
\author[p]{J.~Mates,}
\author[k]{Y.~Mei,}
\author[b]{P.~Merkel,}
\author[b]{T.~Mohayai,}
\author[v]{A.~Nomerotski,}
\author[q]{E.~Oliveri,}
\author[f]{K.~Palladino,}
\author[r]{E.~Pantic,}
\author[b]{A.~Para,}
\author[aa]{K.~Perez,}
\author[s]{M.~Pyle,}
\author[q]{P.~Riedler,}
\author[q]{L.~Ropelewski,}
\author[t]{R. Rusack,}
\author[i]{M.~Schleier-Smith,}
\author[as]{I~Shipsey$^*$,}
\author[u]{K.~Scholberg,}
\author[af]{B.~A.~Schumm,}
\author[v]{A.~Slosar,}
\author[f]{W.~Smith,}
\author[w]{B.~Surrow,}
\author[ah]{A.~O.~Sushkov,}
\author[k]{A.~Suzuki,}
\author[ae]{M.~Szydagis,} 
\author[ak]{D.~Temples,}
\author[x]{J.~Thom,}
\author[y]{M.~Titov,}
\author[ao]{L.~Tvrznikova,}
\author[o]{E.~Usai,}
\author[z]{R.~Van~Berg,}
\author[s]{V.~Velan,}
\author[c]{D.W.~Whittington,}
\author[aa]{L.~Winslow,}
\author[ab]{T.~Wongjirad,}
\author[ap]{Q.~Xia}
\author[d]{J.~Xie,}
\author[f]{Z.F.~You,}
\author[q]{A.~Zani,}
\author[d]{J.~Zhang,} 
\author[ac]{R.Y.~Zhu.}

\affiliation[*]{{\bf co-chair CPAD and corresponding editor}}
\affiliation[\ ]{\quad}
\affiliation[a]{SLAC National Accelerator Laboratory} 
\affiliation[b]{Fermilab, P.O. Box 500, Batavia, IL 60510}
\affiliation[c]{Syracuse University, 900 South Crouse Ave., 
Syracuse, NY 13244}
\affiliation[d]{Argonne National Laboratory, 9700 S. Cass Ave, Lemont, IL 60439}
\affiliation[e]{Sandia National Laboratory, 
P.O. Box 5800, Albuquerque, NM 87185}
\affiliation[f]{University of Wisconsin-Madison, Madison, WI 53706}
\affiliation[g]{University of Washington}
\affiliation[h]{INFN, Trieste}
\affiliation[i]{Stanford University}
\affiliation[j]{University of Texas at Arlington}
\affiliation[k]{Lawrence Berkeley National Laboratory, Berkeley, CA, USA}
\affiliation[l]{University of Florida at Gainesville}
\affiliation[m]{University of Tennessee at Knoxville}
\affiliation[n]{Harvard University, Cambridge, MA, USA}
\affiliation[o]{Brown University, Providence, RI, USA}
\affiliation[p]{NIST, Boulder, CO, USA}
\affiliation[q]{CERN, Geneva, Switzerland}
\affiliation[r]{University of California, Davis, CA, USA}
\affiliation[s]{University of California, Berkeley, CA, USA}
\affiliation[t]{University of Minneapolis, 116 Church Street S.E., Minneapolis, MN 55455}
\affiliation[u]{Duke University, Durham, NC, USA }
\affiliation[v]{Brookhaven National Laboratory, Brookhaven, NY, USA}
\affiliation[w]{Temple University}
\affiliation[x]{Cornell University, 144 East Ave., Ithaca, NY 14853}
\affiliation[y]{CEA, Saclay}
\affiliation[z]{University of Pennsylvania, Philadelphia, PA, USA}
\affiliation[aa]{Massachusetts Institute of Technology, 77 Massachusetts Avenue, Cambridge, MA, USA}
\affiliation[ab]{Tuft University, Medford, MA, USA}
\affiliation[ac]{California Institute of Technology, 1200 E California Blvd, Pasadena CA 91125}
\affiliation[ad]{Stony Brook University, 100 Nicolls Rd, Stony Brook NY 11794}
\affiliation[ae]{University at Albany, SUNY 1400 Washington Av Albany NY 12222}
\affiliation[af]{Santa Cruz Institute for Particle Physics, University of California at Santa Cruz, 1156 High Street, Santa Cruz, CA 95064}
\affiliation[ag]{Thomas Jefferson National Accelerator Facility, Newport News, VA, USA}
\affiliation[ah]{Boston University, Boston, MA, USA}
\affiliation[ai]{Universite de Sherbrooke, Sherbrooke, QC, Canada}
\affiliation[aj]{U.~Massachusetts, Amherst, Massachusetts}  
\affiliation[ak]{Northwestern University, Evanston, IL, USA}
\affiliation[al]{Princeton University, Princeton, NJ 08544, USA}
\affiliation[am]{Sandia National Laboratories, Livermore, CA, USA}
\affiliation[an]{University of Colorado, Boulder, CO, USA}
\affiliation[ao]{Lawrence Livermore National Laboratory, Livermore, CA, USA}
\affiliation[ap]{Yale University, CT, USA}
\affiliation[aq]{University of Puerto Rico Mayaguez}
\affiliation[ar]{Oak Ridge National Laboratory, Oak Ridge, TN, USA}
\affiliation[as]{Oxford University, Oxford, UK}

\emailAdd{demarteau@ornl.gov}
\emailAdd{ian.shipsey@physics.ox.ac.uk}

\abstract{
\newline 
For the field of high energy physics to continue to have a bright future, priority within the field must be given to investments in the development of both evolutionary and transformational detector development that is coordinated across the national laboratories and with the university community, international partners and other disciplines. While the fundamental science questions addressed by high energy physics have never been more compelling, there is acute awareness of the challenging budgetary and technical constraints when scaling current technologies. Furthermore, many technologies are reaching their sensitivity limit and new approaches need to be developed to overcome the currently irreducible technological challenges. This situation is unfolding against a backdrop of declining funding for instrumentation, both at the national laboratories and in particular at the universities. This trend has to be reversed for the country to continue to play a leadership role in particle physics, especially in this most promising era of imminent new discoveries that could finally break the hugely successful, but limited, Standard Model of fundamental particle interactions. In this challenging environment it is essential that the community invest anew  in instrumentation and optimize the use of the available resources to develop new innovative, cost-effective instrumentation, as this is our best hope to successfully accomplish the mission of high energy physics. This report summarizes the current status of instrumentation for high energy physics, the challenges and needs of future experiments and indicates high priority research areas. 

}

\vfill
\break 
\vfill
\break

\begin{document}

\begin{flushright}
\small{.}
\end{flushright}

\maketitle
\flushbottom

\medskip
\input Exec.tex
\input Introduction.tex
\input Science.tex

\input Technologies.tex

\input Conclusions.tex
\input ack.tex

\bibliographystyle{JHEP.bst}
\input{biblio.tex}

\end{document}

%% file: Exec.tex
\section{Executive Summary}
\label{s.exec} 

The Coordinating Panel for Advanced Detectors (CPAD) of the Division of Particles and Fields of the American Physical Society convenes once per year to evaluate the research directions in instrumentation in support of the High Energy Physics science mission within the twenty-year P5 vision. The P5 report, issued in 2014, recommended to focus resources toward directed instrumentation R\&D in the near-term for high-priority projects.
As the technical challenges of these high-priority projects were being met, a balanced portfolio of short-term and long-term R\&D was recommended to the extent that funding would allow. The field is now at a point where most high-priority projects of the field are ballistic and vigorous renewed investment in instrumentation is required to position the U.S. for a sustained long-term leadership position in particle physics. 
To gather wide community input to inform the study of future priority directions for instrumentation, CPAD has established a series of annual community workshops. The inaugural workshop was held at the University of Texas, Arlington in 2015, this was followed by workshops at Caltech in
2016 and the University of New Mexico in 2017. The fourth and latest in the series was held at Brown University, December 9-11, 2018. This report summarizes the findings and suggested research directions from that workshop. 

The challenges facing the next generation of experiments are exceedingly demanding. One contributor to this situation is the fact that many of the current generation experiments are reaching their design sensitivity deploying current technologies. Another is that proposed facilities, at all three frontiers, create experimental conditions that are not met by current instrumentation capabilities. An overarching conclusion of the workshop is that support for instrumentation needs a significant boost in funding for the US to maintain a leadership role in particle physics, and that the emphasis has to be on detector development for future experiments with a long-term time horizon with a substantial effort directed towards the exploration of novel ideas. 

The recent investments in quantum sensing create a unique opportunity for particle physics to exploit the ability to control full systems at the quantum level to provide sensitivities that go far beyond what is currently achievable. It is a fortunate confluence of the investment in quantum sensing techniques and the fact that many high priority projects are reaching their design sensitivities that allows the exploration of areas of parameter space that were previously thought to be inaccessible. 

Although different frontiers have different requirements, common themes emerged for future detectors. Many experiments require low-energy threshold and low-background detectors to overcome the current sensitivity limitations. Directional detectors for dark matter and low-energy neutrinos would open up completely new research pathways in the study of rare processes. Much higher granularity is also a common requirement, be it low-mass pixel detectors, large-volume liquid argon detectors or sampling calorimeters. Photodetectors are ubiquitous in the field and cost-effective, large-area devices with pico-second timing could fundamentally change the research approach of various research areas. 

The critical needs from a science driver and a technology perspective are described in this report. Mapping these critical needs onto the potential future direction of the field, noting that the development of new technologies is a decade-long research program at least, leads to the clear conclusion that a significantly increased investment in instrumentation is required to successfully execute the program for the next decade, and to be in a position to mount the next generation of particle physics experiments.



%% file: Introduction.tex
\section{Introduction}
\label{s.intro} 

Instrumentation is the great enabler of science both pure and applied. Instrumentation is
central to the mission and culture of High Energy Physics, which is to explore the
fundamental nature of energy, matter, space and time. The field is in the midst of a golden age with the discovery of the Higgs boson in 2012, gravitational waves in 2016 and with new experiments being planned at existing and proposed new accelerators, deep underground, at some of the most remote places on earth and in space. Together these experiments will study the origin of mass, may explain the matter anti-matter asymmetry of the universe, search for extra spatial dimensions, determine the nature of dark matter and dark energy, and may probe the Planck scale. For the very first time we
may come to know how our universe was born, how it will evolve, and what its ultimate fate will be.

However, we embark on this adventure of discovery with instrumentation that, while representing a towering achievement, often is a scaled-up version of techniques used in the past. We have, for example huge accelerators serving large detectors whose costs outstrip the available funding. 
The results are often projects with very long construction time scales, often necessitating a de-scoping and reduction in physics reach to contain costs.
In addition, the time scales for our experiments and large collaborations may insulate us from instrumentation advances and innovations in industry.

Instrumentation R\&D has the power to transform this situation, from novel acceleration techniques such as plasma wake-field acceleration, to new detectors that provide enhanced capabilities with significantly reduced cost.
However, there has been a serious decline in DOE and NSF funding for instrumentation
research and development during the last two decades at universities and national
laboratories. If this funding trend is not reversed, declining capabilities will surely lead to a dramatic change in how our field functions, and we will confront a different kind of future for HEP: the golden age of discovery will be stalled and its goals unfulfilled. Energy, matter, space, and time will remain enigmas.

All science, but particularly HEP, would clearly benefit from
the development of both evolutionary and transformative detector instrumentation that is coordinated across the national laboratories and with the university community and international partners and with other disciplines. Instrumentation R\&D is inherently necessary to our scientific
future~\cite{P5-report}. 

This report is organized as follows: we consider the detector R\&D that will be needed, first arranged by the P5 Science Driver questions that motivate the R\&D program, and then by the detector, TDAQ, and electronics technologies that are needed to achieve them.  

%% file: Science.tex
\section{The P5 Science Drivers}
\label{s.science} 

\input{./Science/Higgs.tex}

\input{./Science/Neutrino.tex}

\input{./Science/DarkMatter.tex}
\input{./Science/Cosmic.tex}

\input{./Science/Unknown.tex}

%% file: Science/Higgs.tex
\subsection{The Higgs Boson as a Tool for Discovery} 
\label{higgs}
\subsubsection{Introduction}

In 2012, the ATLAS and CMS experiments at the LHC, discovered a new particle whose properties are consistent with standard model (SM) predictions for the Higgs boson. The mass of the new particle has been measured to be 125 GeV~\cite{mHATLAS,mHCMS}. Given this mass, the SM predicts all properties of the Higgs boson without any free parameters. Any deviations would be evidence for new physics. 

Moreover, the measurements of the properties of the Higgs boson have achieved remarkable precision. For example the mass is known to a precision of 0.2\% and measurements of the Higgs couplings have achieved an overall precision at the level of 10\% ~\cite{Hcouplings}. Direct constraints on the couplings to third generation fermions have been obtained from the recent observations of Higgs boson decays to $\mathrm \tau$ leptons~\cite{HtautauCMS, HtautauATLAS}, $\mathrm{t\overline{t}H}$~\cite{ttHCMS, ttHATLAS} production and $\mathrm H\rightarrow \mathrm b\overline{\mathrm b}$~\cite{HbbATLAS, HbbCMS} decays. ATLAS and CMS are now aiming for the couplings to second generation fermions by searching for $\mathrm H\rightarrow \mu^+\mu^-$ and $\mathrm H\rightarrow \mathrm c\overline{\mathrm c}$ decays. ATLAS and CMS are also probing the existence of additional Higgs bosons, which would indicate a more complex Higgs sector than described by the standard model. The Higgs boson thus offers a rich portal to new physics and the Higgs physics programs must aim to measure its properties with the highest achievable precision while also searching for signatures of non-standard Higgs boson production and decay.

The LHC is the facility to study the Higgs boson in the next decade. With the data from the HL-LHC we will measure the Higgs couplings to a few percent accuracy, its mass to 50 MeV, its spin-parity structure, and probe rare decays, measure Higgs tri-linear couplings to 50\%, and search for an extended Higgs sector.

At future high-energy $e^+e^-$~colliders, circular or linear, we can measure the total ZH cross-section, which provides an absolute normalization of the Higgs couplings and gives access to a measurement of the total width and invisible decays. At an $e^+e^-$~collider one can measure most couplings to sub-percent precision and the Higgs tri-linear coupling to 5-10\%.

A 100 TeV pp collider would be able to measure the Higgs tri-linear coupling to 8\%.
All facilities can exploit the Higgs boson as a tool for discovery.

\subsubsection{Critical Needs}

Detectors at all facilities must be able to reconstruct all particles with high efficiency and resolution, and with low fake rate.
The $\mathrm{H \rightarrow \gamma\gamma , ZZ, WW, \mu\mu}$
decay modes require excellent electron, photon, and muon reconstruction.
The $\mathrm{H \rightarrow b\overline{b}}$ decay mode requires b-quark tagging and precision reconstruction of primary and secondary vertices.
The $H \rightarrow \tau\tau$ decay mode requires tracking of charged hadrons, measurement of electromagnetic energy, and muon and electron reconstruction. Measurement of $\mathrm{t\overline{t}H}$ production requires jet reconstruction and b-quark tagging. Searches for Higgs decays to $\mathrm{c\overline{c}}$ require development of inclusive charm tagging techniques.

\medskip
\noindent
{\bf LHC} \\
Higgs physics requires low trigger thresholds to capture the decay products of the W and Z bosons from Higgs decays with high efficiency. For example, for the measurement of Higgs boson properties in the four-lepton final state, CMS employs trigger thresholds for the transverse momentum of the leading and sub-leading leptons of about 20 and 10~GeV, respectively. The detection of Vector Boson Fusion requires acceptance for jets at high $|\eta|$. Techniques to detect boosted Higgs decays and complex final states are important for searches for new physics involving decays to final states with Higgs bosons.

The biggest challenge from the environment at the HL-LHC is the high pileup. On average there will be 140-200 additional interactions per bunch crossing. These are the largest source of hits in the tracking system, they add energy to calorimeter measurements, create additional (fake) jets, degrade the jet energy resolution, increase the execution time for event reconstruction and trigger, and increase trigger rates.

Detector elements will be exposed to high radiation doses and will suffer radiation damage, which degrades the signals and requires continuous calibration. It will also limit the lifetime of the detectors.

The ATLAS and CMS detector upgrades will maintain (or improve) physics event reconstruction in the much higher luminosity environment. The trackers will be radiation tolerant and have high granularity with significantly less material than the current trackers.
They will extend coverage to the forward region ($|\eta| < $4.0) for increased Vector Boson Fusion (VBF) Higgs physics acceptance and pileup mitigation. The calorimeters will be radiation tolerant, have high segmentation, and new electronics. Trigger upgrades include pushing track information into Level 1, boosted object triggers, and increasing the output rate to 7.5 kHz. Muon systems will increase coverage and improve performance to maintain low trigger thresholds. 

Precision timing capabilities promise to provide a new handle to mitigate pileup at pp colliders. Pileup interaction vertices are distributed in space and time. Particles from pileup interactions can be rejected both by the region along the z-axis from which they originate and by their time difference relative to the signals from the hard scatter. This requires a time resolution of 10-30~ps. Both ATLAS and CMS are planning to incorporate timing detectors in their HL-LHC upgrades and have published technical proposals~\cite{ATLASTiming, CMSTiming}. Their exact implementation still requires some R\&D. 

\medskip
\noindent
{\bf $e^+e^-$~facilities} \\ 
Possible facilities include the ILC with center of mass energy of 200-500 GeV, upgradeable to 1 TeV. The beams are structured into trains of 1315/2625 bunches with 530/270 ns bunch spacing and pulsed at 5-10 Hz. This allows trigger-less readout and power pulsing so that no cooling is required, making possible super-light trackers with 0.0015 X0/layer and dense calorimeters.

CLIC offers higher collision energies, from 350~GeV to 3~TeV and beams would be structured into trains of 312 bunches, 0.5 ns bunch spacing, pulsed at 50 Hz. Power pulsing at this pulse frequency is being studied; it is noted that power pulsing at 120 Hz was already used effectively at SLD. The higher energy brings with it large beamstrahlung background, which can be dealt with. 

Circular $e^+e^-$~colliders (CepC, FCC-ee) with collision energies of 90-350 GeV feature continuous operation with 10 ns – 10~$\mu$s bunch spacing. Power pulsing would not be possible and thus the needed cooling would add material to the detectors. 

The ambition to achieve superior precision in measurements of Higgs couplings puts the onus on control of systematics. While the technologies are generally well understood there are aspects that could still benefit from R\&D, such as the impact of particle ID on flavor tagging, how to correct for the neutral hadron fraction in jet calibration, limitations from clustering effects on jet energy resolution, reconstruction and ID of low momentum particles and the stability of alignment and calibration. 

Improvements in flavor tagging capabilities (e.g. better efficiencies using advanced machine learning techniques) are needed to measure the $H \rightarrow c \overline{c}$ decay mode. Superior jet energy resolution is required to distinguish hadronic W and Z decays, for example to measure the Higgs self-coupling. 

\medskip
\noindent
{\bf 100 TeV pp colliders} \\
Higgs physics at a 100 TeV collider presents many challenges. In addition to the demands at the LHC, the FCC detectors will need to handle increasingly boosted decays resulting in the requirement to separate particles with opening angles on the order of one degree. 

For tracking, 5~$\mu$m pixel sizes are required.  Muon p$_{\rm T}$ measurements will be challenging and require combination of muon spectrometer measurements with a high-granular pixel tracking system. A deep hadron calorimeter (12~$\lambda_0$) is required to contain showers. The calorimeters must be highly granular with cell sizes below 2 cm for electromagnetic calorimeters and below 5 cm for hadron calorimeters with longitudinal segmentation for 3D clusters and they must tolerate fluences up to $10^{18}/{\rm cm}^2$.
The dynamic range of the readout has to be extended by a factor 10 and extended coverage to $|\eta| < 6$ is desired as decay products of the relatively light Higgs boson are boosted into the forward and backward directions.

\medskip
\noindent
{\it Comments}\\
The ATLAS and CMS detectors are being upgraded to deal with the HL-LHC requirements. The challenge is pileup mitigation, high $|\eta|$ acceptance, and low trigger thresholds. 

For detectors at the ILC, the basic technologies exist, but a large development effort is still required before a technical design report (TDR) could be written. It is noted that the machine is already at the level of a TDR. Detectors at CLIC and circular colliders face more challenges, especially relating to cooling super-light tracking detectors. The critical issue is to control the systematics in a high performance detector to reach the ultimate measurement precision. 

A 100 TeV pp collider pushes the envelope of technology in many aspects. Here, new ideas are an absolute requirement to meet the challenge of increasing the dynamic range for all measurements.

\subsubsection{Future Direction} 

Areas in which R\&D is needed for LHC detectors include precision timing measurements. For e+e- facilities, optimization of physics performance by integrating superior resolution detectors with improved reconstruction algorithms (e.g. for leptons/photons, jets and flavor-tagging) is important. R\&D into low-mass cooling systems will be critical for CLIC and circular lepton colliders. For very high-energy hadron colliders, high granularity, calorimeter containment, and forward acceptance at reasonable costs are important
R\&D goals.  

Advances in the following areas could prove transformative:
\begin{itemize}
\item Precision timing measurements in all sub-detector systems for pileup mitigation
\item Low mass cooling systems
\item Highly granular low mass tracking detectors 
\item Granular calorimeters for multi-TeV showers 
\end{itemize}
Achieving these will require new technologies, for example 28nm IC technology or monolithic pixels for integrated readout and on-detector processing.

%% file: Science/Neutrino.tex
\subsection{The Physics Associated with Neutrino Mass} 
\label{neutrino}
\subsubsection{Introduction}

The ``Physics of Neutrino Mass'' science driver addresses questions such as: What are the neutrino masses? What is the mass ordering? What is the origin of neutrino mass? Do neutrinos and antineutrinos oscillate differently? Are there new types of neutrinos or new kinds of neutrino interactions? Are neutrinos their own antiparticles? Answering these questions could lead to the discovery of physics beyond the standard model, unravel potential new symmetries, offer constraints to grand unification theories and provide crucial inputs to cosmological and astrophysical models. To search for these answers, experiments over a wide range of technologies and scales are needed~\cite{deGouvea:2013onf}. 

Neutrino oscillation experiments are designed to resolve the two remaining major unknowns of the three-flavor paradigm---  the neutrino mass ordering and the CP-violating phase $\delta$.  Furthermore, they will make precision measurements of all of the mixing parameters, thereby testing the three-flavor paradigm and searching for signs of new physics beyond the standard model.  Long-baseline oscillation experiments require very large-scale detectors (multi-kiloton) and powerful neutrino beams.
Large-scale detectors, made of argon, water or scintillator, are typically multi-purpose.   Some serve as targets for long-baseline neutrino beams. They all also observe neutrinos from astrophysical and other natural sources--- from radioactivity in the Earth, the Sun, the atmosphere, core-collapse supernovae, and more exotic astrophysical sources.
Such observations will bring knowledge both about the neutrino sources and about neutrinos themselves. Short-baseline neutrino experiments search for sterile neutrinos using different sources (accelerators or reactors) and usually require more modest detector sizes (ton scale to few hundred ton scale). Detailed understanding of neutrino-nucleus interactions over a wide range of energies is needed to fully address many of the questions, and dedicated detectors are required for cross-section studies.    

Additional experiments are hunting for neutrinoless-double-beta-decay (NLDBD), the observation of which will shed light on the Majorana vs. Dirac nature of the neutrino~\cite{Dolinski:2019nrj}.  Other experiments are looking for the kinematic signature of non-zero neutrino mass 
While NLDBD and kinematic endpoint experiments fall under the purview of the Office of Nuclear Physics, the results of these experiments will have enormous impact on High Energy Physics. Many of the detector technologies have direct overlap with High Energy Physics experiments.   


\subsubsection{Critical Needs}

\noindent
{\bf New technologies for neutrino oscillation physics}\\
The U.S. neutrino community is focusing significant attention on liquid argon time projection chambers (LArTPCs) as the main detector technology for long-baseline experiments and some short-baseline experiments. The Short-Baseline Neutrino (SBN) Program at FNAL will use three large-scale LArTPCs~\cite{Antonello:2015lea} and the Deep Underground Neutrino Experiment (DUNE) will have four 10-kton LArTPC modules underground in South Dakota~\cite{Acciarri:2015uup}. LArTPCs are also powerful tools for the study of neutrino interactions.

While the LArTPC detector technology will be widely used by the community, there remain several technical challenges associated with these detectors.  If these challenges are not met, the physics reach of the U.S. program will be jeopardized. The challenges of LArTPCs are being addressed by dedicated efforts, synergies between the different experiments are being exploited, and experience is being shared. The key challenges are high-voltage delivery and stable operation, cold electronics design and operation, light detection systems and automated event reconstruction. Novel pixel-based charge readout is also under development.  This approach is promising both for near detectors and potentially for large far detector LArTPC modules. While the ProtoDUNEs~\cite{Abi:2017aow,Scarpelli:2019kah},  dedicated DUNE prototypes located in a particle beam at CERN, are currently addressing a significant fraction of the challenges, further R\&D will be highly beneficial.

Furthermore, given the scale of DUNE and its long-term schedule, it is of utmost importance that detector improvements are investigated to enhance the scientific impact of this ambitious project. Reducing the detection thresholds, increasing the signal efficiencies and reducing the background contamination, and improving the light detection are among the areas under study. New detector technologies could offer increased scientific potential and could be implemented in the later modules.  Improved detector capabilities will enhance also the broader physics goals of DUNE, such as solar and supernova neutrinos (which require sensitivity to neutrinos of a few tens of MeV or less) and baryon-number violating processes.

The U.S. is also engaged in other international efforts using different detector technologies such as water Cherenkov (Hyper-K~\cite{Abe:2011ts}) or scintillating materials (e.g., JUNO~\cite{An:2015jdp}, THEIA~\cite{Gann:2015fba}). These types of experiments will greatly benefit from developments of photon sensors.

Several smaller-scale experiments are also planning to study short-baseline oscillation (at reactors or with novel sources), to search for sterile neutrinos and address the numerous anomalies in this field. R\&D for new detector technology could make a significant impact in these types of future searches.

\medskip
\noindent 
{\bf New technologies for neutrino interaction physics }\\
Detailed understanding of neutrino interactions is critical, both in the few-GeV regime of relevance for interpretation of neutrino oscillation experiments, and in the $<100$~MeV regime of relevance for supernova and other natural neutrino sources.  LArTPCs are useful for the study of neutrino interactions, but dedicated experiments may be needed to provide a deeper understanding of nuclear models. Technology challenges are similar to those for oscillation measurements, where the goal is precision reconstruction of the interaction final state. Several experiments dedicated to neutrino interaction physics are currently underway and others are planned; some of these also address detector R\&D issues.  

Coherent elastic neutrino-nucleus scattering (CEvNS) of astrophysical neutrinos creates a background floor for dark matter experiments, and at the same time represents an astrophysical neutrino signal in large dark matter detectors~\cite{Billard:2013qya}. The measurement of CEvNS using pion decay-at-rest~\cite{Akimov:2017ade}, reactor and radioactive sources provides a standard model test and probe of new physics (and potentially could be of use for sterile neutrino oscillations). Large, low-energy threshold and low-background detector technology is required for these. Development of detectors with the ability to reconstruct the direction of nuclear recoils is desirable for both neutrino and dark matter physics in these detectors.

\medskip
\noindent
{\bf New technologies for neutrino astrophysics }\\
There is unique physics to be learned by using neutrinos from many astrophysical (and geological) sources: the Sun, supernovae, relic supernovae, exotic astrophysical sources, and cosmic big-bang neutrino background. Neutrino detectors sensitive to these sources address particle physics in addition to astrophysics questions.

Astrophysical neutrino sources of interest span an extremely wide range of energies, from sub-eV to EeV. Technologies for astrophysical neutrinos are correspondingly diverse, and there is some overlap with oscillation experiments. For many topics, one needs large, homogeneous detectors (based on liquid argon, water, scintillator, water-based scintillator).

Many of these technologies are limited by the costs of the active mass and of the photodetectors. Good timing resolution is desirable, but not the most important characteristic for this kind of detector; low cost per photon for large-area coverage would have the most impact. Photodetectors that work well at cryogenic temperatures are also highly desirable.

Astrophysical neutrino detectors tend to be multi-purpose, serving also as long-baseline beam targets and baryon-number-violation search experiments (and even as neutrino-less double beta decay (NLDBD) experiments, when doped with relevant isotopes). Therefore, dedicated efforts should be made to ensure that future large-scale experiments take advantage of new detector technologies to expand their physics reach to astrophysical neutrinos.

\medskip
\noindent
{\bf New technologies for neutrino properties }\\
To probe the absolute neutrino mass scale, KATRIN, using Magnetic Adiabatic Collimation combined with an Electrostatic Filter (MAC-E filter) technology, represents the last generation of its type; there are new ideas to go beyond the currently planned precision measurements, such as Cyclotron Radiation Emission Spectroscopy (CRES)~\cite{Monreal:2009za}. 

NLDBD searches are the only plausible near-term way of answering the fundamental Majorana-vs-Dirac question. There are multiple approaches using different isotopes, all of which have quite daunting background challenges. 

Stewardship of both neutrino mass and NLDBD experiments is provided by DOE Office of Nuclear Physics, but there are strong overlaps with the Office of High Energy Physics and synergetic support between the two should be encouraged. The impact of these two topics on the field of HEP could be game-changing and the expertise in HEP instrumentation is directly relevant to these experiments. Therefore, the instrumentation efforts should encompass some of these topics. 

Other physics topics, such as the neutrino magnetic moment, can also be probed with large, low-energy threshold and low-background experiments.

\medskip
\noindent 


\subsubsection{Future Directions}

The first set of R\&D needs are those relevant to DUNE. Remaining technical challenges need to be addressed to ensure that this ambitious project delivers on its exciting physics goals. There are numerous generic R\&D challenges to address in neutrino physics that will result in incremental (but potentially critical for experiment success) improvements. One always wants large mass, low background, segmentation/resolution, high efficiency, good particle identification and final state reconstruction. Further enhancement of the scientific reach of DUNE should therefore be explored as the impact could be highly significant over the course of a decade. Advances in the following areas are required:

\begin{itemize}
    \item{High voltage delivery}
    \item{Cold electronics design}
    \item{Calibration systems}
    \item{Automated event reconstruction}
\end{itemize}

The second set of needs relates to the many complementary and diverse experiments addressing multiple questions in neutrino physics.
New ideas for small scale experiments and generic technology development should continue to be encouraged and supported. We suggest that the agencies find some way to best exploit the expertise of both HEP and NP community researchers to address the NLDBD physics question. The Cosmic Frontier and Intensity Frontier connections for technologies for dark matter and neutrino physics should also be exploited. The following advances could have strong impact:

\begin{itemize}
    \item{Development of large, low-energy threshold and low-background detectors}
    \item{Directional detectors for low-energy neutrinos }
\end{itemize}

The last set of needs, which connects to the two previous ones, is to try to expand the scientific scope of future very-large-scale experiments. For example, a large mass (perhaps ton-scale or greater) sub-keV threshold, low background nuclear recoil detector would be multi-purpose, serving sensitive dark-matter searches as well as for astrophysical neutrino detection and new physics probes via neutrino scattering. 
As an example of a specific effort, large-area, inexpensive, efficient photosensors would be a game-changer for large, homogeneous detectors for neutrino oscillations and neutrino astrophysics. Advances in the following areas could have strong impact:

\begin{itemize}
    \item{Development of low-cost, large-area and efficient photosensors}
    \item{Pixelization of large detectors and low-noise readout electronics}
    \item{Doping of the detector medium}
\end{itemize}


%% file: Science/DarkMatter.tex
\subsection{The nature(s) of dark matter}
\label{darkmatter}
\subsubsection{Introduction}  

The exploration of the nature of dark matter cuts across many technologies.  There is ample astronomical evidence for dark matter from gravitational interactions, but its particle properties are completely unknown, requiring exploration of an enormous range of possible particle masses and interaction parameter space.  Therefore, models are required to reduce the search window, as well as direct our technology choices.  Promising candidates are Weakly Interacting Massive Particle (WIMPs) and Axions, but there may be multiple contributors and other possibilities cannot be neglected. The avenues to search for dark matter are direct and indirect detection, as well as production in colliders, with each one having its own set of challenges. 

Direct detection experiments search for WIMPs through their scattering on various target nuclei. The assumed weak interaction cross-section and tiny recoil energies pose huge challenges of stringent background reduction and rejection. Current experiments typically measure multiple signals in order to discriminate against background and identify the nuclear recoils as the partitioning of energy deposited in heat/phonons, ionization and scintillation light differ depending on the initial interaction.  The recoil energy distribution also depends on the type of nuclei, which can help identify not only the nature of the interaction, but also constrain the mass of the particles.  Hence, complementarity in target nuclei is extremely important in dark matter experiments. 

Eventually, as direct detection becomes sensitive to smaller WIMP nucleon cross-sections, neutrinos from various sources become the limiting background, since they undergo the same type of coherent scattering.  At this point, it will be necessary to subtract the neutrino background using the predicted Standard Model interaction rates and spectra\footnote{The neutrino background is also a signal; see Section~\ref{neutrino}}.  For some WIMP masses and interaction models, the spectral shape will be very similar, and complementary targets will help disentangle both the interaction model and the neutrino background.   Once near the “neutrino floor”, directional detection and annual modulation can provide important neutrino background rejection.  Stability, background control, and careful monitoring are very important to the success of annual modulation in the general-purpose WIMP search detectors. Special-purpose detectors are required for directional detection, which is sensitive to the Earth’s rotational orientation relative to the WIMP “wind”. The greatest advantage comes with head-tail tracking discrimination, which requires gaseous targets and consequently large volumes. There is an active R\&D effort to make this technology practical.  

Axion dark matter represents a very different type of particle, with much lower mass and a different coupling mechanism, requiring less emphasis on extreme reduction of radiogenic backgrounds. Axions are detected by converting them to photons in electromagnetic fields in microwave cavities or in solar axion telescopes.  Some experiments are searching laser light for signs of axion production and other technologies still are being explored. 

Indirect detection looks for stable final-state annihilation products from WIMPs, in the form of high-energy particles. Charged particle products diffuse in the galactic magnetic field before reaching detectors on earth and on satellites orbiting the earth.  WIMP annihilation into neutral particles, such as gammas and neutrinos, provide possible directional information that can also be compared with excess observations from charged particles. Associated astrophysical uncertainties from the most dense dark and regular matter regions, such as the galactic center, can make interpretation difficult, so a variety of indirect search techniques are employed. These include searches from dwarf spheroidal galaxies which have limited standard astrophysical production of high energy particles, searches for antimatter particles especially anti-protons, anti-deuterons and anti-helium which have limited astrophysical production, and searches for spectral gamma ray lines that can be interpreted as direct annihilation signals with gamma energies equal to the dark matter rest mass. There is a healthy program of ground-based and space-based detectors with a multi-messenger approach.

\subsubsection{Critical Needs} 

\noindent 
{\bf Dark Matter Direct Detection } \\ 
Detector development: As the search for lower WIMP nucleon cross-sections and larger WIMP mass parameter space continues, the pinch points are different for the leading technologies. The U.S. program for Generation-2 Dark Matter (DM) covers the expected WIMP mass range with the LZ experiment at higher masses and SuperCDMS (germanium and silicon) at lower masses. A second US-led and funded experiment, XENONnT and the Chinese PandaX experiment will provide multiple searches with the same liquid xenon TPC technology.  These technologies are working on lowering their threshold, a common issue with noble liquids.  Photodetectors with both low radioactivity and improved quantum efficiency for single UV photon sensitivity would be a game-changer. 

The cryogenic crystal technology has already proven a much lower threshold than xenon, but is modular, making it harder to scale up. It is cost efficient in the low mass region, where the sensitivity limits are as yet poorly explored.   It is also more flexible with respect to target and readout technology, possibly sharing multiple technologies within a single cryostat. The SuperCDMS experiment will go forward for the U.S. Generation-2 dark matter experiment at low mass, possibly sharing their facility eventually with European groups (EURECA (germanium) and CRESST (calcium tungstate and sapphire)).  Cryogenic crystal technology led the way in excellent nuclear recoil discrimination, but has relaxed this constraint in return for increased sensitivity to extremely low-energy deposition (and thus low-mass WIMP sensitivity).  Improvements in the technology to re-assert that discrimination, while still maintaining the lowest possible threshold, are the game changer for these experiments. Improvements in the energy resolution will make this possible.  

It is still important to support additional target nuclei, since there is no other way to disentangle the possible WIMP interaction models.  Once the hint of a signal is seen in one technology, it is also important to see confirmation using a different technology with different systematics and background sensitivities. Thus the R\&D programs in liquid argon, silicon CCDs, threshold detectors, and non-organic crystal arrays must be supported.  Otherwise, it would be possible to completely miss a non-standard WIMP interaction or be unable to interpret the appearance of a signal in one of the targets or technologies.  

Furthermore, beyond traditional searches for the nuclear recoil signal from WIMP interactions, recent advances in particle theory highlight new compelling realizations for the origin of dark matter and its detection. For example, the detection of electronic recoils with energies between a few eV to tens of eV allow for the search of dark matter particles with masses in the MeV to GeV range. Novel approaches to detect single ionized electrons in semiconductors and noble liquids offer a new opportunity to explore this mass range. An R\&D program to bring these technologies to maturity, particularly in their scalability and the mitigation of instrumental backgrounds, should be supported.

\medskip
\noindent 
{\bf Background Discrimination:} To reach required sensitivities, all direct detection dark matter experiments need continued improvements in shielding, material assay, and simulation (mostly better low-energy physics processes in the simulation models).  In the area of assay, the most important needs are access to radiopure materials and screening resources, as well as improved surface screening and new techniques in radon mitigation.   While a bilateral agreement was reached between LZ and SuperCDMS to share resources, this leaves out many of the smaller experiments, which are already finding access to screening facilities difficult.  Thus, shared assay resources, as part of a centrally managed (although not necessarily centrally-located) screening infrastructure, would be a game-changer. Surface screening at the sensitivity required for Gen-2 experiments does not yet exist and this R\&D is crucial for all WIMP dark matter experiments.  The same background concerns drive the double-beta decay experiments (Majorana - LEGEND, EXO, CUORE, etc).  A cross-division (NP and HEP) initiative that funds the many identified commonalities in background issues would be extremely helpful for all parties involved. 

Nuclear recoil signals from solar and atmospheric neutrinos will be seen with expected rates and spectra in future WIMP detectors of the aforementioned technologies. Directional detectors can provide discrimination against these backgrounds to WIMP induced recoils, but significant improvements in the target masses and thresholds for head-tail discrimination will be necessary to bring such technology to the forefront of the direct detection field.

\medskip
\noindent 
{\bf Low energy Calibration:} Robust nuclear recoil calibration is difficult, but required for confidence in the nuclear recoil scale.  This is especially true as thresholds are lowered and electron versus nuclear recoil discrimination becomes more difficult. 

\medskip
\noindent 
{\bf Indirect Detection } \\ 
The primary drivers for indirect detection technological advancement are improved angular resolution for point source identification (to reduce astrophysical backgrounds) and greater sensitivity to weak sources. More collecting area is needed, ideally 25~m$^2$sr acceptance, possibly through monolithic technologies producing bigger telescopes with these larger fields-of-view.  Instrumental R\&D will likely focus primarily on scaling existing technologies for use in future instruments, and reducing costs per channel, data volume and rate, and instrument infrastructure. Seed funding for detector R\&D from DOE is necessary to prove ideas for NASA proposals for space-based instruments.

\medskip
\noindent 
{\bf Axion Searches} \\ 
Large scale axion searches currently rely upon microwave cavities, lumped-element resonators and magnetic resonance technologies. Continued improvements in 
 microwave cavities need to be designed and developed: e.g. high-frequency, large-volume tunable systems - and their quantum limited (0.25 to 10~GHz) high-Q RF detectors. Superconducting Quantum Interference Devices (SQUIDs) and Josephson Parametric Amplifiers (JPAs) are essential for readouts. High field magnets are needed to increase axion conversion signals.
 
 Quantum sensors present opportunities to improve the sensitivities of these techniques, through state squeezing, background-free photon counting, and up-converting photons for low-mass axion searches. Cavity improvements with superconducting niobium resonators or resonators with dielectric inserts allowing for multi-wavelength resonances can extend the sensitivity of these searches.
 
Axion searches can also be advanced with new techniques in atomic interferometry and magnetometry, and torsion pendulums, and will be necessary to reach to lightest dark matter candidates.

\subsubsection{Future Direction}

Many of the technology components naturally fit into the other frontier reports, especially collider-based searches, so the technological aspects in this dark matter science driver section are not discussed.
In this section future directions in the areas of direct, indirect and axion searches for dark matter are described.  

\medskip
\noindent
{\bf Direct Detection}
\begin{itemize}
\item Large directional detectors will be necessary to reduce backgrounds near the neutrino floor in certain mass regions where the spectral shapes are similar. 
\item Signals need to be confirmed with different technologies.  Deciphering the physics of the interaction and narrowing down the mass of the particle require different targets.
\item Joint programs between High Energy Physics and Nuclear Physics would help both the NLDBD and dark matter detectors, improve background rejection by sharing information on radiopure materials and assay facility infrastructure. 
\item High purity germanium detectors, radiopure materials, shielding, sensitive photodetectors and low-noise amplifiers are some of the common core technology needs.
\item 
Development of novel low-threshold, low-background and scalable technologies that probe dark matter particle masses between 1~MeV and 1~GeV, a mass range that is inaccessible to currently mature Generation-2 dark matter detectors.

\end{itemize}

\medskip
\noindent
{\bf Indirect Detection} 
\begin{itemize}
\item Pair-conversion space telescope successors to Fermi-GLAST need to be developed and their angular resolution improved and energy ranges expanded, especially to lower energies.

\item Ground based neutrino, cosmic-ray and air Cherenkov telescopes should continue to be supported. Greater understanding of the conventional astrophysics of hot, dense energetic regions will better constrain the backgrounds and astrophysical uncertainties of indirect searches. Such efforts include infilling CTA to better image the entire air-shower and provide better $\gamma$-hadron separation.
\end{itemize}

\noindent
{\bf Axion Searches} 
\begin{itemize}
\item Experiments employing advancing  current technologies of magnetic resonance, lumped element and microwave cavities will reach sensitivity to the QCD axion.

\item Further in the future new technologies utilizing Squeezed states, single-photon-counters and atomic interferometers will allow greater axion parameter spaces to be studied, especially to higher masses.

\end{itemize}

%% file: Science/Cosmic.tex
\subsection{Early and late time cosmic acceleration} 

\def\as#1{{\bf \textcolor{blue}{[{AS:#1}]}}}
\def\zee#1{{\bf \textcolor{red}{[{ZA:#1}]}}}

\label{Cosmic}
\subsubsection{Introduction}

The universe is believed to have undergone at least two periods of accelerated expansion. Early acceleration, also known as inflation, set up the initial conditions for the evolution of the universe in standard $\Lambda{{\rm CDM}}$ cosmology. Late acceleration of expansion started at redshift $z\sim 2$ and continues today, driven by dark energy, which by redshift $z\sim0.4$ became the dominant form of energy in the universe. Pursuit of the understanding of both periods of accelerated expansion is a P5 science driver.


Early time acceleration refers to an epoch of inflationary expansion prior to $10^{-30}$ seconds after the Big Bang when the energy scale of the universe may have been near $10^{15}$~GeV. Signatures of this unique epoch in the evolution of the Universe open a window into the fundamental physics at the energy scales not otherwise accessible by any terrestrial method. We have very few experimental handles on the process of inflation:
\begin{enumerate}
    \item \textbf{Primordial gravitational waves}. Tensor fluctuations in the metric (gravitational waves) generated by inflation have an amplitude proportional to the energy scale of inflation. These gravitational waves would leave imprints on the Cosmic Microwave Background (CMB) temperature and polarization anisotropies. Current constraints suggest that the gravitational wave signature is too small to be measured in temperature anisotropies; therefore, improved measurements of the CMB polarization are the most promising route to constraining the high-energy physics of inflation.  The amplitude of primordial gravitational waves, for which currently just upper limits exist, is experimentally the most important quantity to be measured in the next decade \cite{ShanderaDecadal}.
    \item \textbf{Non-Gaussianity of the primordial fields.} Minimal inflationary models predict that primordial fluctuations are perfectly Gaussian. If inflation involved several fields or if the inflaton had non-minimal interactions, the primordial fluctuations would have imprinted non-Gaussianities, which would be seen as non-zero bi-spectra. These can be most readily measured with the CMB, where we expect significant improvements with up-coming experiments, but large scale structure measurements will become competitive in the next decade \cite{MeerburgDecadal}.
    \item \textbf{Inflationary Relic in the primordial fluctuations}. Non-minimal inflation can also produce a primordial power spectrum that departs from a simple power law. Especially in string-inspired inflationary model building, it is fairly difficult to produce scalar fields with sufficiently flat potentials and techniques that allow this often leave other imprints, such as small oscillations or localized features. These can again be constrained using the CMB or large scale structure \cite{SlosarDecadal2}.
    \item \textbf{Iso-curvature perturbations.} If inflation involved processes that interacted with the standard model fields, they can generically produce isocurvature perturbations. These are perturbations where the ratio between various species (say baryons and radiation) are not conserved over a density fluctuation. These are best constrained using CMB observations. \cite{isocurvature}.
\end{enumerate}

For the late time acceleration, the main programmatic goal is to constrain the dynamics of dark energy evolution, by measuring:
\begin{enumerate}
    \item \textbf{Expansion history.} A measurement of the expansion history enables us to reconstruct how the properties of dark energy varied through time. This allows us to measure the equation of state of the dark energy and distinguish between a simple cosmological constant, a model with a more general, but constant $w\neq1 $ and a most general model with an evolving equation of state. The most developed techniques for measuring the expansion history directly are Baryon Acoustic Oscillations and Supernovae type Ia used as standard candles \cite{SlosarDecadal1}. 
    \item \textbf{Growth of density fluctuations.} For a fixed expansion history, general relativity predicts how the fluctuations grow. Any departure from this prediction would be irrefutable evidence for non-standard gravity. Growth can be measured using redshift-space distortions measuring in galaxy surveys, using non-linear modelling of weakly non-linear scales in 21\,cm intensity mapping, and using weak lensing as probed by galaxy shear or measuring local 2-point function changes in 21\,cm intensity maps or CMB. When combining auto-correlations of shear and clustering with cross-correlations, the sensitivty and robustness can increase dramatically \cite{SlosarDecadal1}. 
\end{enumerate}

The same surveys that can constrain the early and late time acceleration can naturally also measure two other important cosmological observables:

\begin{enumerate}
    \item \textbf{Neutrino mass.} Cosmology is also highly competitive in constraining the neutrino mass. Neutrinos were thermalized in the early universe plasma and we have over 10$\sigma$ indirect detection of this cosmic neutrino background in the cosmological data \cite{Planck2015}. Cosmological observations are sensitive to the sum of neutrino mass eigenstates (strictly speaking individual masses should be measurable, but we are unlikely to reach the required precision in the next decade). All the main experiments to measure accelerated expansion (LSST, DESI, CMB-S4 discussed below) are forecast to be able to detect neutrino mass at several sigma levels using complementary techniques with different systematic errors. This is an excellent and informative addition to the HEP neutrino portfolio \cite{DvorkinDecadal}. 
    
    \item \textbf{Effective number of neutrino species.} This quantity measures the radiation content of the early universe that is not coupled to baryons. Since neutrinos are the only known species that contribute to this, it is parametrized using the effective number of neutrino species. This number has a standard value\footnote{The difference between 3.046 and the naive expectation of 3 for three neutrino species is technical and irrelevant for the current discussion.} of $N_{\rm eff}=3.046$. It is currently constrained to a precision around $\Delta N_{\rm eff}\sim 0.2$, but with future CMB experiments, these limits can improve by an order of magnitude. This would be immensely powerful, since any light degree of freedom that was in thermal equilibrium with the standard model at some point will contribute to $N_{\rm eff}$ at a level that is bigger than $0.027$. Hence this is potentially the most promising probe into new degrees of freedom in the standard model at high energies. The damping tail of the CMB power spectrum is the only known technique to constrain this quantity \cite{GreenDecadal}.
\end{enumerate}

The present-day sensitivities are blurring the traditional distinction between early universe measurements and late universe structure studies. The CMB can probe the late universe through many secondary effects on the photon propagation from the last scattering surface. Similarly, the large scale structure surveys can probe early universe through left-over relics of primordial fluctuations in the evolved galaxy fields. We therefore argue for a unified and coordinated approach to all cosmological probes.

P5 has determined that studying the CMB would add a new critical dimension to the DOE HEP portfolio in the cosmic frontier, since measurements of CMB polarization anisotropies can constrain early stage acceleration and probe the early thermal history of the universe, provide systematic cross checks for characterization of late stage acceleration, and provide an orthogonal handle on the neutrino sector. Therefore, it has recommended a ``Stage IV" CMB experiment (CMB-S4). CMB-S4 cameras require $\sim$500,000 science-grade, photon-noise-limited superconducting sensor pixels operated at sub-Kelvin temperatures to meet the science goals. This is 10 times greater than the count of CMB sensor pixels deployed in all ground-based CMB cameras over the past decade. CMB-S4 is expected to be granted CD0 in 2019 and CD1 in 2021.
This presents a challenge as well as a great opportunity for technology development. 

The situation with late-time acceleration is considerably more developed. Multiple ``Stage IV'' DOE projects will begin constraining late-time acceleration in complementary ways in the coming decade.  Upcoming experiments include LSST (a large photometric experiment starting observations in 2021) and DESI (a large spectroscopic experiment starting observations in 2019).  These two DOE projects will be complemented by other international collaborations, most notably the Euclid satellite in Europe, launching in 2022, and SphereX and WFIRST satellites to be launched by NASA in 2023 and some time after 2025s respectively.

The triad of CMB-S4, LSST and DESI will provide several ground-breaking new constraints in the 2020s. CMB-S4 will measure (or constrain) the presence of the primordial gravitational waves with the precision to detect or rule out single-field slow-roll inflation models that naturally explain the shape of the observed primordial power spectrum. LSST and DESI will constrain the late-time acceleration equation of state to percent level precision. Neutrino mass constraints are in the guaranteed detection regime (assuming terrestrial measurements of the neutrino mass difference) and will be detected using three different techniques: gravitational lensing of the CMB (CMB-S4), redshift-space distortions (DESI) and galaxy weak lensing and clustering (LSST). Moreover, cross-correlations between data from the three experiments will amplify this program. For example, CMB lensing can help break degeneracies with the LSST photometric redshift estimation, and the combination of DESI and CMB-S4 will allow measurements of the kinetic Sunyaev Zeldovich effect, and so on.

Beyond CMB-S4, it is clear from comparing the number of LSST and DESI sources that many more cosmological modes could be probed with surveys that can probe the true three-dimensional structure of the universe. Such surveys could provide additional constraints on both late-time acceleration by making even more precise measurements of dark energy, as well as inflation by improving constraints on non-Gaussianity and measurements of other relics in the primordial power spectrum. This could be achieved using several different methods. The most straightforward builds on fiber-fed optical spectroscopy, namely expanding the DESI technique with a bigger telescope, higher fiber count and expanded wavelength coverage. Two recent concept have been proposed in this area: MegaMapper \cite{MegaMapper} and SpecTel \cite{SpecTel}. A possible alternative is integral-field spectroscopy over a smaller area, but targeting a larger number of objects. In this case several possible technological advances are possible, which we discuss below. Finally, a new set of techniques are emerging that target the same structure in the universe, but dispense with detection of individual objects. These methods, often referred to as ``intensity mapping'', instead rely on spectral information to reconstruct the three-dimensional information on large scales. Because these techniques measure only the aggregated emission on the large scales where the most information lies, they can be in general very cost efficient. Lines both in optical and infrared (Lyman-$\alpha$, H$\alpha$, H$\beta$, H$_2$ molecular lines) and mm/radio wavelengths (carbon monoxide, [CII], 21\,cm hydrogen spin-flip transition) have been considered.  The 21\,cm neutral hydrogen transition is currently the most experimentally advanced with several dedicated cosmological experiments operating or under construction (CHIME, HIRAX, SKA in single dish mode). The proposed PUMA experiment \cite{PUMA} is the DOE-led effort in this area.  In the coming decades, the radio techniques will particularly benefit from enormous investment in radio frequency electronics driven by the telecommunications industry.

\subsubsection{Critical Needs}

The DOE HEP program should support the development of some critically-needed technologies to enable successful deployment of the CMB-S4 experiment and to prepare for the following generation of large scale structure experiments. In particular, we envision:

\begin{enumerate}
\item \textbf{Sensors, readout and supporting technologies for CMB surveys.} CMB-S4 will have massively increased counts of photon-noise-limited superconducting sensors ($\sim500,000$) in its cameras compared to Stage~III experiments. It is essential that the capability to produce and test sensor modules at this scale with stringent quality control is developed. A preliminary assessment shows superconducting sensor production capacity in the US to be inadequate for the needs of CMB-S4, necessitating some investment. Moreover, readout of these sensors in larger CMB-S4-scale cameras requires further advancement of cryogenic signal multiplexers, amplifiers and readout electronics to meet noise, stability and thermal performance requirements while processing greater signal counts. Finally, production capacity and quality control for supporting CMB camera technologies such as optics, filters, anti-reflection coatings and optical modulators is necessary.  

\item \textbf{Industry partner for CCD development.} CCD detectors play a vital role in many DOE supported experiments, but are especially crucial in optical photometric and spectroscopic experiments. Teledyne DALSA Seminconductor has been supplying for 150mm wafer for CCDs for over 20 years. It was the primary DOE partner for development of CCD technology and had an established relation with multiple DOE laboratories.  DALSA is migrating the manufacturing from 150mm to 200mm wafer
diameter, but will not update their CCD processing tools for the new format wafer. It will therefore no longer serve as a partner to the CCD community for manufacture of CCDs for optical astronomy applications. We recommend that steps are taken to find a new commercial partner in CCD
fabrication to maintain capabilities in custom CCD design for astronomy applications.

\item \textbf{Fiber positioners for spectroscopic galaxy surveys.} After the completion of the DESI program in 2024, a next-generation experiment will be strongly motivated. The crucial missing piece of technology is reduced-pitch fiber positioners. DESI currently employs 10\,mm pitch positioners; lowering the pitch to 8\,mm or 7\,mm could increase the fiber density by a factor of 1.5$\times$ or 2$\times$, respectively, increasing the spectrograph multiplex by the same factor. This could be achieved using several competing solutions.
    
\end{enumerate}

\subsubsection{Future Direction}

In preparation for cosmology experiments beyond DESI, LSST and CMB-S4 in the coming decades, we need a long-term roadmap of technology development that will enable  futuristic experiments. Some of the paths to pursue   include:
\begin{enumerate}
    \item \textbf{Ge CCD sensors.} CCD-like devices that are based on germanium rather than silicon could be transformative in enabling very sensitive infrared sensors. The 1eV effective band gap in silicon leads to a cutoff in detector sensitivity at around 10,000 \AA. For the [O~II] emmission line, this limits the reach of spectroscopy to about $z\sim 1.6$. Development of CCD technology based on germanium rather than silicon could potentially have transformative impact on the future spectroscopic surveys of the high-redshift universe.


\item \textbf{Development of integral field spectroscopy techniques.} One possible solution to long term cosmology would be techniques that would allow spectroscopy without fiber positioners, essentially taking the spectra of every pixel on the sky. It is unlikely that could be achieved to sufficient spectral resolution using color sensitive detectors. Nevertheless, technologies for fiber-packing or distribution of light using other means like micro-mirrors and micro-shutters should be explored. 

\item \textbf{Development of fast digital processing for 21~cm surveys.} The Cosmic Visions--21~cm working group has recently released a roadmap to a Stage-II 21~cm cosmology experiment 
complementary to CMB and optical probes. Such an experiment, by digitizing the RF signals from tens of thousands of antennas scattered over kilometer distances, will require data transport and processing at scales surpassing those found in even the most ambitious planned particle physics experiments. For example, detector concepts for the proposed Future Circular Collider envision raw data rates in the range of 800 - 1100 TB/s \cite{SmithCPAD}, but this rate is reduced by $>99.99$\% with multilevel triggering using relatively simple selection thresholds. In contrast the proposed 21\,cm PUMA experiment \cite{PUMA} will generate about 600 TB/s but will need to run sophisticated real-time calibration and interferometric computation in order to generate the time-ordered data maps for intensity mapping, and even faster processing for recognizing fast radio bursts. Development of this concept will require significant development encompassing integrated GS/s digital front-ends, picosecond accuracy clock distribution, and unprecedented levels of real-time computational power.  The R\&D required to meet these challenges  is well within the purview of the National Labs and by leveraging the explosive growth of wireless and computing technologies, offers a compelling avenue for exploring a large volume of the universe with systematics complementary to those of optical and mm-wave probes.

\item \textbf{Development of new detection technologies for spectroscopy.} The current generation of sensors are background- or photon-noise-limited both in optical as well as microwave bands, and therefore cannot be improved. However, significant improvements could be achieved with spectroscopically-sensitive sensors that possess sufficient spectral resolution $R\gtrsim100$ to allow massive and accurate spectroscopy in optical and microwave. There are several proposed ideas, none of which are currently capable of competing with CCDs in optical and superconducting transition-edge-sensor bolometers in the microwave window. In the optical, very low spectral resolution CCDs can be built by being sensitive to the depth of absorption which decreases with increasing photon energy. Microwave Kinetic Inductance Detectors can work from microwave to optical and have been successfully demonstrated in real astronomical applications. However, they 
currently offer insufficient spectral resolution and pixel counts. 
Finally, the first generation of on-chip spectrometers composed of analog filter banks and operating at 200--500~GHz have been demonstrated. Such spectrometers could be used in future intensity mapping experiments relying on molecular lines such as [CII] 158 micron transition. While most of these technologies face significant development hurdles to be overcome, blue sky R\&D along these lines should be encouraged as it might eventually lead to new viable experimental techniques. 

\end{enumerate}

%% file: Science/Unknown.tex
\subsection{Exploring the Unknown - New Particles and Interactions}
\label{Unknown}
\subsubsection{Introduction}

There is clear evidence that the Standard Model offers an incomplete description of nature at the sub-atomic level, thus motivating a broad-based strategy to search for new physics, new particles, and new interactions. In addition to the opportunities discussed in the previous sections, there exists a range of ongoing experiments designed to search for new physics via indirect effects, typically generated by virtual quantum loops sensitive to new particles or new interactions. These experiments explore for new physics in a manner that is complementary to the programs described in sections~\ref{higgs} -- \ref{Cosmic}, either by having sensitivity to phenomena that other experiments are blind to, or by achieving precision that other experiments cannot.
This program includes experiments that explore for new physics using heavy quarks and $\tau$~leptons (e.g. LHCb, Belle-II); rare muon processes involving charged-lepton flavor violation (e.g. Mu2e, mu3e, MEG); accelerator-based searches for electric dipole moments (EDM) of muons, protons, and deuterons; precision measurement of the muon anomalous magnetic moment (e.g. Muon g-2); and rare kaon decays (e.g. NA62, KOTO).
The efficacy and importance of these experiments is evident in the history of particle physics. Searches for $K^{0}\rightarrow\mu^{+}\mu^{-}$ motivated the theoretical introduction of the charm quark to explain why some flavor-changing interactions were suppressed (ie. the GIM mechanism); searches for $\mu\rightarrow e\gamma$ first established the muon as a new type of lepton and later confirmed the two-neutrino model; precision measurements of the Cabibbo-Kobayashi-Maskawa quark-mixing matrix established a Standard Model mechanism for charge-parity (CP) violation, but also made clear that new sources of CP-violation are required to account for the observed matter/anti-matter asymmetry in the universe. These experiments yielded profound insights and were fundamental to the development of the Standard Model. Future experiments exploring the physics of flavor, rare phenomena, and precision measurements offer outstanding sensitivity to a broad range of new physics that may yield results equally as profound.
%
%
To achieve required sensitivities these experiments utilize high intensity beams to make very precise measurements while maintaining exquisite control of background processes, thus placing stringent demands on detector performance. While these experiments share some of the same experimental challenges as those discussed in previous sections, particularly those of Sec.~\ref{higgs}, there exist areas for which custom solutions are needed in order to fully exploit next generation accelerator facilities.

\medskip
\noindent 
{\bf Heavy flavor studies }\\
Within the Standard Model the origin of flavor remains a mystery, motivating the continued pursuit of precision measurements involving heavy quarks. Moreover, the anomalies observed 
in $b\rightarrow s\ell^{+}\ell^{-}$ and $b\rightarrow c\tau\nu$ transitions require further experimental explorations that may yield definitive evidence of new physics. 
The LHCb Collaboration has both short-range upgrades in progress, and long-range upgrades under study. The short-range program involves improvements to the trigger for increased efficiency and the inclusion of additional trigger lines; upgrades to tracking for higher granularity and expanded acceptance; and improvements to the readout of the ring-imaging Cherenkov detector system.
Long-term upgrades may utilize high-precision timing, rather than photon direction at the exit, to measure the Cherenkov angle for particle identification
and highly segmented calorimetry with a fast-timing insertion to time electrons and photons to a few picoseconds. The excellent timing allows one to determine the location of the primary vertex and to better associate photons to secondary vertices.  An LAPPD(Large-Area-Picosecond-Photodetector)-style insertion in the layered calorimeter is proposed for the fast timing. Timing information will be included also in the charged-particle tracking detector system to cope efficiently with the high number of particles per crossing produced within the LHCb acceptance at the target luminosity of $2\times 10^{34}$cm$^{\rm -2}$s$^{\rm -1}$. Moreover the LHCb detector at this luminosity is expected to produce 400-500 Tb of data per second, which will have to be processed in real time and
reduced by at least 4-5 orders of magnitude before recording the remainder to permanent storage. Thus significant computational and data storage challenges are posed by the physics goals of these future upgrades. The research and development towards novel particle identification devices, tracking detectors, and fast and efficient data processing can be exploited in a variety of ``next generation" experiments aiming at further improvements in the sensitivity to new physics of heavy quark flavor experiments.
\break
\medskip

\noindent 
{\bf Rare muon decays} \\
Charged-lepton flavor (cLFV) violating processes are deep probes of new physics providing discovery sensitivity to a broad range of models including SUSY, Extra Dimensions, extended Higgs sectors, and models explaining the neutrino mass hierarchy and the matter dominance of the universe via leptogenesis. The most sensitive probes of charged-lepton flavor violation use high-intensity muon beams to search for neutrino-less 
$\mu\rightarrow e$ transitions. Experiments are underway or planned in the US, in Japan, and in Switzerland: the Mu2e, COMET, MEG2, and mu3e experiments will explore 
$\mu N\rightarrow eN$, $\mu\rightarrow e\gamma$, and $\mu\rightarrow eee$ transitions with
sensitivity to effective new physics mass scales up to $10^{5}$~TeV/$c^2$. Next
generation experiments will exploit upgraded beam facilities for further improvements. This requires upgraded tracking, calorimetry, timing, triggering, and data acquisition capabilities. These are decay-at-rest experiments demanding excellent momentum resolution for electrons in the $10-100$~MeV/$c$ range thus necessitating the development of very low-mass tracker technologies with a total path length corresponding to $<1\%$ of a radiation length. Fast, compact, radiation-hard calorimetry also plays an important role (e.g. liquid Xe or BaF$_2$). Timing to a few picoseconds is needed to address pattern recognition challenges and to mitigate accidental backgrounds in high rate environments. The triggering and data acquisition capabilities will need to improve by about an order of magnitude relative to the currently planned experiments with roughly $\mathcal{O}\left( 10 \right)$~TB/s of data from the front-ends and trigger rejection factors of $>1000$.


\medskip
\noindent 
{\bf Accelerator-based EDM experiments} \\ 
Experiments searching for electric dipole moments (EDM) are excellent probes of new sources of CP-violation. Accelerator-based experiments have been proposed and offer the possibility of measuring EDMs while removing contamination from hadronic contributions that affect atom-based EDM experiments. These experiments depend on observing dipole precession accumulated over billions of turns in an accelerator. Precision beam-position monitors that measure the beam centroid to $< 1\mu$m are needed to ensure that the two counter-rotating beams have the same orbits. Materials that can support very high electric fields, large-volume magnetic shielding, and sensitive magnetometers are also needed. For simulations, software that can handle particle and spin tracking for thousands of particles over billions of turns needs to be developed.

\subsubsection{Critical Needs}

Experiments that search for new physics via indirect effects, typically generated by virtual quantum loops sensitive to new particles or new interactions, play an important and complementary role to experiments searching for direct manifestation of new particles in advancing our understanding of fundamental physics. By utilizing precision measurements and searches for rare phenomena, these experiments explore signatures and/or new physics mass ranges that are inaccessible at colliders and other facilities. These experiments depend on intense beams and high interaction rates, and the following areas of critical need have been identified: 
\begin{itemize}
\item Fast detectors and excellent time resolution are critical development areas for the future since intense beams and high interaction rates are needed to achieve the desired sensitivity. Picosecond detectors are in demand.
\item Picosecond time resolution not only reduces pileup, but also improves event reconstruction by giving an extra dimension in which to match objects (e.g. photons to primary vertex, or hits to tracks).
\item High granularity calorimetry requires large areas of silicon or affordable fast radiation-hard crystals. Amorphous silicon (such as used in solar-cell technology) may suffice for some applications, substantially reducing the cost, while BaF$_2$ offers an appealing crystal solution provided that the large slow component can be sufficiently mitigated.
\item Tracking in cLFV muon experiments has challenges similar to those at the LHC (high rates, high track density), but is done at much lower momenta (10-100 MeV/c). The development of very low mass (i.e. with a total path length of less than 1\% of a radiation length), large area, radiation tolerant technologies is essential for next generation experiments utilizing upgraded accelerator facilities. Synergies with similar applications in intensity frontier experiments can inspire a broad research program in this area (see Sections~\ref{higgs} and \ref{neutrino}).
\item As commercial and slightly customized data transmission hardware becomes more powerful and less expensive it becomes possible to send much more, or even all, data off a detector into a commercial computer cluster allowing one to do all triggering in a software environment. This is an important development that is actively pursued by the LHC experiments and has the potential to revolutionize some searches for new physics. 
\item As the selection criteria required to identify interesting events are becoming increasingly complex, real time processing of large amounts of data is becoming a critical need and an optimization of the computing hardware at different stages of processing is a crucial line of investigation, involving a combination of field programmable gate arrays (FPGA), graphical processor units (GPU), and specialized and more parallel processor and storage devices.
\item Custom-made application specific integrated circuits (ASIC), processing in a highly specialized way the information coming from local detector segments, are critical to the overall detector optimization. A broad research program in this area, modeled on CERN R\&D initiatives, with coordinated efforts of university and laboratory groups may foster innovation and synergies.
\item Next-generation experiments will require improved power converters, specifically high-power, high-efficiency, radiation-tolerant power converters for use on front-end electronics. The power converters will need to operate in high magnetic fields ($>1$T) and, in some instances, in vacuum. 
\item Next generation muon experiments will reach beam intensities that will induce radiation environments comparable to the LHC experiments. Demand for radiation testing facilities will increase. The development of a dedicated facility, with the associated personnel support, for determining radiation tolerance of components would be of broad interest to the high energy physics community (see also Section~\ref{higgs}). 
\item Planned accelerator upgrades will provide another x10-100 in intensity for potential next generation experiments (e.g. mu3e phase 2, Mu2e-II). Advances in detector capabilities will be required to take full advantage of the increased intensities.
\item Accelerators are the instrument not just a source in (some) experiments searching for anomalous electric dipole moments, and precision accelerator instrumentation is critical in some cases. 
\end{itemize}


\subsubsection{Future Direction}

Risks and opportunities lie ahead for this science driver. The opportunities provided by planned accelerator improvements 
and the returns on the substantial investment made, will not be fully realized without significant detector developments to match the accelerator upgrades. Advances in technology (some through commercial development) make precision measurements possible today that simply were not possible in the past. Picosecond timing is a prime example. These advances open opportunities previously unavailable. It is suggested to support R\&D opportunities that address issues specific to the experiments discussed above, when similar LHC R\&D only partially addresses the issues relevant for these experiments. For example, low-momentum tracking in a high-occupancy environment with very low-mass trackers and fast radiation-tolerant crystals for calorimetry are needed for these experiments. It is further suggested that common R\&D advances needed across many experimental areas are supported. Picosecond timing will have numerous powerful applications in detectors and event reconstruction and serve numerous scientific goals. Similarly, investment in the development of high-performance radiation-tolerant power converters will be beneficial for experiments across the HEP program. Research in different computing architectures and data acquisition systems capable of processing massive amounts of data with unprecedented speed is a common need across all science drivers.
Furthermore, R\&D should be supported to develop cost-effective solutions
for large area, high granularity calorimetry. The development of a facility dedicated to characterizing the radiation tolerance of electronics components should be explored.

%% file: Technologies.tex
\section{Technologies}
\label{s.technologies} 

\input{./Technologies/Calorimetry.tex}
\input{./Technologies/Noble.tex}
\input{./Technologies/MPGD.tex}

\input{./Technologies/Silicon.tex}

\input{./Technologies/Photodet.tex}

\input{./Technologies/ASIC.tex}
\input{./Technologies/TDAQ.tex} 
\input{./Technologies/SC.tex}
\input{./Technologies/QS.tex}
\input{./Technologies/ML.tex}

%% file: Technologies/Calorimetry.tex
\subsection{Calorimetry}
\label{s.calorimetry} 
Calorimeters along with upstream trackers fulfill a number of crucial tasks in collider and fixed target experiments. 
They can be used to distinguish $Z\to q\overline{q}$ decays from $W\to q\overline{q}$, which is a design criterion of the ILC detectors, and
they are essential for the measurement of missing energy. 
The measurement of the energy, location and time of electromagnetic and hadronic showers remains central to all future experiments at the energy frontier and at any future collider that may be built. 

As we look towards the next decade the field of calorimetry for high energy physics has evolved in many ways from the perspective of 2009. In that time there have been several paradigm shifts in calorimetry. The first led by the ILC collaborations under the aegis of the CALICE collaboration is large-scale imaging calorimeters. The choice of silicon sensors and scintillator tiles being the technology of choice for the collaborations designing calorimeters for future e$^+$e$^-$ colliders, as well as the new CMS endcap calorimeter. A second paradigm shift has been the universal adoption of the use of Particle Flow Algorithms (PFA) for object reconstruction. PFA has led to the design of the next generation of calorimeters as part of a complex system of inter-connected detectors, rather than as stand-alone devices, in which the tracks found in the tracker are followed inside the calorimeter.

There has been a third paradigm shift that began only recently, which is the use of high precision timing detectors on a very large scale to distinguish events from within one bunch crossing at the HI-LHC. ATLAS and CMS are designing MIP detectors to measure the time of a single particle with a precision 30~ps. In CMS the lead-tungstate electromagnetic calorimeter is being upgraded to measure the arrival time of EM showers to the same precision, and for the CMS endcap calorimeter the goal is to measure the time in each cell with a precision of 50~ps. High precision timing measurements of the incident electrons, photons and
jets will certainly play an increasing role in the event reconstruction in the near future, as the technology matures. 

Calorimeters fulfill a number of crucial tasks, such as precision measurements of the four-vectors of
individual particles and jets, measurements of mass peaks (such as from $Z\to
e^+e^-$, or $Z\to q\overline{q}$ vs $W\to q\overline{q}$), and of the energy flow in the events
(missing energy), identification of jet substructure from boosted decays, pileup rejection, measurements of the time-structure within the shower development, and the particle's time-of-arrival.

At the energy frontier the principal challenges to detector subsystems is the survivability in the radiation field and the high rates, and the ability to detect and accurately measure electromagnetic showers in the presence of large backgrounds. These devices often need precision timing in the range of a few tens of picoseconds, while having a high tolerance to ionizing radiation. Future lepton collider calorimeters are required to have excellent energy resolution both for single particles and for jets. Good containment for jets up 
to p$_T$ of about 3~TeV will be required for calorimeters at the HL-LHC. Good transverse segmentation for resolving boosted particles imposes requirements on the minimal cell sizes, while requiring also that the Moli\`{e}re radius of the detector is small so as to minimize the effects of overlapping events in the calorimeter. 
Besides these shifts there still remains the need for precision EM calorimetry which can be only met by fast high-light-yield, and low-cost crystals, like the Mu2e experiment at Fermilab or the PANDA experiment in Germany. 

For imaging calorimeters the sheer volume of data produced requires the use of high-end FPGAs for signal processing and trigger decisions, high speed optical data links (10~Gbs and higher) and unprecedented computing power.

\subsubsection{Imaging Calorimeters}

The CMS endcap calorimeter (CE) has a very large number of electronics channels ($>$6M) requiring of order 200~kW of cooling power and operation at $-30$~$^{\circ}$C to mitigate the effects of radiation damage to the silicon. The timing of EM showers has to be measured with a precision of 50~ps for showers above approximately 2~GeV. The ILC and CLIC calorimeter designs have even more channels than the CE and, because of the low interaction rate, can operate with power pulsed only during the period of collisions. They are also planning on using silicon pixels with areas of 0.25 cm$^2$ or smaller, which poses significant problems with the connection between the sensor and the on-detector electronics and the subsequent signal extraction. For the CE, which must operate in the regions closest to the beam pipe, the integrated neutron dose is $10^{16}$ neutrons/cm$^2$.

\subsubsection{Crystal and Homogenous Calorimetry}

There are several future crystal electromagnetic calorimeter implementations such as COMET, Mu2e, g-2, HERD and PANDA. A variety of crystals are under consideration including but not limited to cerium doped lutetium oxyorthosilicate (Lu$_2$SiO$_5$ or LSO) and cerium doped lutetium-yttrium oxyorthosilicate (Lu$_{2(1−x)}$Y$_{2x}$SiO$_5$ or LYSO), BaF$_2$, PbF$_2$, BGO, BSO and PWO. Most of these crystals are well characterized and have been tested for radiation-hardness. A new concept called “Dual-Gate Calorimetry” has emerged, where the slow and fast components of the hadronic interactions are attributed to different physics components. The weighted sum of these individual detector signals is projected to result in a high resolution and linear hadron calorimeter with high precision timing. For best exploitation of such concept, fast and dense inorganic scintillators are needed such as BGO or BSO.

Future HEP experiments at the Energy and Intensity Frontiers require dense, brighter, faster and radiation hard crystals. With high density ( 7.1 g/cm$^3$), bright light (200 times of PWO), fast decay time (30 to 40 ns) and superb radiation hardness, LSO/LYSO crystals are a natural choice. One drawback of LSO/LYSO crystal, however, is its high cost due to high raw material cost and high melting point. Cost-effective fast crystals, such as CsI, BaF$_2$ and PWO are often given preference. The ultrafast scintillation light ($<0.6$~ns decay time) from yttrium doped BaF$_2$ crystals, with suppressed slow scintillation component, promises to be an ultra-fast calorimeter. To facilitate reconstruction, a longitudinally segmented crystal electromagnetic calorimeter would provide a photon pointing resolution and a good position resolution, in addition to an excellent energy resolution at a level of a few percent/${\sqrt{\rm E}}$. UV-transparent crystals and glasses may achieve unprecedented jet-mass resolution with dual gate readout for a homogeneous hadron calorimeter concept at future lepton colliders.

R\&D for future crystal calorimeters thus will proceed along the following lines:
\begin{itemize}
    \item Investigation on the radiation damage effects induced by gamma-rays, neutrons, and charged hadrons in various crystals to be used in future HEP experiments.
    \item Development of ultra-fast crystals with sub-nanosecond decay time.
    \item Development of dense, UV-transparent, and cost-effective crystals for the homogeneous hadron calorimeter concept.
    \item R\&D on detector design and compact UV-sensitive photo-detectors, such as SiPMs, with sufficient dynamic range, for a segmented ultra-fast crystal calorimeter.
\end{itemize}

\subsubsection{Precision Timing for Calorimeters}

The ability to detect and accurately measure electromagnetic showers in the presence of large backgrounds is a major challenge faced in calorimeter designs for the HL-LHC and other future collider experiments. These devices often need precision timing in the range of a few tens of picoseconds, while having a high tolerance to ionizing radiation. The spatial overlap of tracks and energy deposits from these collisions will degrade the event reconstruction, and increase the rate of false triggers. The missing transverse energy resolution will deteriorate, and several other physics objects performance metrics will suffer. Precision time stamping of all particles will add a new handle in discriminating the hard scatter from pileup interactions.

One way to mitigate pileup confusion effects, complementary to precision tracking methods, is to perform a time of arrival measurement associated with calorimeter measurements, allowing for a time assignment for both charged particles and photons. Such a measurement with a precision of about 30~ps, when unambiguously associated to the corresponding energy measurement, will significantly reduce the inclusion of pileup particles in the reconstruction of the event of interest. The association of the time measurement to the energy measurement is crucial since both the time-of-arrival and energy of the particle are measured in the same active detector element.

Several technologies for performing precision timing measurements within showers have been developed and tested recently. Electromagnetic calorimeter prototypes with precision timing capabilities have been demonstrated in test beams and several technologies have been used: micro-channel plates, inorganic crystals, such as LYSO or PbWO$_4$, and silicon diodes as the active medium. 

Several factors impact the precision with which the time-of-arrival of showering particles can be measured. The principle sources of uncertainties in the measurement of the time of an interaction are, among others fluctuations in the transit time in the sensor, Landau fluctuations in the signal, slewing in the preamplifier, and inter-channel variations in the clock distribution system. 
Large signal-to-noise and fast rise time of signals plays a crucial role in achieving good performance. On the single sensor level the timing precision for single charged particle detection is on the order of 30~ps in semiconductor-based sensors. The time of arrival of electromagnetic showers in calorimeters could be measured with significantly better precision using the large number of synchronous particles in a shower, and showers spreading over a large number of independent sensors. An example of this the high-granularity calorimeter upgrade of the CMS experiment, which will feature unprecedented transverse and longitudinal segmentation for both electromagnetic and hadronic compartments. Multiple samples of the shower with timing measurements in each layer can be combined to improve the time-of-arrival measurements. Bright inorganic scintillators and silicon detectors are very promising in the development of high precision timing measurements in calorimeters.  Clock synchronization across various cells of the calorimeter has to be maintained in order to be able to combine multiple cells contained in the shower. Design of low-cost and radiation-tolerant electro-optical transceivers and front-end electronics is therefore critical to obtain a system-level high-precision. 

\subsubsection{Common Development Needs}
In addition to the needs specific to calorimetry there are problems that are common to other aspects of caorimetery. One of these is the problem of a delivering a significant level of electrical power, typically at around 1.5~V or less to the detector. This is almost identical to the same probelm faced in trackers, for exampe. In the case of tracking systems having a low mass is critical, whereas in calorimetry the critical concern is the lack of space between the layers of calorimeter, which have to be narrow to keep the Moliere radius small. Concomitant with this is the need to remove the heat from the detector. The recently-developed radiation-tolerant DC-DC converters have been used or are planned to be used in calorimeters. There is however much room for improvement and the field would benefit by significant improvements in this area. Likewise bi-phase $\text{CO}_2$ cooling has begun to be used in several detectors by the community, and, as in power delivery systems, further development in cooling methods would have a broad impact on the field.

\subsubsection{Critical Needs}
\begin{itemize}
    \item Picosecond time resolution. The use of timing to reduce the effects of pileup at the HL-LHC and other high-rate experiments, to associate showers with collision vertices, to distinguish EM showers and to separate EM from the hadronic components within a hadron shower.
    \item{Modern image processing technology, both hardware (GPUs) and software (image processing and deep learning) to reconstruct complex events both in real time and offline.}
    \item{Low-cost, high-light-yield, fast and radiation-tolerant organic and inorganic scintillators.}
    \item{Further advances in Silicon Photomultiplier (SiPM) technology. Improved UV detection, larger dynamic range though smaller pixels, direct coupling to, or integration with readout electronics.}
    \item{Low-cost radiation-tolerant electro-optical transceivers, operating at rates of 10~Gbps or more.}
    \item{Continued development of GEANT to match the new information being used in calorimeters}
\end{itemize}

%% file: Technologies/Noble.tex
\subsection{Noble Element Detectors}
\label{s.noble} 

This report is based on the scope of presentations at the CPAD 2018 workshop at Brown University, Rhode Island, which took place on 9-11 December 2018.  The Noble Elements session hosted 20 talks and received nine one-page white papers which inform this report. We have also attempted to represent topics in noble detector instrumentation that were not presented at this workshop, but understood to be important and potentially game-changing for the future development of the field. Some parts of this report are not fully completed (for example Sec.~\ref{sec:DM}). 

\subsubsection{Noble Element Detectors for Accelerator Neutrino Physics}

{\bf Present Status} \\
Liquid Argon Time Projection Chamber (LArTPC) technology has been adopted as the approach of choice for accelerator neutrino experiments in the United States. These detectors exploit two signatures: first, charges moving through noble elements produce copious ionization that can be drifted to a readout plane and used to image events in 3D; Second, they produce strong ultraviolet scintillation light.  The physics reach of these programs is maximized by optimal collection of both of these signatures with high efficiency and minimal information loss.

The TPC was invented by Nygren in 1974, and the first large TPC in
a physics experiment began operating at PEP4 in 1981. By 1977, Rubbia proposed that the liquid argon TPC (LArTPC) could be the basis for an excellent neutrino detection technique. The ICARUS collaboration pioneered the LArTPC technique over a 30 year R\&D program culminating in the T600 detector, which has run as a cosmic ray detector, solar neutrino detector, a long baseline detector at CNGS, and is now preparing for its latest incarnation as a detector for short baseline physics at the Fermilab Booster Neutrino Beam.

The first US-based liquid argon TPC for neutrino physics was the ArgoNeuT test stand, which took neutrino interaction data in the NuMI beam at Fermilab starting in 2009. The ArgoNeuT detector has since be re-purposed as the LArIAT test-beam experiment, and continues to inform the US liquid argon program via a continuing suite of characterizations of charged particle interactions in argon. In 2016 the 170 ton MicroBooNE experiment began running at a 470m baseline in the Booster Neutrino Beam, with the primary goal of exploring the MiniBooNE short-baseline low-energy anomaly with the same neutrino beam, similar baseline and different technology. The first results from MicroBooNE on this subject are expected in 2019.

The flagship neutrino facility in the United States, DUNE, will send intense beams of neutrinos from Fermilab to four 10 kTon detectors in the Homestake mine in Lead, South Dakota, with recent prototypes of single and dual-phase modules recently constructed at CERN.  The full-scale DUNE detectors will be TPCs of unprecedented scale and cost.  The necessity to implement cost effective solutions for such a large detector can be in tension with the necessity to ensure that a program of such great cost and duration has the maximal achievable physics reach.  

The many sensitive physics experiments based on the noble element
TPC have been enabled by a variety of test stands, engineering tests
and prototypes. This R\&D has been and continues to be vital for the
success of the program. Even as the DUNE collaboration prepares concrete engineering plans, many key aspects of noble TPCs remain only partially understood or optimized (including, but not limited to, high voltage delivery, light collection, internal electronics, etc), and it is likely that ongoing work in prototypes and test stands will continue to be crucial for success.

Beyond the major long- and short-baseline experimental neutrino programs, other special-purpose detectors using argon are employed in neutrino physics. The COHERENT experiment will use a liquid argon target to study coherent neutrino scattering (CEvNS) on argon, a well-predicted standard model process than can serve as a sensitive probe of neutrino magnetic moments, Non-Standard Interactions of neutrinos, accelerator-produced dark matter, and new scalar and vector mediators, among other things. The LArIAT experiment has provided test-beam studies of liquid argon detector performance as well as providing an important test-bed for new technologies such as LArPIX and ArcLight; and detectors using argon gas are envisaged as near detectors for the DUNE program, enabling unprecedented study of the details of neutrino interaction physics.

\medskip
\noindent 
{\bf Major Challenges and Future Directions} \\
We identify as major challenges for large LArTPC experiments the optimally granular collection of ionization charge and development of electronic and readout solutions to support this granularity; efficiency and high quality VUV light detection; delivery of high voltages; and calibration of a large detector system.

\medskip
\noindent 
{\bf Optimally Granular Charge Collection and Electronics} \\
Traditional TPCs for neutrino physics have used a wireplane geometry.  The primary reason for this choice is that it offers a pseudo-3D readout while reducing the required number of instrumented channels.  In a wire-plane geometry, the number of channels scales approximately like volume N$V^\frac{1}{3}$. On the other hand, with a fully 3D geometry such as pixel or pad readout, it scales like N$V^\frac{2}{3}$. In both cases, N is a small integer encoding how many modules the detector is divided into, or whether channels are doubly populated, for example by wrapped wires.  

While the wire-plane geometry offers advantages in terms of economy, it leaves something to be desired in terms of signal quality. The projective geometry makes recovery of a fully 3D image complicated. Although there have been major advances from Deep Learning~\cite{Adams:2018bvi,Acciarri:2016ryt} and ``classical''~\cite{Qian:2018qbv,Acciarri:2017hat} reconstruction approaches, the issue of a full 3D reconstruction of complicated events in liquid argon remains unsolved, despite being a focal point of neutrino physics research in the US for at least five years. Part of the problem is that this readout scheme can be lossy. Certain topologies, expecially showers or tracks moving in directions along or perpendicular to a given wire-plane, suffer losses of information from the protective representation.  Furthermore, wires have a large capacitance and so are noisy, and also somewhat unwieldy to install. It is thus believed by many that the physics case for DUNE could be substantially improved, if the detector could be pixelized.

A pixel geometry, more common to electroluminescent TPCs like those used in Dark Matter of Double Beta Decay experiments, offers full 3D event reconstruction without reconstruction ambiguities~\cite{Qian:2018qbv}. It may in principle greatly lower the threshold for reconstructable events, enabling sensitivity to low energy neutrino physics such as solar neutrinos.  It would surely reduce the level of contamination between certain difficult-to-reconstruct event topologies, particularly those based on electromagnetic showers (e vs $\gamma$ vs $\pi^0$).  It appears a pixelized approach may be a way to maximize the physics reach of the DUNE program.

While it is hard not to see the benefit of a pixelized DUNE far detector, it superficially presents serious burdens in terms of cost and complexity.  With a much larger number of channels required, reading out the system without introducing enormous cost, heat load and single-point failure modes becomes complicated.  Several R\&D approaches are being pursued to address these issues.

The ArgonCube Collaboration proposes to deploy a pixelated readout in a modularized DUNE near detector. A prototype at Bern has operated with success. The LArPIX~\cite{Dwyer:2018phu} collaboration has also operated pixelated readout within the LArIAT TPC~\cite{Asaadi:2018xfh}, reconstructing 3D events in a small detector.  The ability to deploy pixelized readout at small scales is thus developing strongly.

The system aspects involved in deploying a pixelized LAr readout at large scales present a more substantial challenge.  The Q-Pix concept \cite{Nygren:2018rbl} proposes to use free-running integrating Schmidt triggers, using time between resets as a proxy for local dE/dx.  This presents, in principle, a very low-noise and low-heat readout scheme.  It is envisioned that a heartbeat from radioactive $^{39}$Ar will be used to continuously calibrate the system. Readout from a large system is achieved by a network of token-passing units that act as automatons to broadcast time-stamped data from across the pixel plane. With further development, the Q-Pix approach may address in a compelling way some critical aspects of deploying a large pixel plane, suitable for application in DUNE.

At smaller scales, maximization of the utility and independent physics reach of the DUNE near detector is also a topic of investigation.  Beam intensity at the near location is very high, and the primary challenge is to extract the maximal information content from very copious neutrino interaction events.  In order to improve detector spatial resolution and make low energy tracks from neutrino interactions visible, a gas phase near detector is being developed~\cite{Martin-Albo:2016tfh}. The GOAT prototype at Fermilab recycles ALICE TPC modules with a high pressure argon gas mix to achieve fine tracking resolution as a component of the DUNE near detector.  Tests are also planned using optical readout of an electroluminescence signal in the gas.  Much remains to be studied in order to understand the optimal configuration of near detector(s) for DUNE, and their charge readout schemes.

\medskip
\noindent 
{\bf Efficient and High-Quality VUV Light Detection}\\
Scintillation light in liquid argon has a yield of several thousand photons per MeV, with dependencies on electric field, particle type
and purity. The wavelengths is 128 nm, and
both noble liquids and gases are transparent to their own scintillation light.
This is an extremely useful property that allows the detector scales
to become large without significant self-absorption. The majority
of materials used in the windows of photodetectors, however, are not
transparent at these wavelengths. It is thus necessary to either use
specially modified VUV-sensitive devices (for example, quartz-windowed PMTs for xenon light detection, or VUV-sensitive SiPMs such as the Hamamatsu VUV-4 , or to shift the light (often using TPB) to a longer wavelength for detection by conventional devices. 

The large liquid argon TPCs of interest can be divided into two classes:
long-baseline detectors which will be deep underground (the DUNE far
detectors) and short-baseline detectors which are on the surface (MicroBooNE, SBND, ICARUS, etc). The role of light collection in each of these classes of experiment is very different, and so the optimal light collection technology are not equivalent.

The primary role of the light collection system in surface-based detectors is background rejection. Tens of thousands of cosmogenic muons impinge upon the MicroBooNE detector every second, and in every readout frame, ${\cal O}$(10) background tracks are present. Identifying frames with a neutrino interaction in the detector is a non-trivial task. Because it is prompt, scintillation light can be used to tag events which are in time with the beam, thus rejecting backgrounds from out-of-time cosmic rays. This background suppression, which may also include the role of a trigger, is the most important function of light collection in surface based detectors.

The ``standard'' design for a surface-based LArTPC optical system
is well exemplified by the MicroBooNE experiment \cite{Acciarri:2016smi}, whose design is
based on the original ICARUS concept. The baseline design for SBND
is also very similar, up to details. In this system, 32 photomultiplier tubes (PMTs), which have a platinum undercoated photocathode to maintain conductivity at cryogenic temperatures, are mounted behind TPB coated plates. 128~nm scintillation light striking the plates is converted to visible light, some fraction of which is detected by the cryogenic PMT. Seeking time-correlated light deposits over many PMTs allows distinct sub-events to be identified. The geometrical information encoded in the flash can be used to associate each event observed in the corresponding TPC frame to a distinct time, to few nanosecond precision. In this way, on-beam and off-beam sub-events can be classified.

Because the tagging of cosmic rays and neutrino events on an overlaid TPC image is necessarily a geometrical procedure, for surface based TPCs, granularity light collection is vital. Many small sensors are more valuable than a few large ones. Improved granularity also implies a correspondingly larger channel count, which may drive up the cost and complexity of the readout electronics. The ultimate granularity would be achieved by a fully active tiling of VUV-sensitive SiPMs \cite{Bonesini:2018irj}, but such a channel count makes some internal signal processing via ASICs vital, and the design and integration of such a system is non-trivial. 

A compromise between channel count and granularity may be offered
by the ARAPUCA \cite{daMotta:2017aed,Segreto:2018jdx}. This system uses multiple stages of wavelength shifting in tandem with a dichroic filter to trap photons inside a reflective box. This box houses one or a few SiPMs, covering a small fraction of its surface area. The trapping of photons inside the device increases the effective area of each SiPM, enhancing photon yield per channel. Early demonstrations of the concept have shown promising performance, but the devices are still young and appear to still be evolving toward their full potential.

Deep detectors like the DUNE far detector have very low rates of cosmogenic background, and so cosmic background event tagging is no longer a priority. With present proposals, it is also unlikely that sufficient light can be collected to meaningfully contribute to calorimetry. Therefore the main goal of light collection is to provide timing information about events in order to extend the physics program beyond that achievable with the TPC alone.

There are two broad classes of applications for precise timing in
a LArTPC:
\begin{enumerate}
\item Measurement of the interaction time for non-beam events. This allows
a measurement of the drift coordinate, which is important for applying corrections due to finite electron lifetime and (possibly) diffusion in order to properly reconstruct the event. Examples of such physics cases include supernova neutrinos, proton decay, solar and atmospheric neutrinos. Notably this is not necessary for beam-induced events, where the beam-clock already provides this information. 
\item Identification of sub-events in multi-step decays, for example with muons or (perhaps) kaons or pions in the final state. The track kinks where the decays occur may well be indistinct in the TPC image, while the double-pulse structure of the emitted scintillation light can be used to identify the event. Specific cases where such information may be valuable include the proton decay modes $p\rightarrow K\mu$, $p\rightarrow K\mu\pi$, $p\rightarrow K\nu$.
\end{enumerate}
In all these cases, geometrical information is much less important than the total number of photons detected. Thus any strategy that leads to the largest total number of photons collected for a given event is likely the optimal one.

A mature approach to large area light detectors is based on the light guide concept which has been under active R\&D for large liquid argon detectors since 2010 \cite{Moss:2014ota,Howard:2017dqb}. In such detectors, 128 nm light is converted into the visible region in a thin film of wavelength shifter applied to a transparent acrylic bar. Some of the shifted light is emitted into total internally reflected modes, and can propagate along the bar to photodetectors (usually SiPMs) at one or both ends. Coated light guides are now achieving attenuation lengths in the multiple meter range, which meets the target for DUNE-scale devices. On the other hand, their efficiency is low relative to the PMT+plate approach.

Another way to achieve large light yields is by using TPB-coated reflector foils~\cite{Kryczynski:2015pmj,Spagliardi:2017rec}. Such a system has been implemented in the LArIAT test beam experiment and very high light yields are obtained. This could could provide a large enhancement to the optical capabilities of a deep detector such as DUNE.

A scheme originally proposed by ICARUS, and revisited by many since,
is to enhance the yield of useful light in liquid argon by doping
with small quantities of xenon~\cite{Wahl:2014vma}. This offers three advantages:
\begin{enumerate}
\item 128 nm light is wavelength shifted to 175 nm, either via fluorescence
of dissolved xenon (at lower concentrations) or via direct excitation
transfer (at higher concentrations). Recent work on the wavelength
shifting efficiency of TPB shows that, contrary to older reports \cite{Benson:2017vbw},
it is brighter at 175 nm than 128 nm \cite{Gehman:2011xm}, and hence a higher light
yield is expected.
\item If sufficient quantities ($>$10ppm) of xenon are added, excitation transfer
from argon to xenon dimers, which have shorter lifetime, reduces the
timescale of the triplet light, making it easier to read out.
\item Since the reduced time constant implies less time for collisional
de-excitation of argon triplets before radiative decay, higher photon
yields are expected overall, even aside from improved wavelength shifting
efficiency.
\end{enumerate}
Dopants other than xenon can also be considered. In particular, recent
work shows that TPB may be soluble in liquid argon at the 10 ppb level~\cite{Asaadi:2018ixs}.  The problems associated with this solubility may be alleviated by instead using a polymeric wavelength shifter such as PEN~\cite{Kuzniak:2018dcf}.
On the other hand, if concentrations could be controlled, dissolved or suspended TPB could be used to positive effect as a bulk wavelength shifter.

\medskip
\noindent 
{\bf Delivery of Very High Voltage} \\
The very large drift distances required, especially in the case of the dual-phase far detector concept which cannot be modularized in the vertical direction, requires very high voltages to be delivered.  The goal is to operate at what appears to be the optimal drift field for balancing ionization and scintillation yields of around 500 V/cm.  Recent work has~\cite{Auger:2015xlo,Lockwitz:2015qua} illustrated that the reference values of argon dielectric strength of 1 MV/cm from relatively impure argon in the 1960s~\cite{Swan:1961} were overly optimistic.  

Although good design principles are intuitively understood by some, there remains no truly scientific description of what HV strength of a given detector configuration will be.  The dielectric strength is understood to depend on electrode stressed area, electrode volume, surface finish, liquid purity, polarity, pressure, temperature, dopants, and more.   An important ongoing R\&D activity will be the detailed study of HV strength of noble element detectors, with excellent examples including the XeBRA~\cite{Tvrznikova:2018pcz} program at Stanford and the Blanche program at Fermilab~\cite{Lockwitz:2015qua}.  A relatively unexplored topic at this time is the possibility of adding quench components to liquid argon to raise dielectric strength, a common approach in gas detectors.  The role of UV light in the streamer mechanism in liquid is not so clear, so whether this method would hold promise is not yet known.

Technological advances such as the use of resistive film field cages demonstrated at ArgonCube, self-quenching resistive cathode in ProtoDUNE~\cite{Abi:2017aow}, and the use of surge protection to protect resistors in cryogenic HV divider circuits~\cite{Asaadi:2014iva} are all advances that implement protection of detectors in the event of HV discharges. Such measures are prudent while understanding of HV strength of noble liquids remains imprecise, and further HV discharge protection can likely be developed to ensure long term detector safety.

Despite the imperfect understanding of HV in noble detectors, advances have been made based on accumulated experience.  While many previous liquid argon and xenon experiments were not able to meet their design high voltages by a significant margin, the recent ProtoDUNE prototypes have enjoyed greater success, much of it reported at the CPAD meeting. In a promising recent demonstration the field cage of the dual phase ProtoDUNE detector was charged to 500 V/cm over the drift region; breakdown was not observed to set in until a still higher drift field of 700~V/cm.   The LEM planes, however, were not operable at design field of 6.8~kV without high voltage discharge, being operated instead at a lower gain value of 5~kV/cm. The single phase ProtoDUNE detector was filled and charged to 180~kV, achieving the nominal 500~V/cm drift field. This represents a substantial achievement, but there remains much to understand  in the area of high voltage for large noble detectors.

\subsubsection{Noble Element Detectors for Dark Matter Detection}
\label{sec:DM}

\medskip
\noindent 
{\bf Present Status} \\
To push the sensitivity for WIMP detection several orders of magnitude beyond current levels, the next generation of direct detection experiments will continue progress towards larger target mass, with better shielding and active discrimination against backgrounds.
Each experiment must face the problem that linear improvement in sensitivity with increasing detector mass and running time continues only as long as zero or very low background condition are met.

Direct detection of dark matter, via elastic scattering of galactic WIMPs in a liquid argon or xenon target, is a technique making excellent progress towards probing the dark matter parameter space well below the current experiments' sensitivities. The detection properties of noble element liquids are particularly favorable for the rejection of radioactive background that produces Electron Recoils because these interactions produce light output with time constants of microseconds compared to nanoseconds for Nuclear Recoil events. The DEAP-3600 experiment has demonstrated an exceptional pulse shape discrimination against such background, projected to be in excess of $10^9$~\cite{Amaudruz:2018gr,Ajaj:2019wi}.
The DarkSide-50 experiment has demonstrated a two-phase detection technique that provides in addition efficient fiducialization~\cite{Agnes:2015gu,Agnes:2016fz}. Thus, liquid argon provides excellent sensitivity in the region of WIMP masses above $30$\,GeV/c$^2$ using single or two phase detection.  The two phase method also enables sensitivity for lower WIMP masses through the use of the electroluminescence signal alone~\cite{Agnes:2018fg,Agnes:2018ft}.Given the strong potential for the liquid argon technology to push the sensitivity for WIMP detection several orders of magnitude beyond current levels, scientists from all the major groups currently working with this technology (ArDM, DarkSide-50, DEAP-3600, and MiniCLEAN) have joined to create the Global Argon Dark Matter Collaboration (GADMC) to pursue a sequence of future experiments.The DarkSide-20k two-phase detector, first of the planned detectors, is currently in construction at the Gran Sasso Laboratory (LNGS) and aims at a total exposure of 200 ton-year.

\medskip
\noindent 
{\bf Major Challenges and Future Directions} \\
Some of the challenges for noble liquid based direct dark matter experiments are achieving extremely high intrinsic radio-purity of noble liquid and nearby detector components (both bulk and surface), large-area high-detection-efficiency photosensor array, precise calibration of the energy scale, timely selection of radio-pure materials via screening, and timely procurement and purification of large quantities for desired noble gas. 


\medskip
\noindent 
{\bf Radio-pure noble gases} \\
The future argon target for GADMC project will be procured in a timely fashion from the high-throughput extraction of low-radioactivity argon naturally depleted in $^{39}$Ar from underground sources (UAr) via Urania.
Urania is an underground argon extraction and purification plant capable of extracting $250$\,kg per day of UAr.
The total approximately $50$\,t of UAr necessary for this project then needs to be further purified by an high-throughput purification and active isotopic separation cryogenic distillation column called Aria. Aria is a $350$\,m tall cryogenic distillation column, capable of separating isotopes, and it is under construction in Sardinia, Italy.

\medskip
\noindent 
{\bf Large area photosensor array} \\
The need for a larger mass for the next generation noble liquid experiments directly translates in an increased number of light detection devices in order to maintain a good coverage of the detector.
Traditionally and supported by decades of operation, photomultiplier tubes (PMTs) were used.
Silicon photomultipliers  (SiPMs) are indeed promising candidates to replace PMTs in future detectors.
SiPMs have a number of performance advantages over traditional PMTs, including  higher photo-detection efficiency and an excellent single photoelectron resolution, all while operating at much a lower voltage.
SiPMs also have a compact geometry which allows them to be efficiently integrated into tiles that cover large areas.
Additional expected benefits include better radiopurity up to an order of magnitude than traditional PMTs, low costs, and scalable mass production. 

Photodetectors for xenon-based detectors require sensitivity to the 175~nm scintillation light from LXe. Significant improvements to the radioactivity of the photodetectors have been made and though problematic are not always a dominant source of background. The overall photodetection efficiency has been improved by polytetrafluoroethylene or Teflon (PTFE) reflectors in EXO-200 and LZ. DarkSide-20k is a proposed 30-tonne fiducial mass liquid argon TPC that will be outfitted with 200,000 SiPMs grouped into 8,280 single-channel, $25$\,cm$^2$ photosensors that have a photon detection efficiency better than 40 percent, a timing resolution around 10 ns, sensitivity to single photons, and a total rate from dark counts, correlated noise, and electronic noise less than 0.1\, Hz/mm$^2$. The development of the cryogenic front-end electronics for the DarkSide-20k photosensor are detailed in~\cite{DIncecco:2018hy} and a description of the full photosensor is given in~\cite{DIncecco:2018fx}. Photodetectors appear to challenge all of the noble liquid detectors. All are expensive, have significant radioactivity, or have marginal light collection, or some combination of these. Photodetectors are described in more detail in section~\ref{s.photodet}.

\subsubsection{Noble Detectors for Neutrinoless Double Beta Decay Searches}

\medskip
\noindent 
{\bf Present Status} \\
Neutrinoless double beta decay is a hypothetical process with a half life that experimental data suggest is longer than 10$^{26}$ years. Many new physics scenarios introduce neutrinoless double beta decay, which through the Schechter-Valle theorem would imply the neutrino is a Majorana particle. An observation of neutrinoless double beta decay would explain the lightness of neutrino mass, demonstrate lepton number is not conserved, and may be connected to leptogenesis as an explanation of the matter/antimatter asymmetry of the Universe. Accordingly, neutrinoless double beta decay is considered to be an experimental topic of prime importance in neutrino physics. It is presently stewarded by the Department of Energy Office of Nuclear Physics, although it attracts the attention of many high energy physicists, has great thematic overlap with many HEP priorities, and makes use of many similar technologies to neutrino and dark matter experiments.  Searches for $0\nu\beta\beta$ in various isotopes have been made at the 10-100~kg scale.  Here we focus on noble element detectors: time projection chambers using liquid (EXO-200~\cite{Auger:2012ar}, nEXO~\cite{Kharusi:2018eqi}) and gaseous (NEXT-White~\cite{Monrabal:2018xlr}, NEXT-100~\cite{Martin-Albo:2015rhw}, PANDA-XIII~\cite{Chen:2016qcd}, AXEL~\cite{Nakamura:2017tls}) xenon.  \footnote{We are not including detectors using xenon in liquid scintillator in the category of ``noble element detectors'' in the present scope, although there is active R\&D here also.}

The challenge of detecting double beta decay is ultimately a problem of background rejection.  To reach the target half-lives exceeding 10$^{28}$ years, backgrounds must be controlled at a level below 1 per count per ton per year.   The Nuclear Science Advisory Committee has recommended that the target levels of background for the next generation of experiments should be rejected at the level of 0.1 count per ton per year in the ROI for the coming generation of experiments.  The worlds best performance in any technology by this metric is from $^{76}$Ge diodes~\cite{Agostini:2018tnm} , and is around 4 in these units. The best demonstrated performance in xenon is from EXO-200, and is around $\sim$140 (although a fit over spatial distributions somewhat complicates the interpretation of the background level).  100~kg-scale gas detectors such as NEXT-100 project a value of around 7, with this demonstration expected in early 2020. All detector technologies thus require technological advances before backgrounds can be controlled at the required level at the 100~kg scale.

It is useful to distinguish between two classes of backgrounds: 1) The intrinsic background from the two neutrino decay mode; 2) Backgrounds arising from material radioactivity in the detector.  The former background will be fully rejected at the level of 0.1 count per ton per year in the ROI given energy resolution of better than ~2\% FWHM to distinguish the two-neutrino mode from the neutrinoless mode. For the letter class of backgrounds there are four approaches to further reduce backgrounds: improved energy resolution; improved radiopurity and shielding; improved topological reconstruction; and daughter tagging.  Since the ultimate goal of neutrinoless double beta decay collaborations is an ultra-low background detector, these can be considered to be four primary challenges for the field.  Further challenges may include separation of sufficient isotope to fill a very large experiment, calibration, and stable high voltage delivery.

\subsubsection*{Major Challenges and Future Opportunities}
{\bf Energy Resolution} \\
Energy reconstruction is a critical handle in neutrinoless double beta decay searches. Not only must be two-neutrino mode be rejected, requiring a resolution of better than 2\% FWHM (and advisably even better, given the possible presence of non-Gaussian tails), but contamination from radioactive backgrounds must be minimized by using the smallest energy window possible. Demonstrated energy resolutions at Q$_{\beta\beta}$ are 2.9\% in liquid xenon~\cite{Auger:2012ar} and 1\% in xenon gas~\cite{Renner:2018ttw}.  The demonstrated resolution is achieved in different ways between the liquid and gaseous experiments. 

In high pressure xenon gas experiments such as NEXT, energy is reconstructed from the ionization signature alone, which has a very small Fano factor of ~0.15 due to the lack of significant recombination.
The primary technological challenge is to amplify the detected ionization signal without introducing significant fluctuations. One approach is to use the process of electroluminescence, which involves providing a bias of 2 kV / cm / bar over a 5~mm to 1~cm EL gap. Achieving very large and stable electroluminescent planes in xenon gas is non-trivial, and is the subject of ongoing R\&D.  As well as traditional mesh-plane approaches, recent work by the AXEL collaboration in a honeycomb electroluminescence cell shows promise for large scale application~\cite{Ban:2017nnm}.

The PANDA-XIII collaboration proposes to use a Xenon-TMA gas mix with MicroMegas to detect ionization charge.  The optimal energy resolution that has been achieved to date in such a configuration is 3\% FWHM ~\cite{Galan:2015tgl}.  If energy resolution in such devices can be substantially improved, this would be a major advance.

In liquid xenon, recombination is substantial and this introduces fluctuations, forbidding a precise energy measurement using ionization alone. However a partial cancellation of these fluctuations can be  achieved by detecting the anti-correlated scintillation signal. Using this strategy, EXO-200 has demonstrated energy resolutions of $\sim$2.9\% FWHM~\cite{Auger:2012ar}.  The nEXO collaboration is undertaking R\&D on novel light collection strategies using densely packed planes of VUV SiPMs~\cite{Sun:2018tkw}, and novel charge collection strategies using pixelated charge readout pads~\cite{Jewell:2017dzi}, to further improve this energy resolution, at large scales. 

It is notable that the energy reconstruction strategies in both nEXO and NEXT require high efficiency collection of 175~nm light.  In both liquid and gaseous xenon, as in many other noble element-based physics experiments, large area, low-radioactivity and low-cost photon detectors for VUV light would represent a substantial advance.  

\medskip
\noindent
{\bf Material Screening and Radiopurity} \\
All double beta decay collaborations have undertaken extensive campaigns of material screening and assay.  A rule of thumb is that no amount of assay is ever enough, and while radiopurity databases such as {\tt Radiopurity.org} have improved the coherence of the community, sharing of pertinent information between collaborations remains inefficient.  Improved access by all collaborations to radioassay facilities and better coordination between $0\nu\beta\beta$ and dark matter collaborations with similar needs would surely advance the field.   Continued development of low radioactivity materials and handling techniques will continue to enhance the sensitivity of low background experiments, through a process of iterative R\&D that will surely continue for decades.

\medskip
\noindent
{\bf Topology} \\ 
Access to topological information is a major advantage that noble TPCs hold over many other double beta decay technologies.  In liquid xenon, an important handle is single vs multi-site discrimination, which helps to reject multiple scattering gamma ray and neutron backgrounds, and also constrains backgrounds in a multivariate fit.  Xenon gas allows additional topological discrimination, distinguishing one-electron from two-electron events by counting Bragg peaks~\cite{Ferrario:2015kta}.  Advances in reconstruction techniques including Deep Neural Networks have been used by both liquid and gaseous xenon experiments to further improve background rejection~\cite{Delaquis:2018zqi,Renner:2016trj}.  Enhancing topological information in the presence of diffusion at larger detector scales via the introduction of diffusion-reducing additives is an active area of R\&D within the gaseous double beta decay detector community~\cite{NEXTHelium,Azevedo:2015eok,Henriques:2018tam,Felkai:2017oeq,Henriques:2017rlj}. PANDA-XIII, on the other hand, plans to use this diffusion to measure the Z position of events to reject cathode backgrounds, due to the presence of quench gas that absorbs scintillation light. Achieving optimal topological capabilities will also require dense, multiplexed readout planes, which must be radio-pure and have manageable heat load, a goal that has much commonality with pixel schemes for liquid argon TPCs.

\medskip
\noindent
{\bf Daughter Tagging} \\
It has long been recognized that efficient detection or ``tagging'' of the barium daughter ion in neutrinoless double beta decay would enable a new class of ultra-low-background experiments in $^{136}$Xe~\cite{Moe:1991ik}. Barium tagging is a complex and multi-facetted topic, with much R\&D still to be untertaken.  We refer elsewhere for details~\cite{Brunner:2014sfa,Mong:2014iya,Jones:2016qiq,Bainglass:2018odn,Byrnes:2019jxr,Sinclair:2011zz}. Recent progress by the NEXT and nEXO collaborations has demonstrated single ion sensitivity using single molecule fluorescence imaging at transparent surfaces~\cite{McDonald:2017izm} and atomic fluorescence in frozen films under vacuum~\cite{Chambers:2018srx}, respectively.  R\&D is ongoing in both collaborations to integrate such detection techniques into a double beta decay medium.  

The challenges of this integration are substantial, and distinct between technologies. Gas-phase experiments favor an in-situ approach to detection, whereas in liquid, insertion of a robotic probe to remove the barium ion to an external scanning system is the preferred approach~\cite{Brunner:2014sfa}.  An efficient barium tagging technique may revolutionize neutrinoless double beta decay searches and substantially change the experimental landscape, since if this tagging were suitably efficient, the combination of precise energy resolution to reject the two neutrino mode with a barium tag would be sufficient to deploy a highly sensitive experiment at arbitrarily large scale, in principle.  Significant R\&D is still required before the ultimate form or active medium for a barium tagging $0\nu\beta\beta$ experiment is clear.

\medskip
\noindent
{\bf High Voltages and Long Drift Lengths} \\
Construction of larger detectors necessitates delivery of higher voltages.  The nEXO collaboration has a continuing program of R\&D on high voltage delivery to liquid xenon, as well as a program to study the effect of longer drift lengths needed for a multi-ton detector, using a switched purity monitor.  NEXT has studied and published the HV strength of polymers in xenon and argon gases to inform the development of supports for large EL planes~\cite{Rogers:2018lle} and is pursuing a continued program of HV R\&D for high pressure gases. Demonstration of HV solutions for future larger detectors will be critical for deployment of ton-scale and larger $0\nu\beta\beta$ experiments.

\medskip 
\noindent
{\bf Calibration} \\
Calibration of very large detectors is challenging, especially in cases where self-shielding prevents the use of external gamma sources. Internal gamma sources may be deployed robotically, although it is non-trivial to imagine how this scheme could be implemented without disturbing electric fields.  Alternatively, dissolved calibration isotopes such as $^{83m}$Kr may be used for maing spatial maps, though not at the energy renage of relevance for $0\nu\beta\beta$~\cite{Rosendahl:2014qqa,Martinez-Lema:2018ibw}.  Ensuring suitable flow of fluid to calibrate the full volume is likely to be a complex matter.  For detectors aiming to realize energy resolutions of better than 2\% FWHM, regular and extensive calibrations will be vital, and so this is an important topic of R\&D for any monolithic approach to $0\nu\beta\beta$ at the ton-scale or beyond.

\medskip
\noindent
{\bf Isotope Enrichment} \\
Next-generation experiments will require one to several tons of isotope, depending on technology.  It is conceivable that a future generation may need an order of magnitude more still. Enrichment will certainly be a costly and time-consuming process, and the capabilities to enrich such quantities of $^{136}$Xe do not presently exist in the United States. Developing of enrichment capability in the US may be important for the future of the $0\nu\beta\beta$ program.

%% file: Technologies/MPGD.tex
\subsection{Micro-Pattern Gas Detectors}
\label{s.mpgd} 

Since its foundation, the RD51 collaboration has provided important stimulus for the development of Micro Pattern Gas Detectors (MPGDs) and focused on a broad networking effort to share and disseminate the "know-how" and the technologies, and to promote generic R\&D: a seminal activity for the enlargement of the application portfolio~\cite{rd51}. Important consolidation of the some better-established MPGD technologies has been achieved within the CERN RD51 collaboration, often driven by the working conditions of large collider experiments; in parallel, the portfolio of MPGD applications in fundamental research has been enlarged in other science sectors.  Envisaging the needs for R\&D activities on MPGD, RD51 is structured in several working groups focusing on various aspects of gas-detector technologies: detector physics and measurements, new technologies, simulations and modeling, electronics, production techniques, industrialization, and common testing facilities. This structure facilitates interactions within the community and effectively focuses efforts and resources. Over the years, the number of collaborating Institutes from Europe and other continents, like from China, India, Japan, USA, has greatly increased - reinforcing the RD51 world-wide vocation and enhancing the geographical diversity and expertise of the MPGD community; today it comprises of 90 institutes in 25 countries. Highlights of the future R\&D scenario, based on the ongoing projects and plans to explore new materials and technologies, are summarized in the recent proposal, dated May 2018~\cite{rd51a}. The Research Board extended RD51 for a further five years beyond 2018, following the LHCC recommendation stating: “The LHCC considers the working mode of RD51, with a small but focused core team and corresponding infrastructure at CERN, attracting contributions and bright ideas to be explored from collaborators around the world, to be an excellent setup~\cite{rd51b}.”

\subsubsection{MPGD Applications}
During the past 10 years, the deployment of MPGDs in operational experiments has increased enormously, and RD51 serves a broad user community, driving the MPGD domain and any potential commercial applications that may arise. Nowadays, several MPGD technologies, such as: Gas Electron Multiplier (GEM), Micro-Mesh Gaseous Structure (MicroMegas, MM), THick GEMs (THGEM), also referred to in the literature as Large Electron Multipliers (LEM), GEM-derived architecture 
($\mu$-RWELL), Micro-Pixel Gas Chamber ($\mu$-PIC), and integrated pixel readout (InGrid) are being optimized for a broad range of applications. Technological and scientific advances have led to improved detector performances - also when scaling up these technologies to large-area detectors. During the last five years, there have been major MPGDs developments for ATLAS, CMS, ALICE and COMPASS upgrades, towards establishing technology goals and technical requirements, and addressing engineering and integration challenges. The consolidation of the better-established techniques has been accompanied by the flourishing of modern ones, often specific to well-defined applications. Novel technologies have been derived from Micromegas and GEM concepts, hybrid approaches combining different MPGDs, gaseous with non-gaseous multipliers; others are based entirely on new concepts and architectures. Thus, the potentiality of MPGD technologies became evident and the interest in their applications has started growing not only in HEP, but also in hadron/heavy ion /nuclear physics, photon detectors and calorimetry, neutron detection and beam diagnostics, neutrino physics and dark matter detection, X-ray imaging and $\gamma$-ray polarimetry. Beyond fundamental research, MPGDs are in use and considered for applications of scientific, social and industrial interest; this includes the fields of medical imaging, non-destructive tests and large-size object inspection, homeland security, nuclear plant and radioactive-waste monitoring, micro-dosimetry, medical-beam monitoring, tokamak diagnostic, geological studies by muon radiography.

\subsubsection{RD51 Legacy, Expertise, Infrastructures and Dissemination} 

The main objective of the RD51 collaboration is to advance technological development and application of MPGDs. It is a worldwide open scientific and technological forum on MPGDs, and RD51 has invested resources during ten years in forming expertise, organizing common infrastructures and developing common research tools; it continues to advance the MPGD domain with scientific, technological, and educational initiatives, such as Academy-Industry matching events, MPGD training sessions and an international MPGD conference series. The progress in various R\&D projects has been made possible by open access to facilities and research tools. 

In addition to the support mechanisms and facilities tools, the RD51 portfolio is diversified, including: maintenance and development of simulation software dedicated to gaseous detectors, development of SRS~\cite{rd51-readout}, a complete read-out chain designed to operate in a laboratory context, also expandable to large read-out systems, realization of affordable laboratory instruments dedicated to MPGD developments, and more. Last, but not least, the RD51 community has open access to the instrumentation, services and infrastructures of the large and well-equipped Gas Detector Development (GDD) laboratory at CERN, continuously hosting several parallel R\&D activities. In addition, the common test beam infrastructure at the H4 test-beam area at SPS, available usually three times a year during the periods of beam availability for RD51, allows several groups to investigate in parallel their R\&D projects.     

\subsubsection{RD51 R\&D Program on Advanced MPGD Concepts}
The better understanding of the physics processes, originating from experiment-validated model simulations, paved the way towards novel detector concepts. A clear direction for future developments is that of the resistive materials and related detector architectures. Their usage improves detector stability, making possible higher gain in a single multiplication layer, a remarkable advantage for assembly, mass production and cost. The frontier of fast and precise timing is moved forward by novel developments for picosecond-time tagging or filtering measurements, e.g. for high-luminosity colliders. Very encouraging trends are obtained by coupling gaseous detectors and Cherenkov radiators to take advantage from the prompt radiation emission. A variety of novel opportunities is offered by MPGD hybridization, a strategy aiming to strengthen the detector performance combining the advantages offered by a variety of approaches. It is obtained both by combining MPGD technologies and by coupling gaseous detectors to different detection technologies, as is the case for optical read-out of gaseous detectors. Contributions to the MPGD concepts from up-to-date material science are required for several domains: resistive materials, solid-state photon and neutron converters, innovative nanotechnology components. The development of the next generation of MPGDs can largely profit of emerging technologies as those related to nanomaterials, MicroElectroMechanical Systems (MEMS), sputtering, novel photoconverters, 3-D printing options, etc.  This is just a partial list of fascinating R\&D lines that MPGDs will see in the years to come.

\subsubsection{Overview of US-based MPGD activities}
The US community has been for a long time engaged in the development and application of MPGD detectors. A brief overview of some of those activities are listed below:
\begin{itemize}
    \item 
Particle physics (LHC, ILC): one of the research efforts refers to GEM-based readout for a hadronic calorimetry at the ILC. Several US groups are also involved in the construction of GEM detectors for the CMS muon system upgrade for HL-LHC;
\item 
Nuclear physics: ongoing experiments at JLAB, such as the SuperBigBite for Hall A and future efforts at SoLID, are based on GEM-based tracking, while CLAS12 uses MM-based tracking system. At RHIC at BNL, the STAR experiment employed triple-GEM detectors for a forward tracing; the future sPHENIX experiment will use GEM-based TPC readout. Several US institutions were involved in the construction of a GEM-based readout for the ALICE TPC. MPGDs are also used at the FRIB facility. The nuclear physics community has been also engaged in activities of technology transfer to industry for GEM foil commercialization with support by US SBIR (Small Business Innovative Research). 
\item 
US-based Electron-Ion Collider: several ongoing funded MPGD 
R\&D programs are based on: GEM, MM or $\mu$-RWELL as TPC sensors (synergies with sPHENIX); low-mass large-size GEM or $\mu$-RWELL trackers; zigzag shaped charge collecting anodes coupled to GEMs for optimized space resolution; a five-layer GEM photon detector for a window-less RICH detecting photons in the very far UV domain ($\approx$~120~nm); hybrid MPGDs (THGEMs + MM, synergies with COMPASS) for a single photon detection with miniaturized pads; coupling of gaseous photon detectors to innovative photo-converters by hydrogenated nano-diamond powders; a GEM-based detector acting both as  TRD and as tracker. 
\end{itemize}

\subsubsection{Summary and Conclusions}
Future RD51 activities aims at bringing a number of detector concepts to maturity, initiating new developments and continuing the support to the community in order to preserve and enrich the present scenario, including: networking activity, with focus on training and education, further development of dedicated simulation tools, advances in electronics tools, continuation of the collaborative interactions with strategic CERN workshops. RD51 and MPGD success is related to the RD51 model in performing R\&D: combination of generic and focused R\&D with bottom-top decision processes, full sharing of “know-how”, information, common infrastructures. This model has to be continued and can be exported to other detector domains.

%% file: Technologies/Silicon.tex
\subsection{Silicon Detectors}

\subsubsection{Silicon-based Detectors in Cosmology}
\label{s.sicosmo}

In the coming decade, detectors consisting of hundreds to thousands of silicon sensors will be deployed in diverse experiments to understand the fundamental nature of dark matter and dark energy.
These detectors face common challenges in terms of scalability, readout and mitigation of noise sources.

\subsubsection*{Astronomical CCD cameras}

Scientific silicon charge-coupled devices (CCDs)~\cite{janesick} are the sensors of choice for astronomical telescope cameras~\cite{decam,lsstcam}.
Ambitious imaging surveys over large regions of the sky~\cite{des,lsst} and spectroscopic surveys of millions of galaxies~\cite{desi} have pushed CCD technology forward in the past decades.
Developments toward lower noise sensors with increased quantum efficiency in the infrared~\cite{ccdtech} provide higher sensitivity to the distant faint and red objects from the history of the early universe, allowing for more expansive surveys.
These cameras have been instrumental in the measurement of the large scale structure of the universe and cosmological parameters~\cite{desres}, providing insights into the properties of dark energy and dark matter.

The latest project is the 3.2 gigapixel camera for the Large Synoptic Survey Telescope (LSST)~\cite{lsstcam}, which will start operations in northern Chile in 2022.
The camera's large focal plane is divided into 21 modules, consisting of nine 16 megapixel CCDs closely tiled on a plane, together with an integrated package of electronic, mechanical, and cryogenic support components~\cite{lsstraft}.
A new development is the use of a dedicated CMOS video-processing application-specific integrated circuit (ASIC) to read the 144 channels of each module (16 per CCD), which is necessary to satisfy the stringent compactness and low-power requirements while achieving a read time of 2 seconds per frame~\cite{aspic}.
The construction of the LSST camera demanded a major and exhaustive effort in quality control: all module components (CCDs, electronics boards, etc.) were extensively tested at Brookhaven National Laboratory (BNL)~\cite{lsstraft}, first individually~\cite{lsstccd} and then once assembled into a module.
Finally, a test in a camera integration stand will be performed at SLAC National Accelerator Laboratory~\cite{lsstint} to conclusively demonstrate the required mechanical, cryogenic, electronic and optical performance.

\subsubsection{Silicon-based Detectors for Dark Matter Detection}
\subsubsection*{CCD arrays to directly search for dark matter}

Arrays of scientific CCDs have also been deployed in underground laboratories to directly search for interactions between dark matter particles and silicon atoms~\cite{damic,damichp}.
The large thickness (hundreds of micrometers) of fully-depleted sensors, with the original purpose to increase infrared sensitivity, leads to a significant target mass (10 grams) per device, while their extremely low readout noise and leakage current allow for very low energy thresholds.
The SENSEI collaboration has shown readout noise as small as 0.06\,$e^-$ with a ``skipper'' CCD designed by Lawrence Berkeley Laboratory~\cite{skipper}, which resolves individual charges by averaging over a large number of repetitive, uncorrelated measurements of the pixel charge.
The DAMIC collaboration has measured leakage currents as small as $<$$10^{-21}$A\,cm$^{-2}$ in similar devices~\cite{damichp}.
Single charge resolution has also been demonstrated in depleted field effect transistors (DEPFET)~\cite{depfet} and quasiparticle-trap-assisted electrothermal-feedback transition edge sensor (QET)~\cite{qet} silicon detectors.
The extremely low instrumental noise of these new technologies will allow for the search for interactions that produce only a few charges in silicon (from a few electronvolts of deposited energy), with sensitivity to dark matter particles with sub-GeV masses from their interactions with both silicon nuclei~\cite{lewinsmith,pradler,migdal} and valence electrons~\cite{essig,hochberg}.
Small prototype detectors have yielded first results~\cite{sensei,cdmser}.

Further development of CCD technology is attractive because of its scalability and broad applications in cosmology, imaging and radiation detection.
Current skipper CCDs are small (2 megapixels, 1 gram) and their readout rate is slow (1 millisecond per pixel), which severely limits their application for imaging and spectroscopic surveys.
Readout rate can be increased by increasing the number of readout amplifiers per CCD, decreasing the amplifier noise so that the same performance can be achieved in a shorter time, e.g., with a lower capacitance floating gate, and implementing electronics with higher functionality (e.g., ASICs or FPGAs) for signal readout.
Further increases in size would be most beneficial for direct dark matter searches.
The DAMIC-M detector~\cite{damicm}, which is scheduled to start operations in the Modane Underground Laboratory in France in 2022, will feature a tower of 50$\times$36\,megapixel skipper CCDs\textemdash the largest that can be fabricated from 150-mm wafers\textemdash for a silicon target mass of 1\,kilogram.
A particular challenge for DAMIC-M are the stringent constraints in its design and construction to limit backgrounds from cosmic rays and natural radioactivity.
The proposed background-mitigation strategy includes detector shielding, careful screening and selection of materials, proper handling of detector components to minimize cosmogenic activation and surface contamination, and background rejection strategies based on the high spatial resolution of the detector~\cite{spatcoinc}.
Furthermore, the single-charge resolution makes the detector sensitive to infrared photons inside the cryostat, requiring extreme care in the heat dissipated and luminescence by the large number of active components on the CCDs and nearby electronics.

The future of scientific CCDs beyond the early 2020s is uncertain because the conventional CCD technology on 150-mm diameter wafers is fast becoming obsolete.
Thus, it should be an R\&D priority to migrate thick-CCD fabrication to CMOS foundries that process 200-mm wafers (e.g., using CCD-in-CMOS technology~\cite{ccdincmos}).
The more advanced CMOS process nodes would also help significantly in realizing higher-performance lower-noise readout amplifiers, broadening the application of skipper CCDs in imaging and spectroscopy.
Furthermore, the possibility to fabricate CCDs on few-millimeter thick wafers at nanofabrication research facilities or modern specialty foundries should be explored. Wafer bonding technology can also be used to build sensors with large sensitive volume by bonding CCD wafers to thick float zone substrates.
Next-generation dark matter searches will require hundreds of these very massive (100 gram) CCDs to be deployed in an extremely low radiation environment.
Such engineering effort would be at the scale of the LSST camera but with particular focus in the use of low-radioactivity materials for the CCD package and support structure.
A strong R\&D program to mitigate radiation backgrounds for modular silicon targets in the 10-kilogram scale would be beneficial for any proposed experimental technology (CCD, DEPFET or QET) to directly search for sub-GeV dark matter.

\subsubsection*{Germanium CCDs}

Beyond larger, faster and lower-noise silicon CCDs, recent progress in the fabrication of germanium CCDs offers another possibility for the future.
Germanium devices have a higher detection efficiency (better response) in the infrared and high-energy X ray bands, and better sensitivity to certain dark matter particle candidates.
High-quality gate dielectrics are now available to realize the metal-oxide-semiconductor (MOS) gate structure in germanium~\cite{delabie,ballenger,matsubara}.
MIT Lincoln Laboratory has recently demonstrated promising performance of 128$\times$128 pixel arrays fabricated from 200-mm diameter germanium wafers using the same tools to build silicon CCDs~\cite{leitz}.
Current limitations in performance arise from bulk trapping, likely from trace levels of metallic contamination.
The development of metal gettering processes in the fabrication of germanium devices could significantly improve their performance and size, as it did for silicon CCDs~\cite{getter}.

\subsubsection*{Si(Li) detectors to indirectly search for dark matter}

The tracking calorimeter of the General Antiparticle Spectrometer (GAPS) experiment~\cite{gaps1,gaps2} is an upcoming large silicon detector to indirectly search for dark matter.
GAPS is the first experiment optimized specifically for the detection of cosmic antideuterons and antiprotons with energies $<$0.25\,GeV/n as signatures of dark matter annihilation or decay~\cite{antideu}.
To achieve sensitivity to cosmic antinuclei in this unexplored energy range, GAPS uses a novel particle identification method based on atom capture and decay~\cite{adcd}.
The instrument consists of ten planes of 1000 10 cm-diameter, 2.5 mm-thick lithium-drifted silicon Si(Li) sensors~\cite{sili}.
Although much lower resolution (in energy and space) than the CCD arrays used to directly search for dark matter, Si(Li) detectors provide the active area ($>$10\,m$^2$), stopping power and performance required for the proposed particle identification technique, all within the significant temperature, power and cost limitations of an Antarctic long-duration balloon mission.
Recent technological advances have focused on the lithium evaporation, diffusion and drift methods and on the use of high-quality crystalline silicon to allow for a uniform active detector region.
In addition, significant progress has been achieved in the suppression of the leakage current in these large-area, high-temperature (230\,K) devices.

\subsubsection{Ionization pixel detectors with integrated-circuit readout}
\label{s.ionic}

Hybrid pixel detectors with ASIC readout offer the flexibility that may be necessary for certain detector applications.
Recently, a thick fully-depleted photodiode array was bump bonded at BNL~\cite{tpxcam1,tpxcam2} with the Timepix3 readout chip developed by CERN~\cite{timepix}, providing the same quantum efficiency as a CCD imager but with much higher time resolution~\cite{tpxcam3}.
Each pixel in the 256$\times$256 array is triggered and readout independently, and is capable of recording the time of arrival (TOA) of the signal with 1.5 nanosecond resolution and its time over threshold (TOT) to measure its amplitude with a resolution of 100 $e^-$.
By further placing a microchannel plate (MCP) with a phosphor screen in front of the pixelated device, it is possible to count with high time resolution single ions and electrons for applications in ion and electron imaging~\cite{ionim}.
Single photon counting was also demonstrated by placing a photocathode in front of the MCP, i.e., an ``image intensifier,'' with applications in quantum information science (QIS)~\cite{tpxqis}.

As the ASIC development progresses, e.g., with Timepix4 or conventional CMOS active-pixel readout, the size of the pixel arrays, and the time, energy and spatial resolution of the devices are expected to increase significantly.
Future developments include the use of silicon sensors with internal amplification~\cite{spadfight} to improve single-photon detection efficiency, and the interface of the readout ASIC with novel sensor materials, e.g., germanium, gallium arsenide (GaAs)~\cite{gaasnir}, mercury-cadmium telluride (HgCdTe) and amorphous selenium (aSe)~\cite{aserev, karim}. 
The properties of these materials would lead to pixelated devices with higher sensitivity in the infrared and hard X ray bands, with applications in astronomy, X ray science~\cite{ulvestad} and medical imaging~\cite{antonuk}.
Hybrid pixelated aSe detectors could also be used to image with high energy and spatial resolution electron tracks in aSe, providing a strategy for particle identification to suppress backgrounds from natural radioactivity in a next-generation search for neutrinoless $\beta\beta$ decay with $^{82}$Se~\cite{selena}.

\subsubsection{Silicon Detectors for Collider Experiments}

Projects are now underway to build the detectors suitable for the harsh conditions of HL-LHC running, and planning is ongoing to develop detectors for possible future projects such as the HE-LHC, ILC,  large circular e+e- and pp-colliders, and more.

A vertexing and tracking instrumentation research program for the coming decade will have to address the need for exquisite tracking precision (spacial and timing), all in the face of  extremely high event occupancy. Individual events will have to be identified within beam crossings that may involve more than 1000 events in a 10~cm
crossing region. 
Ideally, one would like to develop a device that will measure both space and time coordinates of a particle at a precision around 10~$\mu$m and 10~ps. 

Radiation damage continues to be a daunting challenge, for all components from front-end electronics to sensors. For example, the requirements for the HL-LHC tracker upgrade are a radiation tolerance up to 1MeV neutron equivalent fluence of 2.3~10$^{16}$ n$_{eq}$/cm$^2$~\cite{HL-LHC CMS}.

We will need continued development of extremely low mass ($<$1$\%$ in pp) devices, and control cost and complexity. Pixel detectors cover more and more area, and bump-bonding in hybrid silicon detector systems is the cost driver in currently used technology.
 
Progress in ASIC  technology  is  essential  for tracking devices, as we continue to take advantage of the scalability to large number of channels,  and  the  increasing  complexity  of  the  digital  back-end,  to  construct  large systems.

Relevant and promising technologies include development of sensors based on Low Gain Avalanche Diodes (LGADs), CMOS active pixels, 3D integration of sensor and readout layers, 3D pixels, and radiation hard ASICs. 

\subsubsection*{Development of fast timing sensors based on LGADs}

Ultra-fast detectors based on Low Gain Avalanche Diodes (LGADs) are promising candidates to address timing and spatial resolution \cite{INFN2}. They are n-on-p sensors with internal charge multiplication due to the presence of a high field region below the junction, formed using a deep "reach through" implant. Both ATLAS and CMS upgrades will include a LGAD-based timing Detector in front of the end cap or forward calorimeters to help mitigate the effect of pile-up by tagging primary vertex locations. The chosen sensors are arrays of 50 ${\mu}$m thin LGADs~\cite{seiden}. Current generations of LGADs suffer from low fill factor and only moderate radiation hardness. Studies to improve radiation hardness focus on optimization of the implant that defines the gain layer. As devices are irradiated dopants are removed, reducing gain. To combat this effect, carbon doped implants, bonded wafers and engineered silicon substrates with sharply graded epitaxial layers are all being developed. AC coupling of the anode implant ~\cite{INFN} can solve problems associated with high fields near the pixel edge. Structures to mitigate the fields limit the fill factor. AC coupling eliminates these inter-pixel structures, allowing for smaller pixels, and providing detectors with gain that can be used in a variety of applications.

\subsubsection*{Monolithic Silicon Pixel detectors (MAPS)}

Pixel detectors using  optimized  
CMOS  monolithic  pixels are low-mass and low-cost technologies, because of an active industrial market. CMOS Pixel Sensors allow integration of the full signal processing circuitry on the sensor substrate with no need for bump bonded interconnect. Charge is collected in the underlying epitaxial layer and diffuses or drifts to the collecting electrode.  Full depletion is needed for fast time response and radiation hardness. The geometry of a small collecting node in a larger pixel makes it difficult to fully deplete MAPS sensors, requiring specialized designs. Custom processes are also needed to reduce parasitic charge collection by transistor n-wells. CMOS sensors are naturally limited to the $\approx 2 \times 3 ~cm$ size of a CMOS reticule. Foundries can provide ``stitching'' technology to combine these reticules into larger sensors. Good relations with the IC foundries are essential for the development of these sensors.

LF-Monopix~\cite{monopix}, a fully depleted monolithic active pixel sensor prototype, is a proof of concept of a fully monolithic sensor capable of operating in the environment of outer layers of the ATLAS Inner Tracker upgrade in 2025 for the High Luminosity Large Hadron Collider (HL-LHC). This type of device has a lower production cost and lower material budget compared to presently used hybrid designs.

\subsubsection{3D Si sensors}

3D silicon sensors, where electrodes penetrate the silicon substrate fully or partially~\cite{3D pixel}, have been studied, for example as a solution for the innermost layers for the HL-LHC detectors. The deep electrodes provide for full charge collection from the full detector thickness with reduced drift distance, mitigating the effects of charge traps and providing fast charge integration. 3D detectors are more radiation hard than conventional sensors, but have large pixel capacitance and fabrication is relatively complicated and expensive. Active edge technology uses similar electrode formation techniques to eliminate dead regions around the edges of the sensor. Sensors fabricated with active edges could be tiled to provide seamless large area coverage.

Efforts continue to develop improved methods and tools, such as simplified processes and performance enhancing innovations such as the introduction of multiplication regions.

\subsubsection{Substrate engineering}
3D sensor technology is an initial example of the possibilities available by combining traditional planar fabrication with capabilities enabled by nanotechnology.  It is well established that thin sensitive regions are required for radiation hardness as well as for fast timing. Such thin wafers are not compatible with standard fabrication equipment. Structures such as Silicon-on-Insulator (SOI), Silicon-Silicon (Si-Si) bonded wafer stacks or deposition of high resistivity epitaxial layers have been developed to allow thin sensitive substrates on thick support wafers.  Alternately wafers can be physically thinned and then annealed using lasers or newly developed microwave techniques. Both of these technologies control the thermal budget and limit the diffusion of dopants inherent in traditional thermal annealing. MIP and x-ray response of the sensors can also be tailored by shaping profiles of dopants by graded epitaxial layers or by bonding wafers of varying resistivities. 

\subsubsection{3D integrated IC and small pixel sensors}

3D integration of sensor and readout layers use a set of technologies and processes which allow for fabrication of devices with vertically stacked interconnection. These technologies, initially demonstrated for HEP and x-ray applications, are now in large-scale commercial production for cameras. They have the capability to connect multiple layers of  electronics to pixels with a few micron pitch. This raises the possibility of small pixels ($< 25\mu m$) combined with sophisticated readout electronics. Low capacitance and noise allow for electronics that can use prompt induced current, rather than collected charge, for reconstruction of the pulse shape, mapping the pattern of charge deposition in the sensor. A detector combining these technologies can provide $\approx 10$ picosecond time resolution, track angle measurement, and few micron hit position in a single layer.
3D is also a highly effective technology to reduce form factor and allows for 
heterogeneous integration without the costs associated with smaller feature size~\cite{3D IC}.

\subsubsection{Direct access to industrial vendors and foundry processes}
In the development of silicon detectors for cosmology and collider experiments, it has become absolutely imperative to collaborate as closely as possible with the semiconductor industry. Progress is dependent on direct access to industrial vendors and foundry processes. This is not an easy task even for mid-size groups, as submissions are extremely costly and often involve atypical needs. Development and testing requires many low-volume iterations. We need mechanisms for members of the community to pool resources and person-power to gain access. Facilities such as the Silicon Detector Facility (Fermilab), the Semiconductor Detector Laboratory (Berkeley Lab) and the Microsystems Laboratory (Berkeley Lab, contact: Stephen Holland) are important centers of technology knowledge and infrastructure that need continued support. MIT-Lincoln Labs (contact: Vyshi Suntharalingam) is an important intermediary partner. It is also essential to integrate the University research community to ensure training of the next generation of experts.

%% file: Technologies/Photodet.tex
\subsection{Photodetectors}
\label{s.photodet} 

The detection of photons is fundamental to the detection of particles. It cuts across the programs at DOE HEP and extends far beyond. Relevant technologies are the Noble Detectors, Silicon, Quantum Sensing and Cryogenic Detectors. Furthermore, the analysis of photons in a variety of detectors using Machine Learning could be of interest. The discussion here focuses on the particular applications and requirements for these detectors and asks the question of how we push beyond the current limits of photodetection.

The photomultiplier tubes(PMT) has been the workhorse of photodetection for over fifty years. It provides robust, low noise, detection of single photons with nanosecond timing. They range in size from 1~cm up to 50~cm. Solutions have been developed for operation at cryogenic temperatures and in high magnetic fields; however these extreme environments remain challenging.  The R\&D major efforts can be broadly categorized as follows:
\begin{enumerate}
\item MCP based detectors as a direct replacement for the PMT that has improved timing and magnetic field response.
\item Arrays of SiPMs for cryogenic detectors.
\item Novel solid state detectors.
\item Photon sensors at the quantum limit.
\end{enumerate}

The LAPPD (Large Area Picosecond Photodector) is intended as the direct replacement for the PMT. It makes use of microchannel plate (MCP) technology and therefore its flat profile drastically improves the timing to  $<$100~picoseconds and makes it more immune to magnetic fields. The current design is an 8-inch square so provides the large area needed to directly compete with the equivalent large form-factor PMTs. Incom Inc. has started to produce LAPPDs on an industrial scale with LAPPD~\#37 achieving both $10^7$ gain and a photocathode efficiency of 24\%.  Looking towards the next generation of LAPPD, nanocomposite materials can be tuned to optimize the electrical properties of the MCP.  Incom Inc. and the University of Chicago groups are producing devices with a more robust capacitively coupled anode while the University of Chicago is also making good progress on an air transfer assembly process that could lead to a more economic production compared to the vacuum assembly process. 

The LAPPD is now ready for early adopters to start working with the device. The first experiment using LAPPDs, a time of flight detector, was operated in the electron test beam at Fermilab. The device is also now being adapted to construct a highly pixelated detector that can operate in a 2-3~T magnetic field at the future EIC collider. 

Silicon Photomultipliers (SiPMs) are solid state detectors. They are small but economical, and when cooled, have dark noise levels competitive to PMTs. They find natural application in detectors that already require cooling like the liquid nobel TPCs. The detectors have been driving the development of arrays of SiPMs and cross-talk and afterpulse effects are now within specifications. One of the challenges has been producing devices that are themselves sensitive in the VUV, therefore limiting the need for secondary wavelength shifters. A group from TRIUMF is working in this area and the Sherbrooke group is assembling arrays of 3D digital SiPMs. Their design shows promise for both its low power consumption and impressive $\sim$10~ps timing.

Looking beyond the SiPM, two groups from Dartmouth are working on novel solid state detectors. An ultra-low noise CMOS-based single photon detector with excellent imaging capabilities is being developed based on a single photon quantum-dot detector that directly couples the quantum dots to the amplification circuit. Sandia National Laboratory is working on a single photon detector based on carbon nanotubes, which was inspired by biological systems. 

The above technologies are operated at temperatures ranging from  liquid nitrogen temperatures up to room temperature.  In general, great advances are now being made with detectors based on superconductivity. Photodetectors are an important subset of these detectors, especially those with applications for cosmology. The nature of superconductivity, allows these detectors to push measurement to the point that quantum effects become the dominant effect limiting the sensitivity of the technique. Two groups are working on superconducting-based single photon detectors. The Yale group has a nanowire detector with a fully integrated nanophotonic readout. This design allows the device to obtain low timing jitter, 18.4~ps, and high absorption efficiency, 99.9\% at 775~nm. The MIT group has a similar nanowire-based detector, which makes use of quantum optics to measure a buildup of single-photon states in a system coupled to a series of dielectric layers due to the interaction of low-mass dark matter.

Advances in photodetector technology naturally push developments in the other components of the detector. 
Work is ongoing to demonstrate that ultra-fast timing can be used to resolve the development of hadron showers and therefore improve the resolution of hadron calorimeters. These advances will require fast and radiation hard scintillation materials. Work is also ongoing on the performance and selection of fast and ultra-fast scintillating crystals.  Related to this work, capillaries filled with liquid wavelength-shifter show promise as a light guide for intense radiation field regions. A very novel development for a promising new dark matter detector is the development of a chamber containing super-cooled water,  which images the phase transition induced by nuclear recoils.

The efforts in this area is representative of the many different technologies being harnessed to improve photodetectors. Since photodetectors are the foundation of most detectors the critical nature of this work cannot be oversold. We are beginning to see new detector technologies emerge from this work with many more advances on the horizon as these detectors move from R\&D to steady production and even large-scale commercialization.

%% file: Technologies/ASIC.tex
\subsection{Microelectronics}
\label{s.asic} 

The landscape of particle physics in the US is at this moment driven by the P5 report. A common element in the pursuit of the five science drivers is the need for substantial gains in sensitivity, speed, data rate or radiation hardness, to name a few. The study of the Higgs as a tool for discovery at the energy frontier drives the need to process large amounts of data of highly granular detectors with very low power consumption that have to withstand unprecedented levels of radiation. Dark matter detectors are exploring a wide range in mass of dark matter candidates that require ultimate sensitivity. Studies of the early and late universe require integrated solutions of RF frequency de-multiplexing. With liquid argon as the base technology for large-volume neutrino detectors, understanding the nature of the neutrino has led to the development of novel cryogenic electronics. And, exploring the unknown through, for example, experiments at the intensity frontier is requiring ultra-fast timing electronics. The stringent requirements of the high energy physics environment make the off-the-shelf components not usable and demands the development of custom integrated circuits, usually referred to as ASICs (Application Specific Integrated Circuits). 

\subsubsection{Present Status}

Particle physics is a customer of commercial integrated circuit technology. The field is not and cannot be in the business of developing new IC technology. Why then, can high energy physics not simply purchase ASICs from commercial design vendors? As noted earlier, particle physics needs to apply the technology well outside the commercial operating conditions it was intended for. Radiation, cryogenic temperature, extreme reliability, very low power, real time operation, sensitivity to particle ionization, etc. require detailed understanding and continual monitoring of device function that is not included in simulation modules provided by the IC manufacturers. Without in-house expertise and ownership of design in main commercial IC processes, hand-in-hand with the high energy physics use case expertise, it is not possible to produce suitable devices. No vendor could have ever produced the high radiation tolerant chips for ATLAS and CMS, for example, without the many FTE-years of investigation and characterization of radiation effects that have been carried out in the HEP community. 

Partnerships with commercial vendors are possible and even necessary for some applications, for example monolithic pixel detectors, but there is not a general partnership model that can be used in every case. The need for in-house IC design capabilities to enable new science is as important today as it was in the 1980’s.

Many ASICs for high energy physics experiments are designed by a small team in single institutes. This is reasonable for an ASIC not complex enough to warrant partitioning of the design. Yet even in these cases, there is rarely a case where circuit blocks, technology items 
(for example, cold or radiation models or guidelines), and even actual prototypes, are not collaboratively shared, as such sharing increases everyone’s efficiency. And even if nothing at all is shared, it is common to call upon colleagues from other institutes for design reviews. 
Through collaboration, the experience and lessons learned in one ASIC are applied to the design of another. Direct communication between designers is essential for this to happen. What has been described so far is ``design community’’ collaboration, with still one individual institute being responsible for an individual ASIC.

An even more collaborative model is now becoming more prominent, which is that of a design collaboration. Just as large experiments call for multi-institute collaborations, complex ASICs using more advanced technology call for multi-institute design efforts. The 130~nm FE-I4 pixel readout chip, operating in the ATLAS inner layer since 2013 was designed by a collaboration of 12 designers working at 5 institutes and one commercial vendor, in 5 countries. The present RD53 collaboration, now charged with delivering the pixel readout chips in 65~nm for both the ATLAS and CMS upgrades, has about 50 designers from 20 institutes. The readout chips being produced are systems on a chip with 500M transistors, with extensive prototyping of individual blocks and a sophisticated verification flow to generate high confidence submissions that do not require costly iterations.  

While RD53 is presently the extreme example of HEP collaboration, a more general adoption of the design collaboration model at varying scales is expected, as the success of such collaboration clearly shows the benefit of combining expertise and capabilities from various institutes from the start of the design process. The success of design collaborations among many countries also has shown that there are no serious logistical problems with remote IC design collaboration.

\subsubsection{Future Needs}
\medskip
\noindent 
{\bf Colliders} \\
Approximately 20 ASIC designs are presently in advanced stages of development and prototyping and over the coming years will be produced in quantities of up to 50k chips for use in detector upgrades of ATLAS and CMS for HL-LHC. These upgrades will be installed around 2025. The main technologies for these designs are 130~nm and 65~nm CMOS. Impressive progress has been made over the past decade on harnessing the power of these commercial technologies for the special needs of LHC detectors, including of course operation in a high radiation environment. The HL-LHC detector upgrades, particularly for the inner detectors, are only possible because of this success. There would not be a HL-LHC without these technologies and a HEP detector community that has developed the ability to use them. 

The delivery of the HL-LHC upgrades is by no means the end of the ASIC needs for the LHC program, which will continue at least until 2035. Beyond the need to maintain the ability to react to problems, there will be further detector upgrades on a 2030 time scale. The innermost tracking layers in particular suffer the most damage from radiation and at the same time have a large impact on the physics capabilities, making their replacement to recover or increase performance highly desirable. Their small size also makes such upgrades financially feasible The exploration of pixel front-end chip design in 28~nm CMOS is expected for this purpose. Smaller pixel and higher readout rates would boost the physics reach for the second half of the HL-LHC program. 

In addition to the inner layers, data transmission upgrades may also be possible. Radiation-hard, high-speed readout is a limitation of the present designs. Improved data transmission chips may be desirable of even needed to alleviate bottlenecks.

Looking beyond the HL-LHC, candidate colliders rely on the ability to produce ASICs that in many cases have performance requirements beyond those being developed for the HL-LHC.  However, the main research need is to maintain the ability to harness commercial technologies for HEP detectors. Increasing performance will always be desired, but even to simply replicate the HL-LHC detector functionality in the future demands an active IC design community and continued characterization of the favored commercial processes. Many HL-LHC ASICs are in the 130~nm node not because they had to be, but because it is not possible to freeze time and use an IC design process forever. The processes with the most commercial and academic users and the best support will always be the best choice for HEP design, and those will always change with time. 

Beyond mere scaling and preserving design capabilities, several developments over the past decade did not reach sufficient maturity to be used in the HL-LHC upgrades or barely reached the point of enabling experimental subdetectors. These are:
\begin{enumerate}
\item Monolothic pixel sensors for high rate and radiation (as opposed to those in ALICE for lower rate and radiation)
\item Integrated photonics and ASICs for data transmission
\item Readout of position sensitive devices with $\leq$~50~ps time resolution
\end{enumerate}

ASIC R\&D in these directions will continue developing and will lead to superior capabilities for future colliders. These include: lower mass, higher trigger rates and even trigger-less readout, and pixel detectors with fill four-dimensional tracking.

\medskip 
\noindent 
{\bf Neutrino Detectors} \\
The Deep Underground Neutrino Experiment (DUNE) is, by far, the largest U.S. initiative in the Intensity Frontier. An intense neutrino beam from Fermilab is aimed at four 10,000 ton scale cavities, 4850 feet underground in the Sanford Underground Research Facility near Lead, South Dakota. The first of the four caverns will be instrumented with a single-phase LAr TPC and the second cavern may be instrumented by either a single-phase or dual-phase TPC, while the instrumentation of the third and fourth chambers will be decided later. 
Unlike previous Lar TPCs, the DUNE single-phase detector is large enough that there would be a disabling noise penalty if the front-end amplifiers were located outside the cryostat. The Anode Plane Assemblies (APAs) in DUNE single-phase have collection (vertical) wires 6~m long with a lower APA hung below an upper APA. In addition, the cable route out of the membrane cryostat and through the insulation layer is also on the order of 6~m. Thus, even in the best case an extra 12 or more meters of cable (with higher capacitance per meter than the free-standing collection wire) is added to the already significant capacitance of the 6~m collection wire. By placing at least the first stages of the preamplifier inside the cryostat all that load capacitance is removed and, at least in theory, a white noise floor of around 400 electrons can be achieved – remembering that there is also a reduction in the preamplifier first transistor noise at 80~K relative to the transistor noise at room temperature. In addition, if the digitizing electronics and data concatenation logic is also placed at the end of the wires, the cable plant for the ~350,000 channels of a DUNE cavern can be significantly reduced.

Given these incentives, it is natural to investigate building custom integrated circuits to operate in the liquid argon. There are, however, several challenges beyond simply creating a good design. The first problem is that 
“hot carrier effects”~\cite{DUNE-hot}, which can result in charge lodged in an oxide layer of a device and cause the failure rate to increase at cryogenic temperatures. Secondly, commercial IC foundries do not provide any models of device behavior outside the industrial range of roughly -40 to +125~C so reliable simulations of circuit behavior at cryogenic temperatures is not possible. 

The hot carrier problem has been studied for some time and rules for achieving high reliability for long periods have been worked out~\cite{Radeka}. The basic idea is to limit the electric field in the transistor channel to a value significantly less than can be used for reliable operation at room temperatures. This means either reducing the voltage rails, increasing the gate length or both. For custom analog design this just means choosing transistor dimensions with slightly modified rules. For digital design it is necessary to either build a new digital library with increased gate lengths (no minimum length transistors) or use the vendor library but at reduced VDD. Studies of failures of devices and circuits built using these rules and subject to higher than nominal VDD allow projection of operational lifetime at nominal VDD and expected chip lifetimes $> 10^4$ years have been measured. For most purposes we now know how to design devices that will not degrade at cryogenic temperatures due to the hot carrier effect.

However, the lack of reliable models at cryogenic temperatures is not so easily solved. One can take a vendor’s model set and extrapolate the behavior from the vendor’s guaranteed region towards much lower temperatures. That has been tried for some designs, but the behavior of those devices at liquid argon temperatures has not always been well predicted by the extrapolated models. A more difficult but also more effective path is to extract basic device models at liquid nitrogen temperatures (because nitrogen is cheaper and easier to obtain and boils at slightly lower temperature than argon) and then verify those models by comparing simulations and measurements of various complex circuit configurations. Recently such modeling has been done for a 65~nm process~\cite{COLDATA}
and shown to accurately predict circuit behavior.

The design and testing of ASICs for DUNE has been going on for some years. The "near baseline" design of custom CMOS front-end chip and ADC was tested in ProtoDUNE at CERN using a commercial Altera FPGA as the data aggregator. This configuration was shown to have excellent noise performance – in the range of 500 electrons RMS – even for the longer induction wires in ProtoDUNE. However, the ADC used in that test, version 8 of a novel “domino” design showed poor performance in ProtoDUNE (as the previous versions had in the 35ton test at FNAL) and a new design, based on a more standard ADC topology is presently under development. The first version of this ADC was not a complete success apparently because communication between the three groups executing the design was not optimal but a new version should be able to meet most or all of the requirements. In addition, an entirely new ASIC, COLDATA, is being developed for the data aggregation task and the front-end chip, LArASIC, is being re-spun and still shows some bias stability problems with large input charges. 

As a backup, an integrated device including front-end, digitization and data aggregation on a single substrate, the CRYO chip, is also being developed and is under test now with good success in most areas but an unfortunate choice of a flip flop cell must be corrected to achieve required performance on all channels. The baseline and backup designs should provide DUNE with the luxury of being able to select the most effective solution for the first cavern. For the later caverns and, more immediately the Near Detector, a good deal of activity is taking place in trying to replace the “2D” crossed wire plane readout with a pixelated readout giving true “3D” information. Even with a very stringent power limit of ~50~$\mu$W per channel both a fairly conventional charge integrator – discriminator design, the LArPIX and a novel charge counting design, the Q-PIX are both progressing well with version 1 of the LArPIX having already measured tracks in a small test TPC.

For DUNE, well designed custom integrated circuits are an absolute necessity – both to meet the unusual environmental conditions and to satisfy the stringent power and signal to noise requirements. Successful progress will depend upon clever designs, good device models and active participation by several groups. There are grounds for optimism on all counts, but we do not yet have a qualified solution for even the first cavern and much still remains to be accomplished.

\medskip
\noindent 
{\bf Cold Electronics} \\
One of the main drivers for the vast and diverse field of QIS has been the development of qubits and quantum computing. Most qubits operate below 100~mK, while measurement signals used for readout and conditional control are sent through an expensive and meter-scale amplification chain up to room temperature, requiring attenuators and filters between the different stages of refrigeration~\cite{YanFei}. Even for the case of dense frequency-multiplexing, both the number of interconnects and the thermal load to the quantum system scale linearly with the number of qubits. Controlling qubits from room temperature also involves long latencies, which limit the number of computations available for a given qubit coherence time.
Thus, the current approach involves inefficiencies in size, power, cost, and speed that will prove increasingly challenging and soon incompatible with quantum information systems beyond hundreds of qubits~\cite{Andersen}. This limitation is further compounded by the large level of qubit redundancy required by the emerging quantum correction algorithms~\cite{Vander}. 
Hence readout and control electronics able to operate from 4~K down to 100~mK, at the interface between the classical and quantum regimes, could sidestep most of the inefficiencies inherent to the two-way communication~\cite{Dzurak}. In addition to the area of QIS, the characterization and development of cryogenic IC processes and ICs will readily benefit low-noise applications for HEP, dark matter searches and astrophysics, examples of which were given in previous sections. Besides the improvement in thermal noise, amplifying the signal close to the source improves the signal-to-noise ratio.

While specialized technologies such as single-flux-quantum (SFQ) have unique capabilities with mature fabrication facilities as well as tool support~\cite{MIT-LL}, modern CMOS processes have emerged as one of the most promising contenders for cryogenic electronics. Unlike more established cryogenic transistors, such as JFET, HEMT, or superconducting devices based on Josephson junctions, CMOS transistors offer established design automation infrastructure and the possibility to integrate billions of devices on the same chip. MOS transistors are fully functional at cryogenic temperatures, but their performance is different with respect to the standard temperature range. Improvements include increased mobility, and hence larger maximum current, higher sub-threshold slope, lower leakage and lower thermal noise. As a drawback, the threshold voltage increases, thus leading to less voltage headroom, and flicker noise performance and device matching degrades.

Advanced CMOS technologies (approximately 65~nm and below) are not affected by critical cryogenic non-idealities, such as current kink or hysteresis. However, since the models provided by foundries are only valid above -40~C, accurate cryogenic modelling is required even for nanometer CMOS technologies to enable the design of complex cryogenic circuits~\cite{Cryo-cmos}.  
Several CMOS processes have been studied for cryogenic operation (e.g.~\cite{Akturk}). A European collaboration has recently prototyped most of the analog and RF building blocks required for control and readout of large multi-qubit quantum systems directly at 4.2~K~\cite{Cryo-cmos1}. Although several groups have investigated cryogenic ASICs, no equivalent large-scale initiative or consortium exists in the USA. 

The latest generations of SiGe Heterogenous Bipolar Transistors (HBTs) have also been proven to operate below 100~mK with promising performance~\cite{Bardin, Cressler}.
The lack of accurate cryogenic models prevents the exploitation of the latest state-of-the-art IC technology and the resulting performance gain. For example, a recent survey of cryogenic ADCs highlighted how they typically underperform room temperature ADCs by a factor of ~200 in terms of energy efficiency, despite the performance boost potentially available at 4~K due to the use of past node technologies and the unavailability of reliable cryogenic models for their design~\cite{Rotta}. Cryogenic models of various complexity are being developed~\cite{Beckers, Incandela}, but are all process specific (and with the exception of~\cite{Enz}, also temperature specific). Other prominent challenges include the non-interference with sensing elements, and achieving the required low noise without exceeding the tight power budget set by the cooling capacity of the cryogenic environment (which scales exponentially from a few W at 4~K to a few hundred $\mu$W at 100~mK). Because of the sensitive nature of QIS, there are also stringent policies pertaining to intellectual property protection that should be carefully evaluated when charting future road maps.
Specifically, the development of custom ICs for cryogenic applications requires:
\begin{itemize}
\item 
{\bf Process:} it is important that the community has access to established options for cryogenic electronics, but also explores promising developments such as the cryogenic CMOS. In this regard, it is important to have access to advanced, low-power IC process nodes (e.g. 28~nm bulk or SOI, FinFets, RF SiGe HBTs, etc.): an expensive endeavor that justifies a level of coordination across institutes and groups on par to what has been done for irradiation studies and characterization of technologies for the LHC, or for the cryogenic electronics for DUNE. Following an explorative phase where several technologies are evaluated, it might be appropriate to consolidate efforts on a few processes, the long-term availability of which is paramount. A discussion should be established to guide this process in the near future. \item 
{\bf EDA tools:} advanced nodes require advanced (i.e. expensive) design tools, but these are absolutely necessary to design with any confidence. They also require more stringent measures for IP protection, which certainly places a burden on the administrative resources, and on the sharing among groups.
\item 
{\bf Development of custom design kits:} 
    \begin{itemize}
    \item {\bf Device models:} custom model libraries describing the parameters and the behavior of transistors at the desired temperature must be created. These models are process specific. Isothermal models for a target temperature (e.g. LN2, 4K, 2K etc.) should suffice, since it is much more difficult to capture the temperature dependence of many of the main parameters, and given that temperature is very well controlled for all cryogenic applications. Including noise modeling and mismatch analysis would be highly desirable but requires systematic studies.
    \item 
    {\bf Standard cells and IP libraries:} similarly to what was done for DUNE LAr TPCs electronics and for the HL-LHC, the development of custom standard cells libraries with accompanying timing files are a must for automated logic synthesis and design.
    \item 
    {\bf Design methodology:} a community know-how and guidance, identifying for example techniques that avoid premature failures (e.g. due to hot-carrier degradation) and other prominent or novel effects in cryogenic operation (e.g. resilience to increased mismatch as mentioned above)
    \end{itemize}
\item
{\bf Resources:} the learning curve grows exponentially as geometry scales. This translates into additional time and resources.
    \begin{itemize}
    \item {\bf Training:} the engineering community require, for the large part, to be introduced to the peculiar nature and challenges of QIS, and with the rapidly evolving state-of-the-art. There is a recognized need for basic training, as one of the objectives of the National Quantum Initiative Act is to train the workforce required to advance the field of QIS. This could be achieved through workshops and seminars. 
    \item 
    {\bf Testing expertise and equipment:} testing circuits at cryogenic temperatures also requires unique skills and instrumentation. It would be beneficial to establish or strengthen collaboration between groups with such expertise (e.g. dark matter searches, CMB) and ASIC design teams. National labs could play a pivotal role in this.
    \end{itemize}
\item 
{\bf Collaborations:} additionally, a closer integration and collaboration between groups to bridge specific areas of excellence and expertise is envisioned, across agencies and with other international QIS-leading institutions. Examples include RF IC design for qubit control and readout, silicon photonics to couple to waveguides and optical transducers, packaging expertise for good thermal performance, device physics modeling and simulation, etc.
\end{itemize}

In summary, there is exciting potential but also distinctive needs and challenges that should be addressed for leveraging the ASIC design expertise in HEP community for the design of high-density, low noise, state-of-the-art VLSI integrated circuits for cryogenic experiments and QIS. 

\subsubsection{Coordination and HEPIC Workshops} 
With increasing complexity in the design of ICs due to more complex functionalities as well as more challenging technologies, multi-institute design efforts are becoming common practice in the HEP community. Compared to other fields there are established collaborations between groups with skills distributed in a way that rarely overlap. In an effort to increase even more the level of collaboration and coordination among HEP IC groups, a bi-annual series of workshops named HEP-IC bringing together US scientists and engineers involved in developing integrated circuit electronics for particle physics and related applications has been established. During these workshops, participants from US laboratories and Universities have an opportunity to learn about the latest activities and developments of the various groups, trends in IC design and fabrication, and new technologies that are needed to enable future scientific goals. The workshops promote collaborations, exchange of ideas on innovative circuit techniques, and provide a forum to discuss new approaches to partnerships and optimizing the use of technical resources. The last meeting was organized at SLAC in October 2017 bringing together attendees from five National Labs, ten Universities as well as several industry partners. Topics related to novel development efforts, trends and future directions were covered in round-table discussions as well as possibilities of collaborations and partnerships across research agencies and industry. The opportunities for training new generations of designers was addressed as well. During the event seven key findings were identified and recommended to the HEP community:
\begin{itemize}
    \item 
Continue to encourage strong physicist-IC designer links in the US as a vital part of innovation and also emphasize the importance to the educational/training mission. Labs are a great place where physicists and engineers work closely together
\item 
Provide ASIC R\&D to establish knowledge in new technologies. Long term benefit for future advanced developments (e.g. R\&D on 28~nm and 3D-IC)
\item 
Basic literacy on IC technology should be included in the education of physics students to facilitate the communication between physicists and engineers, which is especially true for
analog circuits for detectors.
\item 
Facilitate communication among designers, continue organizing HEP-IC workshops of US IC designers.
\item 
Investigate practical options for a designer at institute A to work a small fraction of time on a project at institute B on which institute A is not involved. Most of the labs have already visiting positions although more communication is required to broadcast opportunities (open positions, fellowship, internships)
\item 
Complete and maintain an up-to-date catalog of existing ASICs
\item 
Explore the feasibility of an inter-institution organization, such as HEPIC.ORG, aiming at removing legal barriers to sharing design blocks, increase communication and foster more collaboration.
\end{itemize}

%% file: Technologies/TDAQ.tex
\subsection{Trigger and Data Acquisition}
\label{s.tdaq} 
\subsubsection{Introduction}
Future HEP accelerators will produce luminosities that will challenge today’s TDAQ designs and implementations. There are plans for an HE-LHC with luminosity around $L = 2\times 10^{35} cm^{-2}s^{-1}$, and for an FCC-hh capable of up to $L = 3\times 10^{35} cm^{-2}s^{-1}$. The field of particle physics needs to develop new trigger and data acquisition technologies to be able to cope with extremely large instantaneous data volume, to optimally mine the data collected, and to alleviate the demands on offline resources. 

The LHC is expected to start the high luminosity (HL-LHC) operation phase in the middle of 2026, to ultimately reach the peak instantaneous luminosity of $L = 7.5\times 10^{34} cm^{-2}s^{-1}$, corresponding to approximately 200 inelastic proton-proton collisions per bunch crossing (pileups), and to deliver more than ten times the integrated luminosity of the Run 1-3 combined.

An important challenge is to extract the data at high bandwidth from the tracking detectors without adding a prohibitive burden of material due to the large number of fiber drivers and the power and cooling they would need. Hermetic detectors, such as CMS and ATLAS provide no path outside of the active region for the data and trigger fibers and drivers which must survive in a harsh radiation and magnetic field environment. Non-hermetic detectors such as LHCb can exploit fiber pathways which do not fall within the active detector region to operate without a Level-1 trigger, instead absorbing a large increase in data volume directly into software-based triggers running on next generation CPUs and GPUs, which by necessity must be some distance from the detector and its radiation and magnetic field environment. In either case, Higher Level Triggers would have to process at least 10x as much data. The overlap of detector front end signals from different collisions and different beam crossings will increase by an order of magnitude, requiring more information (such as the z-vertex from tracking data and potential precise timing information) and much more sophisticated algorithms to separate them out. 

The ALICE and LHCb experiment will be triggerless (i.e., no hardware trigger level) after the Phase-I upgrade. This allows the experiments offloading easily the compute-intensive software tasks, such as track reconstruction, to commodity hardware platforms. However for both ATLAS and CMS experiment, Phase-II upgrade will still rely on the multi-level trigger system comprising both custom hardware and software algorithms on commodity CPUs, to reduce the output rate from 40 MHz down to $\sim$10 kHz for permanent storage. Whether triggerless or not,  Higher Level Triggers would have to process at least 10x as much data as the present LHC detectors. The overlap of detector front end signals from different collisions and different beam crossings will increase by an order of magnitude, requiring more information (such as the z-vertex from tracking data and potential precise timing information) and much more sophisticated algorithms to separate them out. 

If the crossing frequency of the FCC-hh is 25 ns, the anticipated maximum FCC-hh collision pile-up (PU) would be gaussian-distributed with a mean of 1000 and a peak of 1100 at the start of each fill, depending on the $\beta^*$. Although the instantaneous luminosity may drop by half in 2-3 hours, the TDAQ systems must be able to sustain the peak rate. Higher crossing frequencies might alleviate the PU per crossing but may not allow the separation of data from individual crossings. The high occupancy will necessitate high granularity detectors, yielding anticipated large data rates such as about 800 TB/s from the tracking detectors and 200 TB/s from a combined LAr/Scintillator tile calorimeter and higher from a silicon calorimeter. Readout options in such a detector include trigger-less, single level triggered, multi-level triggered and regional readout. If there are no significant advances in radiation tolerant optical fiber technologies, this implies more than one million detector readout optical fibers at 10 Gb/s and a 10 Pb/s event builder network. This will have significant implications for the detector material budget to extract the data and provide power and cooling as well as for the off-detector farm’s requirements for CPUs, power and cooling.

The high anticipated PU will prove very challenging for the key technique of vertex identification for reducing this background with a Level-1 track trigger. At the FCC-hh, the average distance between collision vertices would be 170~$\mu$m in space and 0.5~ps in time, whereas at the HL-LHC, this is 1~mm and 3~ps. For the FCC-hh this distance is less than the multiple scattering in the beampipe at $\eta < 1.7$.  

For future neutrino experiments such as DUNE, and the next generation dark matter experiments such as LZ and SuperCDMS, the trigger requirement may not be as stringent, but the volume of data to be produced will be at PB scale as well, which will require significant DAQ infrastructure and computation resource.

\subsubsection{ASIC}
Custom Application Specific Integrated Circuits (ASICs) are widely used in HEP experiments, mainly for detector front-end readout and trigger electronics as well. These ASICs are designed for the much constrained environment, to deal high radiation (10-100 Mrad or even higher TID, and SEUs), to produce low power dissipation particularly for high channel count detectors, and to reside closer to the sensor with small physical size for better signal/noise performance. The noticeable change in recent years has been the widespread adoption of the commercial deep sub-micron CMOS technology over the specific radiation-hardened process which was the only viable solution for colliding beam experiments.

The TDAQ system of HEP experiments is generally moving towards FPGA based custom hardware if needed and/or software running on commodity PCs, though ASICs may still have advantages for some latency critical tasks. The Hardware Track Trigger (HTT) system in the Phase-II upgrade of ATLAS experiment is designed to use Associative Memory (AM) technology. The AM based pattern matching was demonstrated by CDF experiment successfully and being deployed by the ATLAS Fast TracKer (FTK) project. The HTT aims to provide tracks for the High Level Trigger (HLT) quickly, hence reducing substantially the processing requirements at the HLT. The HTT receives requests to find tracks, along with the related tracker data from the HLT, then performs track finding in two stages. The first-stage processing finds track candidates in two steps, with the first step finding groups of tracker  hits  that  match  precomputed  patterns  stored  in  the AM ASIC, then the second step processing the matched hit combinations with linearized track-fitting algorithms in FPGAs, to extract the tracking parameters.  The second-stage processing takes the track candidates found in the first stage, then extrapolates each track to the additional tracker layers unused in the first stage and performs a full track fit to improve the track parameter resolution. The second-stage processing is implemented in FPGAs. The capacity of the new AM ASIC chip is expected to be three times larger than the one used in the FTK project (3×128k patterns), with a processing speed of 250 MHz words per input bus. The full HTT system will have about 700 ATCA boards and around 1500 mezzanines, holding total of more than 2000 FPGAs and about 14k AM chips. Such a system will be very challenging to design, build, commission and operate.

\subsubsection{FPGA}
The Field Programmable Gate Array (FPGA) is offering breakthrough advantages to address many challenging applications, such as 400G packet processing, wireless remote radio units, data centers, and high performance computing (HPC). The current and next generation FPGA products use the tailored approach to target different applications, with different process technologies, architectures, and integration options. HEP experiments utilize the capacity of powerful FPGAs to tackle time-critical tasks in TDAQ system, particularly in the earlier trigger stages. Custom devices in different formats (ATCA, $\mu$TCA, PCIe, etc) are built with combinations of FPGAs and high speed optical links, to be used in data taking or to be deployed for future experiments. Continuous increase of Configurable Logics, memory, DSP slices, and I/O speed (transceiver up to 32~Gbps) in the FPGAs makes possible rapid decisions by identifying features with sophisticated algorithms on a large fraction of detector data, or even on the full event. System on Chip (SoC) FPGA devices bring even more benefits of flexible implementations, better communications, and easier device control and diagnostics, by providing higher integration, lower power and higher bandwidth communication between the processor and FPGA fabric, and a large set of peripherals.

The development of the Track Trigger at the ATLAS and CMS experiments represents a challenging case. An FPGA-based track finder using a fully time-multiplexed architecture is planned by the CMS experiment for the Level-1 trigger in its Phase-II upgrade. Two firmware approaches are being explored with commercial FPGAs. The tracklet based approach implements the track finding using a traditional road search technique, which is commonly used in software-based track finding algorithms. Another approach reconstructs track candidates using a projective binning algorithm based on the Hough Transform, followed by a combinatorical Kalman Filter. The entire system will require less than 200 ATCA boards, each with a Xilinx Virtex Ultrascale+ FPGA and up to 64 optical links, depending on link speed (16 or 25~Gbps). The full chain functionality has been demonstrated with hardware of ATCA cards, each holding a Xilinx Virtex-7 FPGA and twelve Avago Technologies MiniPODs at up to 12.5~Gbps per link. The demonstrator can process events at 40~MHz, with up to an average of 200 pileups, whilst satisfying the latency requirement of $<$ 5 $\mu$s. The ATLAS experiment is also starting to investigate alternative technologies for HTT. A quick study shows that a system of 24 Global Common Modules (GCMs), which is being designed for the L0 Global Trigger in the Phase-II upgrade, should be able to handle the data throughput foreseen for the HTT system. Each GCM is an ATCA board with two large Virtex Ultralscale+ and one Zynq Ultrascale+ FGPA. But whether such an FPGA-based system is viable needs detailed studies of tracking algorithm firmware on targeted FPGAs, to demonstrate the latency and resource usage. It will still be very time and resource consuming to prepare the data and to tackle the combinatorics, even if the system can be largely parallelized.

High-level synthesis (HLS) becomes increasingly useful for the design of high-performance and energy-efficient heterogeneous systems. It allows designers to work at a higher-level of abstraction by using a software program and create digital hardware RTL code, while hardware implementations can be easily refined and replaced in the target device. This will significantly accelerate the design process and reduce overall verification effort.

\subsubsection{GPU}
The architecture of the GPU was initially driven by the need for 3D graphics for video games, and then used as massively parallel processors for HPC workloads. Currently, GPUs are becoming the engine for machine learning training, particularly Nvidia with the Pascal and Volta GPUs at the Tesla accelerator lines. Furthermore, improving machine learning is driving the changes of GPU technology. With the experience of the $\mu$BooNE experiment, the DUNE experiment has started exploring 3D data reconstruction using deep machine learning on GPUs for the LAr TPC, and is planning to scale it up to use the GPU-based HPCs. 

GPGPU has been used successfully for implementing tracking algorithms by the PANDA experiment in online software and by the ALICE experiment in their HLT system. Utilization of Nvidia and AMD GPGPUs at ALICE has either accelerated the track reconstruction (by about a factor of 10 compared to a CPU-based solution running on four CPU cores), or has introduced a large cost saving. GPGPUs have been studied by the ATLAS experiment as a potential commodity hardware accelerator to which suitable compute-intensive processing tasks (such as tracking in silicon detectors, topological clustering in calorimeters, etc) can be off-loaded from the main CPU in the HLT. Studies based on current technologies (Nvidia Kepler, Pascal) have concluded that GPGPUs and CPUs could be used to increase the throughput of the HLT at roughly the same cost and with similar power and space requirements. The relative cost-effectiveness of these technologies in the future depends on the relative evolution of CPU and GPGPU in terms of price, performance and packaging. The CMS experiment has also investigated the possible software solution on GPGPUs (Nvidia) for the Track Trigger in the Phase-II upgrade. Dedicated computation schemes (spinning kernel) and data transfer controls are being developed to mitigate the normally involved latency overheads. The Hough transformation is used for track finding with respect to the Track Trigger specifications. The benchmark performance does not satisfy the latency requirement but may be of the same order of magnitude as the FPGA approach.

\subsubsection{CPU and Software Improvement}
Besides the collider experiments, the future neutrino experiments, the next generation dark matter experiments, and the next generation observational cosmology instruments will also face significant computing challenges because of the large data volume that is produced and the MC simulation needs. Surveys of the main market trends for hardware show that Moore’s Law has slowed down. The current expectation for annual price-performance improvement over the next few years is about 15\% for CPU servers. Therefore more processing power can still be acquired every year for the same cost, but unlikely at a level enough to match the increase of the needs for the next 8-10 years.

In the HL-LHC phase the need of the experiments is driven by the expected increase in event rate (between a factor of 5 and 10 is foreseen) for both data and Monte Carlo and by the increase in event complexity. ALICE and LHCb will already address these challenges in Run 3. For ATLAS and CMS the earlier naive estimations indicate that approximately 20 times more resources are needed with respect to the currently available ones, though a recent detailed study by ATLAS shows that the factor may be smaller. For the TDAQ system, the situation could be more problematic. For example, all the commodity computing power for the ATLAS trigger system likely needs to be accommodated within a fixed rack-space located in the existing trigger computing infrastructure. An extrapolation based on the evolution of the compute-power in the ATLAS TDAQ farm over the past ten years results in an estimated computing capacity of 1.5 kHS06 per dual-socket server, while without the custom HTT the CPU requirement for HLT tracking under Phase-II conditions is about 40 MHS06. Such a server farm is impossible to be accommodated in the existing infrastructure.

Improvements in the software are certainly expected over the next decade. The HEP community has started R\&D in various areas for reducing the computing needs, such as using enhanced vectorization techniques to better utilize the newer CPUs, using computing frameworks to allow easier utilization of many cores, using accelerators with efficient software on heterogeneous architectures, and modern software technologies like OpenCL.

\subsubsection{Emerging Architectures}
While the performance of CPU/GPU cores and FPGAs will continuously be improved with smaller silicon package sizes (10~nm, 7~nm) by the major vendors (Intel, AMD, Nvidia, Xilinx), the most affordable computing platform (commodity server PCs) is developing towards systems hosting multi-core CPUs with an increasing core-count and heterogeneous hardware architectures. Furthermore, high-capacity, high-speed memory technology will fill gaps in the memory hierarchy to provide bandwidths closer to the silicon die. A range of leading interconnect technologies (PCIe Gen 5, HyperTransport, QuickPath, etc) will increase the scalability and bandwidth available, down to the silicon-level package and die interconnects.

Given the fast evolving market it is difficult to predict what the emerging architecture will be, but it will likely be a diverse mix of scalar, vector, matrix and spatial architectures deployed in CPU, GPU, accelerator and FPGA sockets, enabled by a scalable software stack, integrated into systems by advanced packaging technology. In 2018 Intel unveiled the 3D packaging technology Foveros, which is expected to extend die stacking beyond traditional passive interposers and stacked memory to high-performance logic, such as CPU, graphics and AI processors for the first time. Xilinx has also started extending from the traditional FPGA space and announced the heterogeneous architecture Everest, which is based on the Adaptable Compute Acceleration Platform (ACAP) 7~nm technology, and will include innovations such as the Software Programmable Engine, faster I/O, and a fast on-die fabric. So, we will see hardware platforms having CPUs with GPGPUs and/or FPGAs all integrated. Most importantly, the major vendors all emphasize easy-to-use, high-level developing software frameworks and tools. Therefore, these technologies should be evaluated once they become available, in case a major cost performance breakthrough occurs which would make it worthwhile changing the technologies currently being proposed and demonstrated for the TDAQ architecture of future experiments. 

%% file: Technologies/SC.tex
\subsection{Superconducting Detectors}
\label{s.sc} 

The development of superconducting detectors has flourished over the past decade. The relevant energy scale of superconductivity is $\sim$100 $\mu$eV, allowing devices based on superconductors to open up a new class of experiments that are able to probe low-energy particles that are otherwise inaccessible with alternative technologies. Example applications of superconducting detectors relevant to the HEP directive include low mass dark matter detection experiments, precision measurements of the cosmic microwave background, neutrino  mass  studies  via  single-beta  and  double-beta  decay  spectroscopy, and  hidden photon/axion-like  dark  matter  searches.  For these applications,  superconducting  detectors  often form  the core experimental component and provide a technique that is complementary to other experimental methods.

Superconducting detector technology covers a number of sensor architectures and implementations that have been developed for a wide range of applications. 

\subsubsection{Superconducting detector application: cosmic microwave background}

The clearest example of the unique capabilities of superconducting detector technology is the deployment of the various experiments designed to measure and characterise the cosmic microwave background (CMB). The unparalleled sensitivity provided by superconducting transition edge sensor (TES) bolometers has enabled precise measurements of the anisotropies in both temperature and polarisation, and have set exquisite constraints on the cosmological parameters that form the current picture of how the Universe began and evolved. The focus of the next decade is to detect the smoking gun of the theory of inflation: the signature of inflationary gravitational waves imprinted into the polarisation of the CMB. 

All of the current, ``Stage-3'', experiments are based on arrays of TES bolometers. Radiation is guided to the bolometer through an antenna-coupled transmission line where it is dissipated by a resistor located on the bolometer island. A superconducting film (the TES) with a tuned critical temperature biased in the sharp transition between the superconducting and normal states, serves as an ultra-sensitive thermistor permitting measurement of extremely small temperature fluctuations. The intrinsic sensitivity of an individual TES bolometer now routinely approaches the background photon-noise limit, and to improve the sensitivity for a given experiment, more detectors are required. The evolution of CMB experiments has been swift; within a decade experiments have evolved from tens-of-pixels, single colour, hand assembled detectors, to fully monolithic arrays with thousands of multi-colour, dual-polarisation pixels. Advancing arrays of TESs to such large-scale has been a major driver for the CMB instrument community and has resulted in numerous examples of successfully deployed experiments currently in the field taking data (SPT-3G, AdvACT, BICEP3, CLASS) that demonstrate the substantial advances that have been accomplished in this field.

Multiplexing large numbers of detectors will be critical to future CMB experiments, as well as in other HEP applications. A relatively new technology that has potential applications ranging across a broad band of operating wavelengths is the kinetic inductance detector (KID). Arrays of KIDs offer a promising alternative detector technology that requires minimal cryogenic readout hardware and can offer unparalleled multiplexing ratios that are critical to realising large format detectors arrays for long wavelength applications. KIDs operate by sensing the change in the surface impedance of a superconducting thin-film upon absorption of a photon. The film is patterned into a planar resonant microwave circuit that is used to monitor the impedance of the superconductor, and offers built-in frequency domain multiplexing without the need for complex cold multiplexing components. Continued development of KID technology could pave the way for applications that require massively multiplexed arrays of ultra-sensitive, low-threshold superconducting detectors. 

\subsubsection{Superconducting detector application: dark matter}

Recently, there has been growing interest in exploiting the unique properties of superconducting detector technology to develop a new class of dark-matter detection experiments. Low temperature, conventional superconductors offer the promise of high sensitivity and ultra-low detection thresholds that are unavailable with alternative approaches. As an example, SuperCDMS employs TESs to detect WIMP interactions that occur in a crystal substrate through the measurement of phonons originating from nuclear recoil. Experiments currently under construction will serve as a proof-of-principle demonstration to verify the required sensitivity and detector performance. 

For axion-like DM, superconducting technology has emerged as one of the leading candidates to begin to probe the vast range of unexplored axion-masses. Experiments based on resonant cavities take advantage of the effective gain of the resonant cavity and provide exquisite sensitivity in a narrow band of axion masses. The resonant frequency of the cavity is then tuned to probe a range of axion masses. New experimental concepts, such as the optical haloscope, have been proposed to widen the mass range of axion searches, including both higher and lower masses, beyond the range accessible in resonant cavity experiments. DM interactions are converted to single-photons with a signal power that can be detected using superconducting technologies, and present a promising path forward to realising wide-band axion-like DM searches. 

One candidate technology that is gaining significant interest for applications in direct DM searches of this type is the superconducting nanowire single-photon detector (SNSPDs). SNSPDs are a relatively well established technology, routinely operating at optical/IR wavelengths. A narrow ($\sim$100 nm) superconducting wire is patterned into an absorber and biased with a current close to the superconducting critical current. When a photon is absorbed in the wire, it temporarily suppresses the superconducting state local to the absorption event, causing the wire to become resistive. One of the primary advantages of the SNSPD is the high detector count rate and low dark count rate, both critical for DM experiments. Current research directions continue to improve the detection efficiency at longer operating wavelengths, as well as investigate methods for multiplexed readout.

A number of other detector architectures are currently being proposed for applications in first-generation prototype dark matter experiments. Popular candidates under investigation are TESs, KIDs, and magnetic calorimeters. The next decade will determine which particular technology is most appropriate for a given experimental approach. 

\subsubsection{Superconducting detector needs}

\begin{itemize}
  \item High volume fabrication and quality control of large-format arrays for the next generation CMB-S4 experiment. Coordination of process control, array characterisation and quality assurance between the relevant institutions will be critical to the successful delivery of the CMB-S4 detector payload.
  \item Development and demonstration of the first generation of ultra-sensitive superconducting detectors for detection of ultra-light/wave-like DM particles. 
  \item Continued ‘blue-skies’ development of existing and new superconducting detector technologies to continue to maximise output from prior investment in detector development
\end{itemize}

\subsubsection{Superconducting detector readout technology}

Low temperature detectors, such as superconducting transition edge sensors (TESs), kinetic inductance detectors (KIDs), and magnetic calorimeters (MagCals), superconducting nanowire single-photon detectors (SNSPDs), etc., provide some of the best available noise performance for HEP experiments, allowing the detectors to be optimized for low threshold measurements, low dark count measurements, or simply sensitive measurement of weak signals. In fact, the noise of these detectors has improved to the point where it is often below the background fluctuation of the signals of interest. To perform increasingly sensitive measurements in these cases therefore requires arrays of detectors, forming a pixelated camera or increasing the absorbing volume of the experiment.

For example, the proposed HEP cosmology CMB-S4 experiment is the ambitious extension of a series of smaller experiments: APEX-SZ (320 pixels), SPT-SZ (960), ACT (1,024), Ebex (1,476), Keck array (2,500), Advanced ACT (5,934), SPT-3G (16,140), Simons Observatory ($\sim$70,000, in progress), CMB-S4 ($\sim$500,000, proposed). The scale of this experiment will reduce the detection limits for inflationary gravity waves to r $<$ 0.002, over an order of magnitude improvement below current limits, and measure with high significance the sum of the neutrino masses. This new class of experiments pose the technical challenge of scaling up from tens of thousands to hundreds of thousands or even millions of detectors. And while the detector fabrication on this scale is a challenge, it is straightforward compared to the cryogenic readout. To read out these new arrays without prohibitive heat load and cryogenic complexity demands dramatic improvement in multiplexing techniques.

All experiments using TES detectors use superconducting quantum interference devices (SQUIDs) to amplify their current signals, and multiple mature SQUID readout technologies have been deployed on HEP experiments: SuperCDMS, SPT-3G, etc. Time-division SQUID multiplexing (TDM) is the most mature, having been deployed to roughly 30 experiments around the world. For scale, the multiplexed readout of the sub-mm astronomy SCUBA-2 experiment uses $\sim$1,600 wires for the $\sim$10,000 pixel array, rather than the $\sim$60,000 wires that would be required for non-multiplexed readout. TDM is limited, however, in its ability to scale much beyond 40-fold multiplexing factors.

Frequency-domain multiplexing (FDM, fMux) uses cryogenic circuit elements to place each detector in a distinct MHz resonator, with all resonators in parallel on a common bias line and read out through a common SQUID amplifier. A sum of tones can thus interrogate all detectors on the line simultaneously. FDM has been demonstrated with up to a 68-fold multiplexing factor on the SPT-3G instrument, reading out $\sim$16,000 detectors with $\sim$1,900 wires. While the headline multiplexing factors for MHz FDM are not as large as for the microwave techniques we discuss next, the signals may be carried on twisted pairs rather than coaxial cables, and should not impose any greater heat load on a cryogenic system. In fact, at 200-fold multiplexing factors, the cryogenic complexity of experiments with many tens of thousands of pixels should still be manageable, as demonstrated by SPT-3G, and the technology is well-proven.

To advance the state of FDM to hundreds of channels per twisted-pair requires an increase in the bandwidth available to MHz FDM and a narrowing of the per-channel bandwidth. To achieve this requires understanding and reducing stray impedance in a superconducting circuit extending substantial distance between cryostat temperature stages. Reducing these strays should directly extend the frequency to which resonances can be placed. Reduced parasitics will permit development of MHz FDM with lower resistance ($\sim$200 m$\Omega$) TES bolometers, importantly reducing readout noise, but also narrowing the resonances, which allow more detectors to be multiplexed in the same bandwidth.

An extension of this technique to GHz frequencies began with the development of kinetic inductance detectors (KIDs). The resonators provide natural frequency domain multiplexing as with MHz FDM, but access multiple GHz of bandwidth (e.g. the 4-8 GHz band) available using coaxial cables and broadband components.

The readout of the next generation of TES arrays will also require this bandwidth increase and there must therefore be a vigorous effort to advance microwave frequency multiplexing technology. Microwave SQUID multiplexing allows KID-style multiplexing of TESs by coupling each resonator to a TES through an RF-SQUID. This readout has been demonstrated in multiple labs, for example a fielded 4-coax 256-pixel high rate gamma-ray array operating for over a year at LANL. In preparation for the Simons Observatory, a microwave SQUID multiplexer refurbishment of a section of the BICEP-2 array has recently been fielded to the South Pole and is currently making measurements of the CMB.

\subsubsection{Superconducting detector readout needs}

\begin{itemize}
    \item High volume fabrication and quality control of cryogenic multiplexing architectures for the CMB-S4 experiment. Coordination of process control, array characterisation and quality assurance between the relevant institutions will be critical to the successful delivery of the CMB-S4 detector payload.
    \item Characterisation and on-sky demonstration of the control of the underlying systematics introduced by different readout schemes.
\end{itemize}

\subsubsection{Megapixel superconducting detector arrays}

The requirements on the number of detectors has continued to increase over the past decade, and it is reasonable to expect that detector arrays consisting of order 1 million detectors will soon be called for. Production at such large scale poses a significant challenge for detector fabrication, quality control, and readout technology, and requires preparation and development to begin now to tackle such a formidable increase in experimental capabilities. 
For CMB-S4, detector arrays will be fabricated at close to production-style scale. The delivery of 0.5M detectors will ensure that a combination of universities, national labs, and industrial partners have the expertise and knowledge to scale CMB-S4 for future experiments. Furthermore, on the medium-length timescale of CMB-S4, with continued investment in technology development, new ideas and technologies will mature and have the ability to improve performance, as well as simplify and reduce costs. This continued development is an essential component in engendering enabling technologies for future opportunities within HEP.

Arrays of kinetic inductance detectors are a promising candidate for such large scale arrays. Each detector doubles as a high-Q superconducting microwave resonator which enables in-built frequency multiplexing that eliminates the requirement for any additional cold multiplexing hardware, while exploiting increased readout bandwidth at GHz readout frequencies. Development of KID arrays is on-going, and is being pushed by the sub-mm astronomy community where detector count increases with increasing frequency. A number of sub-mm instruments (MUSCAT, TolTec, BLAST-TNG) are set to deploy KID-based experiments this year, and will provide valuable demonstrations and verifications of the technology. 

To enable the readout of the next generation of TES arrays will also require this bandwidth increase and there must therefore be a vigorous effort to advance microwave frequency multiplexing technology. Microwave SQUID multiplexing is still a relatively new technology. To advance microwave SQUID multiplexing to the level of technical readiness necessary for CMB-S4 and beyond, will require development of multiplexer designs, fabrication processes, and microwave packaging and interconnects to enable robust and repeatable assembly of high-quality, large-scale readout chains for future instruments. In particular, there is a substantial challenge of placing thousands of microwave resonators with frequency accuracy better than one part in ten-thousand, while also achieving internal Q values of better than 200k. Systematic studies of fabrication variability and controllability are needed. Post-fabrication tuning techniques may also be needed, along with wafer-scale screening tools.

While microwave multiplexing of TESs expands the readout bandwidth by 10-100X, the efficiency of utilizing that bandwidth remains quite low. It should be possible to achieve substantial improvement in multiplexing factor in two different regimes: For bolometers, hybridizing the microwave SQUID multiplexer with slower multiplexing techniques should provide readout channels with narrower bandwidth than allowed by the natural limits on resonator quality factors. For microcalorimeters, mitigating crosstalk and compensating for transients at high modulation rates should enable the accurate tracking of faster input signals at higher count rates.

The most banal advances in microwave readout, in the room-temperature de-multiplexing electronics, may actually be the most impactful. Multiple groups across multiple institutions have been working to improve the room-temperature readout of microwave resonators using software-defined radio (SDR). Without reliable, low-noise, low-cost SDR, the proposed large-scale detector arrays will not be practical. SDR is common in industrial applications, but the readout for these HEP applications will require specific features and optimization. For example, dynamically tracking the resonators in a feedback loop benefits the readout of these microwave resonators, maximizing signal-to-noise and minimizing the load on the dynamic range of the broadband components.

Finally, there is an opportunity to leverage telecommunications advances, in which the technology is rapidly getting both better and cheaper. For example, companies are beginning to produce radio-frequency system-on-a-chip devices (RFSoC) in which multiple high-speed digital-to-analog converters (DACs) and analog-to-digital converters (ADCs) are integrated in the same package with a powerful field-programmable gate array (FPGA). Readout with such devices could allow far simpler and cheaper SDR implementations and should be actively pursued by the field.

\subsubsection{Superconducting detectors: tomographic intensity mapping}

As mentioned in Section~\ref{Cosmic}, the concept of a stage~5 cosmology experiment is already being discussed, and it is imperative that the superconducting community begins preparing. Conceptually, such an experiment is currently not well-defined, but one direction that is gaining traction and interest is the measurement of large-scale structure as a function of redshift using tomographic measurement techniques. Tomographic line intensity mapping (IM) ---using a relatively coarse beam to measure a spectral line integrated over many unresolved sources--- is a promising technique for probing large cosmological volumes.  Compared to galaxy surveys which require emission to be above a flux limit, IM measures \emph{all} of the line-emitting sources.  Redshifts are encoded in the frequency dependence of the 3D data cube.  IM with [CII] in particular has been identified as a probe of numerous, faint galaxies that likely powered the EoR, and should be feasible with next-generation instruments.

To access the high-redshift universe, technology development at long wavelengths is critical to future experiments. Ultra-compact, on-chip spectrographs, operating at 100-600 GHz, constructed from superconducting circuits are a promising technology that could enable such a new class of cosmology experiments. The first generation of on-chip spectrographs are currently being deployed and the results of these will be an important milestone and will allow the community to assess the prospects for maturation of this technology development. Continued support for development of such fledgling technologies will be hugely beneficial for future HEP directions and experiments.

%% file: Technologies/QS.tex
\subsection{Quantum Sensors}
\label{s.qs}

There is a general consensus emerging that we are at the dawn of a second quantum revolution. The first quantum revolution enabled the exploitation of the quantum nature of matter to build devices, which has led to a wide range of applications ranging from the development of the laser and transistor to the global positioning system and magnetic resonance imaging. The second quantum revolution is characterized by the ability to engineer large quantum systems to gives control of the quantum nature at the individual level. It provides the opportunity to use modern tools to manipulate and control coherent quantum systems to impact computation, simulation, communication, sensing and measurement.

 Quantum sensing plays a unique and central role in the new quantum revolution. It is providing many of the early successes of QIS and is advancing the development of new quantum devices. Because of the ability to control the full system at the quantum level, quantum sensors allow to go beyond the Standard Quantum Limit (SQL) and provide measurement sensitivity not feasible with conventional methods. Given that many high priority projects, such as the search for dark matter and the search for new physics at the LHC, are reaching their design sensitivities and not uncovering the anticipated breakthroughs, quantum sensing technology has been fully embraced by particle physics and is already having an impact. The adoption of quantum sensors are opening up a range of tabletop physics experiments that require precision measurements that permit exploration of  previously inaccessible regions of parameter space.  

\subsubsection{MAGIS: Dark Matter and Gravitational Wave Detection with Large Scale Atom Interferometry}

Atom interferometers exploit the quantum mechanical properties of matter to make a variety of highly precise measurements~\cite{Kasevich}. By leveraging recent advances in atom interferometry technology~\cite{AI1, AI2, AI3}, the MAGIS collaboration aims to build a 100-meter-tall strontium atom interferometry apparatus (MAGIS-100) to search for ultralight, wavelike dark matter in the mass range 10$^{-22}$~eV to 10$^{-11}$~eV and to serve as a prototype gravitational wave detector in the ‘mid-band’ frequency range of 0.1~Hz to 10~Hz, which is complementary to the frequency ranges probed by LIGO and the planned LISA interferometer~\cite{P-1101}.  Additionally, MAGIS-100 will test the linearity of quantum mechanics for massive particles delocalized over macroscopic distances of up to 10 meters for multiple seconds, a significant improvement over current results~\cite{AI1}.  The planned location for MAGIS-100 is the NuMi access shaft at Fermilab.
MAGIS-100 will be capable of operating in three different modes, which can probe complementary dark matter mass ranges and coupling types~\cite{Graham-AI1, Arvanitaki-AI, Graham-AI2}.   The three detection modes respectively entail measuring time-varying, equivalence principle violating forces in an atom interferometer that simultaneously interrogates two different atomic species (dual species interferometer~\cite{AI3})~\cite{Graham-AI1}; using a pair of atom interferometers separated over a baseline (gradiometer configuration) to measure time-variations of atomic transition frequencies~\cite{Arvanitaki-AI}; and measuring phase shifts between atomic states with different nuclear spins~\cite{Graham-AI2}.  For gravitational wave detection, the gradiometer configuration will be used.
Several research and development directions will be pursued to enhance the sensitivity of MAGIS-100 to dark matter and gravitational waves.  Technologies to be developed include atomic beam splitters with increased momentum transfer, ultra-cold strontium sources with increased atom flux, and spin-squeezed~\cite{Hosten} strontium sources for entanglement-enhanced quantum metrology~\cite{P-1101}.  It is anticipated that MAGIS-100, in combination with these research and development efforts, will pave the way for a kilometer-scale apparatus (MAGIS-1000) that will be capable of detecting known sources of gravitational waves.  In combination with other instruments, such as MIGA~\cite{Canuel}, the MAGIS collaboration plans to contribute to a global network of atomic gravitational wave detectors.

\subsubsection{Searches for new physics with differential optical clock comparisons}

Optical clocks are currently the most stable and accurate timekeepers in the world, with accuracies and precisions equivalent to neither losing nor gaining a second over the entire 14-billion-year age of the universe \cite{ClockReview}. This unprecedented level of precision offers sensitivity to new and exotic physics. Differential comparisons between optical clocks are attractive for this purpose because they mitigate the detrimental impacts of clock laser noise and systematic shifts, which are the dominant limitations on clock stability and accuracy, respectively. 

A unified field theory that can describe all fundamental forces is believed to require Einstein's theories of relativity to fail at some scale, and many of the proposed beyond the Standard Model theories that account for dark energy require modifications to relativity. It is therefore natural to ask whether gravity behaves as Einstein predicted. Optical atomic clocks offer an attractive tool to explore this question thanks to their remarkable frequency precision, and the sensitivity of comparisons between clocks in different frames to relative differences in the passage of proper time \cite{AMOBSMReview}. Differential comparisons between clocks at different altitudes in Earth's gravitational potential and in inertial and accelerating reference frames can be used to constrain modifications to relativity. 

The overwhelming evidence for the existence of dark matter and dark energy and the observed imbalance between matter and antimatter hint at the existence of as yet undiscovered particles and interactions. Isotope shift measurements of atomic clock transitions offer a precision probe of new beyond Standard Model particles. Deviations from linearity in the isotopic frequency shift ratio of two narrow atomic transitions in 3 or more isotopes of the same element can be used to constrain spin-independent couplings of light boson fields to electrons and neutrons without requiring precise calculations of internal atomic structure \cite{IsotopeShift}. In addition, certain proposed dark matter candidates such as ultra-light scalar dark matter and topological dark matter are expected to cause a transient or oscillatory relative frequency shift between two spatially separated clocks, or between two nearby clocks utilizing transitions with different sensitivities to changes in $\alpha$, which would be revealed through a differential clock comparison \cite{AMOBSMReview}.  Included in this last category are comparisons between optical clocks and ultra-stable optical cavities, which have recently shown promise in the search for dark matter in the mass range $10^{-19} - 10^{-15}$~eV~\cite{ClockCavity}.
 

\subsubsection{Quantum engineering of spin systems as detectors of physics beyond the standard model}

Intrinsic spin angular momentum of atoms has been the central concept in a wide range of scientific disciplines, ranging from biology and chemistry to materials science and fundamental physics. Recent progress in the field of quantum information science has produced novel approaches to establishing accurate quantum control over spin systems, and many applications to metrology and sensing. These techniques enable spins to be used as a new generation of sensitive detectors for physics beyond the standard model.

The search for axion dark matter is one of the areas where spin ensembles are a promising detector technology. Axions, originally introduced to resolve the strong CP problem in quantum chromodynamics (QCD), and axion-like particles, are strongly motivated dark matter candidates. Nuclear spins interacting with axion-like dark matter experience a torque, oscillating at the axion Compton frequency. Nuclear magnetic resonance (NMR) and precision magnetometry techniques can be used to search for the effects of this interaction. As an example, the Cosmic Axion Spin Precession Experiments (CASPEr) can search for axion-like dark matter in a wide mass range, with sensitivity at the level of the QCD axion.

To maximize their science potential, spin-based detectors need to consist of macroscopic ensembles of spins sensitive to new physics effects, with high-fidelity experimental control over spin quantum state and quantum coherence, and with accurate readout of spin dynamics. Current experiments use $^{207}$Pb nuclear spins in ferroelectric polar crystals, with enhanced sensitivity to axion-like dark matter. How to improve the fidelity of quantum state preparation of this nuclear spin ensemble is being explored by hyperpolarization using light-induced transient paramagnetic centers, and by using spin coherence properties using pulsed decoupling techniques. Achieving significant improvements in these areas has the potential to expand the reach of experiments, such as the CASPEr search for axion dark matter, by covering a greater range of QCD axion masses.


Trapped, laser-cooled ions are another mature experimental platform with demonstrated quantum control capabilities. Crystals of laser-cooled, trapped ions behave as atomic-scale mechanical oscillators with tunable oscillator modes and high quality factors ($\sim10^6$). Using trapped ions as a quantum sensor can enable detection of extremely weak electric forces ($< 1$~yN) and electric fields ($< 1$~ nV/m), which could be useful in the search for detecting dark matter.

Earlier work has demonstrated the detection of coherently driven amplitudes larger than the zero-point fluctuations of the trapped-ion oscillator~\cite{Biercuk, Sawyer}. More recently, a technique to measure the center-of-mass (COM) motion of a two-dimensional, trapped-ion crystal of $\sim100$~ions with a single measurement sensitivity below the zero-point fluctuations has been demonstrated~\cite{Gilmore}. A time-varying spin-dependent force is employed that couples the amplitude of the COM motion with the internal spin degree of freedom of the ions~\cite{Sawyer}. When the frequency of the spin-dependent force matches the frequency of a driven COM oscillation a spin precession proportional to the driven displacement occurs. Thus, sensing of weak electric fields can be performed by mapping the motion of the ions to their spin state and performing a measurement of this spin state. Using more ions provides an advantage: larger ion numbers means less projection noise and smaller zero-point motion, which can lead to better sensitivity. Because ions are charged particles, they respond sensitively to electric fields – making them an ideal sensor for weak electric fields. 

Previous work demonstrated measurements of displacements as small as 50~pm in $\sim100$~seconds of averaging, 40 times smaller than the ground-state wavefunction size~\cite{Gilmore}. Recent experimental advancements – phase stabilization of the optical potential – have improved this sensitivity and will allow for using techniques such as spin squeezing~\cite{Bohnet} and parametric amplification~\cite{Ge} to make further improvements. Additionally, ground-state cooling via electromagnetically-induced transparency (EIT)~\cite{Jordan} enables performing these measurements resonantly with the COM mode (1.6~MHz) of the ion crystal, where we predict electric field sensitivities less than 0.5~nV/m in a second over a range of frequencies from $\sim50$~kHz to 5~MHz. Electric fields of this size may be produced by some dark matter candidates. In particular, axion and hidden photon dark matter in the neV (MHz) regime has not been experimentally explored at this level.

\subsubsection{Molecules as Quantum Sensors for New Physics}

According to all known physical laws, there should be equal amounts of matter and anti-matter in the universe -- yet there is no free anti-matter whatsoever.  The existence of this asymmetry requires new particles and forces, outside of the Standard Model, that violate CP symmetry.  One of the ways that this unknown physics would manifest itself is as CP-violating electromagnetic moments of fundamental particles, which are amplified in the extremely large electromagnetic fields inside polar molecules.  Searching for these moments using modern, quantum techniques enables exploration of energy scales that now exceed the reach of particle accelerators, and are placing significant restrictions on new CP-violating physics and baryogenesis models.
 
Maximum sensitivity requires the molecules to be prepared in a superposition of quantum states having equal yet opposite effects due to the CP-violating interaction.  This, in turn, requires that the molecules be under quantum control -- not just of their internal states, but their external states (to a degree) in order to avoid decoherence.  Next-generation searches making use of squeezing, interaction engineering, and many body effects will require even finer quantum control, and in more complex systems.  A long-standing goal has been to implement laser cooling and trapping of molecules to achieve full quantum control.  Molecules are very difficult to laser cool compared to atoms, yet are worth the effort due to their significantly increased sensitivity and robustness against systematic effects -- especially in molecules that can be fully polarized in small fields.  Searches with full quantum control could increase sensitivity by orders of magnitude, and reach into the PeV scale  --  well beyond the reach of direct searches.  Additionally, these molecules are sensitive to a variety of exciting physics, including precision electroweak physics, axions, dark matter, and more.

Polyatomic molecules (with at least three atoms) are a promising platform for these next-generation searches~\cite{Kozyryev}. They generically have internal degrees of freedom that enable full polarization in small fields and significant robustness against systematic errors, and suitably-chosen species can be laser cooled -- features which are mutually exclusive in diatomic molecules sensitive to CP-violating physics.  For example, polyatomic molecules containing Yb atoms, such as YbOH, are sensitive not only to the electron electric dipole moment (EDM), but also to the CP-violating nuclear magnetic quadrupole moment (MQM), which is significantly enhanced in the quadrupole-deformed Yb nucleus.  Because the polarizable degrees of freedom relate to the mechanical motion of the polyatomic ligand, we can utilize sensitive and robust molecular states featuring essentially any atom, including those with octupole-deformed nuclei such as Ra, Th, Pa, \textit{etc}., which have extreme sensitivity to CP-violating nuclear Schiff moments.  Atoms without laser-coolable electronic structure, such as Th or Pa, can be built into a laser-coolable species by building molecules additional laser-coolable metal centers~\cite{Rourke}, enabling experiments with quantum control of exotic isotopes.  As quantum control techniques in molecules continue to advance, molecule-based quantum sensors will continue to explore higher and higher energy scales for new fundamental physics.

%% file: Technologies/ML.tex
\subsection{Computing and Machine Learning}
\label{s.ml} 

This section is largely based on the scope of presentations given in the Computing and Machine Learning Parallel sessions at the CPAD 2018  workshop. However, as there are many important efforts and approaches in the extremely wide areas of computing and machine learning in high-energy physics that were not represented at the conference, this report only gives a partial view, and a more limited scope related to detector technology research and development. A more complete picture of research and development efforts in high-energy physics related to computing for the 2020s is given in~\cite{computing}. A good broad overview of machine learning applications, research and development directions in high-energy physics can be found in the High-Energy Physics Machine Learning Community White Paper~\cite{cwp}. 

As detector technology advances, the amount and complexity of information produced by particle physics detectors increases. Therefore, unsurprisingly, particle physics has had a long history of developing both the algorithms and computing devices that can fully utilize this ever-increasing influx of information. In particular, in recent years there have been many examples of machine learning algorithms applied successfully across different types of particle physics experiments. These algorithms learn to model the data and can be used to make predictions. In recent years machine learning algorithms have overtaken more traditional approaches based on expert hand-tuning and found natural applications in a variety of areas in particle physics, such as physics analysis, detector simulation and reconstruction, triggering, data quality monitoring, to name a few. 

These algorithmic developments have been an excellent complement to advances in detector technology, leading to new levels of sensitivity and state-of-the art results with an excellent return-on-investment from this research and development direction. For example, the application of deep learning algorithms by the present neutrino experiments such as NOvA, have produced results equivalent to having 30\% additional exposure, achieving more science per dollar. Similar-level results have been obtained by the LHC experiments and make machine learning and in particular deep learning, a very promising direction to address the physics challenges faced by the high-energy physics experiments in the next decade, such as High-Luminosity LHC and DUNE. 

\subsubsection{Deep Learning for Detector Reconstruction} 

In recent years, there have been many new approaches introduced for detector reconstruction that are based on modern machine learning algorithms, in particular based on deep learning neural networks. We will describe several examples from both the intensity and the energy frontiers. In particular, machine learning algorithms dedicated to image analysis from the field of computer vision have seen major advancements.

Although particle detectors do not produce ``natural" images, they produce information with a well-defined  geometrical relationship. For example, arrays of detectors  capture the time and/or energy of particle hits at particular physical locations. Such data can often be transformed into images,  enabling the use of state-of-the-art computer vision algorithms, such as convolutional neural networks. Such algorithms learn how to represent images as a complex, hierarchical set of patterns and use them to make accurate predictions or decisions.  In the context of particle physics experiments, applications of computer vision might include, but are definitely not limited to, the prediction for the type of particle of interaction occurring in some portion of the detector. For example, local groups of detector hits can be associated to different types of particles or interactions.

Deep learning applications in high-energy physics are not limited to images. Uses of other network architectures have found applications where other symmetries of the problem can be exploited, and in particular where both pattern recognition and timing information is relevant, for example for applications in particle tracking. Algorithms such as recursive, recurrent and graph neural networks have been used extensively for this class of problems. In particular, graph networks, that do not require a regular geometrical representation are a powerful tool for problems where problems do not lend themselves very well to image construction, such as detectors with irregular geometry. At the same time, applications of fully-connected networks are likewise common and produce good baseline results. In what follows we describe several examples of deep learning-based detector reconstruction that showcase their potential.  

\subsubsection{Data Reconstruction using Deep Neural Networks for LArTPCs}

For many current and future neutrino experiments, the detector of choice is a liquid argon time projection chamber (LArTPC). These detectors produce high-resolution images of the trajectory of charged particles which traverse it. Because the data of the LArTPCs can naturally be represented as an image, many of the advances in the field of computer vision can be adapted for event reconstruction. Deep convolutional neural networks are currently applied
to data from experiments with liquid argon time projection chambers.

Convolutional neural networks can be used to classify individual pixels into two classes: those that belong to particles that produce showers and those that belong to particles that produce tracks. The first demonstrations were reliant on simulation but now there are examples using real LArTPC data. In addition to 2-dimensional images from LArTPCs, there is currently research and development work going into LArTPCs whose readout directly provides the location and quantity of charge deposited in the detector in three dimensions. Initial results on simulated images of 3-dimensional convolutional networks performing track and shower classification on individual 3-D voxels were presented at CPAD.

\subsubsection{End-to-end Deep Learning for Particle and Event Identification}
Another promising research and development direction in deep learning detector reconstruction focuses on combining low-level detector representations with modern deep learning algorithms. This approach, termed end-to-end deep learning, starts with the lowest possible detector representations, such as the reconstructed hits and timing information, and trains different deep learning algorithms, ranging from convolutional to graph networks, to identify individual particles and events. The end-to-end deep learning approach to particle identification attempts to recover some of the information lost by more traditional particle identification algorithms that rely on high-level information, such as particle shower-shapes for example. End-to-end deep learning reconstruction has demonstrated state-of-the-art performance in particle identification in the case of electromagnetic showers \cite{e2e1}, quark and gluon jets and entire collision events \cite{e2e2}. End-to-end applications to jet and boosted jet identification were presented at CPAD based on CMS Open Data \cite{opendata}. 

\subsubsection{Identification of Double-beta Decay Events} 

Convolutional neural networks are also useful in the context of liquid scintillator detectors, such as those used to identify signal events
in experiments searching for neutrino-less double beta decay. Studies based on convolutional neural networks show promise in rejecting backgrounds from $^{10}C$ 
(a product of cosmic muon spallation) and for single gammas. 
Instrumentation is advancing for scintillator detectors in several ways. Examples include fast electronics aiming for <100 pico-second timing and LAPPDs aiming for 1 mm spatial resolution. Extensions of applications of convolutional neural networks and other deep learning techniques will likewise be important to take full advantage of the new information these devices will provide. 

\subsubsection{Computational and Real-time Inference Challenges}

A common feature of modern machine learning algorithms is that they often employ highly-parallel computation. This is because they are typically trained on large data sets necessitating the need to
process as much data as quickly as possible. Algorithms employed by high-energy physics experiments are being implemented to run on on a variety of computational acceleration devices such as GPUs, FPGAs, and large CPU and HPC clusters to speed-up the training.

As a result, some algorithms, in addition to being highly effective, can be very fast if executed on the appropriate device. Speeds of certain algorithms are potentially fast enough to be used for triggering along with offline analysis. In the next sections, several such examples focused on computing optimization, event streaming and next generation real-time triggering systems are presented. A forward-looking algorithm to tracking reconstruction algorithm that will use quantum computers is also described.

Though not extensively discussed at CPAD, efforts to employ High-Performance Computing (HPC) centers and big-data techniques from industry are on-going. Such solutions provide experiments access to resources such as GPUs and other resources such as cloud computing and optimize data analysis. 

As the distribution of hits in detectors such as LArTPCs is typically sparse, sub-manifold convolutional neutral networks can be a useful approach to data reconstruction. Sparse submanifold convolutions are a variation from convolutional neural networks optimized for sparse data, initially developed by Facebook research. When applied to sparse data, convolutional neural networks will spend many of their computations on empty regions where the result is zero. By representing the data as a sparse matrix, the number of computations and the amount of memory used can be significantly reduced.

LArTPC data, either in the form of 2-dimensional images or 3-dimensional charged deposits represented as 3-dimenstional voxels, is sparse, making LArTPC data a good candidate for this technique. Studies in training 3-dimensional convolutional networks for voxel-wise classification indicate that the time to train the network is reduced by more than an order of magnitude. The amount of memory used in training is also reduced by a factor of 50, significantly affecting development time. When it comes to inference, the increased speed and memory usage will likewise reduce the time for processing data. This technique is broadly useful beyond LArTPC data to other detector technologies that produce sparse detector hits and where convolutional networks have shown strong potential for particle identification across many experiments.

\subsubsection{Deep Learning on FPGAs for CMS Level-1 Trigger and DAQ}

The High-level Syntesis (HLS) language simplifies the design and translation of high-level algorithmic code into FPGA devices. The HLS4ML effort focuses on implementations of modern deep learning algorithms on FPGAs for high-energy physics applications. Several applications of fully-connected deep networks have been demonstrated that focused on two sample problems of classification (identifying boosted jets) and regression (estimation of muon momentum by the CMS Level-1 endcap muon trigger system). Current implementations require network compression to reduce the number of hyper-parameters and units for application to present FPGAs. Additional model architectures, such as convolution neural networks and graph network implementations are being presently studied.

\subsubsection{Track Reconstruction in High-pileup Collider Environment}

Track reconstruction is expected to be the main consumer of CPU resources at HL- LHC. The cost of running tracking algorithms over the lifetime of HL-LHC is about the same as the cost of tracking detectors themselves. For that reason, a number of on-going efforts in the high-energy physics community are focused on reducing the time needed to identify and reconstruct tracks in the silicon-based detectors. One of the challenges is that conventional tracking algorithms, based on kalman filters, although accurate, do not scale well with the expected levels of pile-up and resulting event complexity in the high-luminosity environment. In this environment, track pattern recognition is combinatorially intensive. 

One of the ways to address this challenge is to adapt the track reconstruction algorithms to take better advantage of computing hardware and hardware accelerators, such as GPUs and FPGAs and take advantage of parallelism whenever possible. Other approaches involve developing faster reconstruction algorithms, for example avoiding costly matrix operations and adaptation of tracker detector design to aid track reconstruction. As an example of the latter, a grouped tracking detector layer layout is expected to reduce computing costs compared to equidistant-layer tracker layouts. Additionally, machine learning methods based on deep learning algorithms applied to track reconstruction have shown good promise to help address the computational challenge \cite{heptrkx}. 

Looking towards the future, quantum pattern recognition promises super-linear speedup for certain algorithms. Specifically, D-Wave Systems Quantum Annealer (QA) finds in constant time the ground state of a Hamiltonian expressed as a Quadratic Unconstrained Binary Optimization (QUBO).
Following \cite{qubo}, the pattern recognition step of track reconstruction can be expressed as a doublet classification problem. In this algorithm, doublets are combined into triplets and quadruplets, allowing to make better pre-selections and use a richer feature set, such as the track curvature in the transverse plane. The QUBO is used as a disambiguation step, selecting the best set of non-conflicting triplets to make track candidates. QUBOs were generated for a subset of the TrackML dataset and solved using qb-solv and a D-Wave 2X. Those early experiments achieved a performance in terms of efficiency, and TrackML score that exceeds 90\% for occupancies up to 6,500 particles/event. The purity, however, starts to drop for occupancies above 3,000 particles/event.

In future work, more performance studies will be run on D-Wave and QA integrated in a production tracking chain. To address the fake rate problem, more physics properties will be incorporated into the definition of the QUBO. Another research direction worth exploring are hybrid QA/Boltzmann machine algorithms in which QUBO weights are learned from data.

\subsubsection{Integrated Research Software Training in HEP}
As advancements are made in both the algorithms and computational devices, the amount of technical expertise required to develop and apply these tools is quickly growing. Such expertise is also needed to evaluate the results from application of advanced algorithms. Therefore, an effort to develop adequate training material must grow as well. 

During the CPAD workshop several community resources and training efforts were presented. For example, an integrated program for research software training was discussed. One way to accelerate development is to coordinate the training in the aspect of various techniques common between experiments. Additionally, an effort to consolidate resources and collaborate more effectively was shown that incorporates public software repositories and containers in addition to simulated public data sets. These efforts and community tools are meant to help groups quickly develop new ideas and compare with other existing approaches. More details about community training can be found in the Community White Paper on Training, Staffing and Careers \cite{comtraining}.  

The LHC and neutrino experiments are presently providing Open Data Sets and Machine Learning competitions to encourage involvement from both the particle physics community and outside as well. Projects such as FIRST-HEP (http://first-hep.org/) and IRIS-HEP (http://iris-hep.org/) have started to help standardize and organize the training of scientists in new algorithms and tools needed to address the challenges of the upcoming decade.  

\subsubsection{Summary}

As the high-energy physics community enters the era of the High-Luminosity LHC and DUNE experiments, the use of advanced algorithms, such as machine learning, is expected to grow and be used to address a number of challenges that arises from both the computational and physics environment of these next-generation experiments. To make optimal use of these advanced algorithms, one of the challenges today is the need for hardware devices and infrastructure that can take advantage of the algorithmic advances being made. It is critical to co-develop both the detector and sensor technology and the software, computing and algorithmic layers that will be used in conjunction to attain the required sensitivity and physics performance of these next experiments.

%% file: Conclusions.tex
\section{Conclusions}
\label{s.conclusions} 

Throughout its history the field of particle physics has developed unique cutting-edge instrumentation in the quest to understand the fundamental nature of energy, matter, space and time . The detectors and the collaborations that design and build them and the high quality data they record about the universe, constitute the experiment. Over the course of the last three decades the success of high energy physics has led to ever more ambitious projects being proposed and built. These projects, for example the LHC, are large and complex and seek to answer some of the highest priority questions in the field. Given the very long timescales and costs involved, these projects have come to dominate the high energy physics program in the US and in other nations. 

We find ourselves in a golden age of discovery with tangible excitement about the opportunities to uncover new worlds. We embark on this adventure of discovery, however, with instrumentation with which it will be difficult to make significant progress in our journey of exploration without significant renewed investments. Accordingly, and as this report indicates, there is now a clear need to rebalance priorities with renewed emphasis on the development of advanced detector technologies.

Several examples illustrate the opportunity space that exists. Quantum sensors based on superconducting detector and readout technology are at the core of a diverse set of high energy  physics experiments and have a broad range of potential applications, including quantum computing, the defense industry, and cosmology. A compelling aspect of high energy physics quantum sensor R\&D is the interplay between improvements in detector technology and  the opening of a new frontier in HEP science. HEP science provides a well-defined framework for quantum sensor development while improvements in quantum technology enables crucial probes of the universe that cannot be realized in other ways.  In many cases, HEP innovation can then be exported to other non-HEP applications. 

The excellent sensitivity of superconducting detectors enables exploration beyond the limits of silicon technology. As such, they are the favored technology for probing particle physics where the characteristic energy depositions are below ionization thresholds, e.g., nuclear recoils from dark matter particles below 100~MeV, axions and other ultra-light dark matter candidates. Beyond HEP, similar detectors and technology can be used to study quantum systems. A prominent goal of the quantum sensor challenge is to increase the sensitivity of superconducting technology through developing ultra-low threshold ($< 20$~mK) dark matter detectors and sub-quantum limited amplifiers for axion and axion-like particle searches. 

There is a broad range of quantum sensors beyond superconducting sensors that can be applied to probe the fundamental forces of nature. For example, atom interferometers and optical clocks provide a most sensitive probe to study deviations from general relativity and time variation of  fundamental constants. Spin ensembles of atoms can be quantum engineered to provide ultra-sensitive probes of very weak interactions, such as those originating from dark matter.

Photo-sensors are integral to almost every large detector in HEP, from dark matter and neutrino experiments to collider detectors at the LHC. In addition, photo-sensors are widely used outside of HEP thus advances made in this realm tend to have direct and immediate uses in medicine, industry, and other scientific fields. It is often the case that HEP catalyzes industrial R\&D for the development of new instruments, providing expertise and testing facilities in addition to an initial market to help defray R\&D costs. Photo-sensor  performance is characterized by timing resolution, spectral sensitivity, efficiency, radio-purity and cost. Improvement in each of these areas will have enormous impact on a wide spectrum of particle physics experiments. 

Future HEP accelerators will produce luminosities that will challenge current TDAQ designs and implementations. This is already becoming apparent with the highly successful recently completed Run 2 of the LHC. The HL-LHC will have luminosities around $7.5\times10^{34}$, and there are plans for an FCC-hh capable of $3\times10^{35}$. The field of particle physics needs to develop new trigger and data acquisition technologies to be able to cope with extremely large instantaneous  data volumes, to optimally mine the data collected, and to alleviate the demands on offline resources. A key challenge to demonstrate the individual technologies needed in a realistic environment would be a first step towards developing a complete system. 

A perennial problem for particle and nuclear physics experiments is the detrimental effect of the mass and power budget required for a high-precision detector, notably silicon vertex and tracking detectors. CMOS MAPS sensors have several advantages over traditional heterogeneous technologies with separate sensors and readout ASICs. The built-in amplifiers allow a reduction in signal, and therefore sensor thickness. Small feature size technologies make a fine readout pitch possible. The sensor periphery can accommodate a digital circuit for fast data readout and processing in-situ. The data can be serialized for readout, which drastically reduces the interconnect requirements. Instead of bump-bonding or wire-bonding every channel on a sensor, just a few wire-bonds for the data bus connection are sufficient. Furthermore, the CMOS devices are made at standard commercial foundries and therefore can be produced quickly and inexpensively. The integrated nature of the sensors and reduction in the number of interconnects facilitate a large tracker construction in a very short time at significantly reduced cost compared to current trackers. 

High resolution timing information is a powerful discriminant in many experiments. In the context of the LHC experiments, timing will provide a powerful handle to address pile-up. Particle identification is also aided by timing information up to high momenta. Timing signals enable vertex reconstruction in neutrino interactions and dark matter experiments. Having
channel-by-channel timing information available with sub-10 picosecond resolution would be disruptive. The technological challenges to achieve this level of precision are formidable. Without specifying the technology, achieving pico-second timing at the system level is a key challenge for the field. 

Bringing the modern computing and algorithmic advances from the field of machine learning from offline applications to online operations and trigger systems is another major challenge. There are examples of relatively simple machine learning algorithms implemented at both the hardware and software trigger levels of several experiments at the energy and intensity frontiers. However, more powerful algorithms based on deep learning have become the state-of-the art in a number of fields and applications, and bringing them to the trigger decision-making logic, including implementation on FPGAs and other accelerators that allow for the necessary trigger latency requirements, is one of the major challenges, and also opportunities, that the high-energy physics community today should exploit in the next several years. 

The representative examples outlined above are just a handful of the many examples of the need for renewed investment in instrumentation research for particle physics. A Basic Research Needs Workshop would be a most appropriate next step in setting the strategy for future investments in detector technology  development. 

%% file: ack.tex
\section{Acknowledgements}

We gratefully acknowledge Paul Grannis and Abe Seiden for their advice and careful reading of this report.

%% file: biblio.tex
\providecommand{\href}[2]{#2}
\begingroup
\raggedright

\endgroup